%% file: 000_wbsig53.tex
\documentclass{article} 
\usepackage{iclr2023_conference,times}

\input{math_commands.tex}

\usepackage{hyperref}
\usepackage{url}
\usepackage{graphicx}
\usepackage{authblk}
\usepackage{microtype}
\usepackage{graphicx}
\usepackage{subfig}
\usepackage{booktabs} 
\usepackage{dblfloatfix}
\usepackage{color, colortbl}
\definecolor{LightCyan}{rgb}{0.88,1,1}
\usepackage[capitalize,noabbrev]{cleveref}

\title{Large Scale Radio Frequency Wideband \\Signal Detection \& Recognition}

\author[1]{Luke Boegner}
\author[1]{Garrett Vanhoy}
\author[2]{Phillip Vallance}
\author[3]{Manbir Gulati}
\author[1]{Dresden Feitzinger}
\author[2]{Bradley Comar}
\author[1]{Robert D. Miller}
\affil[1]{Peraton Labs}
\affil[2]{Laboratory for Telecommunication Sciences}
\affil[3]{Applied Insight}
\affil[ ]{}
\affil[ ]{\textit{\{luke.boegner,gvanhoy,dresden.feitzinger,rmiller\}@peratonlabs.com}}
\affil[ ]{\textit{\{pvallance,bcomar\}@ltsnet.net}}

\iclrfinalcopy 
\begin{document}

\maketitle

\input{00_abstract}
\input{01_introduction}
\input{02_prior_work}
\input{03_dataset}
\input{04_neural_networks}
\input{05_signal_detection}
\input{06_signal_recognition}
\input{07_conclusion}
\input{09_acknowledgements}
\clearpage

\bibliographystyle{iclr2023_conference}
\bibliography{wbsig53}

\clearpage
\appendix
\section{Appendix}
\input{10_data_appendix}
\input{11_tools_appendix}
\input{12_model_appendix}

\end{document}

%% file: math_commands.tex
\usepackage{amsmath,amsfonts,bm}

\def\eqref#1{equation~\ref{#1}}

\def\1{\bm{1}}

\DeclareMathAlphabet{\mathsfit}{\encodingdefault}{\sfdefault}{m}{sl}
\SetMathAlphabet{\mathsfit}{bold}{\encodingdefault}{\sfdefault}{bx}{n}



%% file: 00_abstract.tex
\begin{abstract}
    Applications of deep learning to the radio frequency (RF) domain have largely concentrated on the task of narrowband signal classification
    after the signals of interest have already been detected and extracted from a wideband capture. 
    To encourage broader research with wideband operations, we introduce the WidebandSig53 (WBSig53) dataset which consists of 550 thousand synthetically-generated samples 
    from 53 different signal classes containing approximately 2 million unique signals. 
    We extend the TorchSig signal processing machine learning toolkit for open-source and customizable generation, augmentation, and processing of the WBSig53 dataset. 
    We conduct experiments using state of the art (SoTA) convolutional neural networks and transformers with the WBSig53 dataset.
    We investigate the performance of signal detection tasks, i.e. detect the presence, time, and frequency of all signals present in the input data, 
    as well as the performance of signal recognition tasks, where networks detect the presence, time, frequency, and modulation family of all signals present in the input data.
    Two main approaches to these tasks are evaluated with segmentation networks and object detection networks operating on complex input spectrograms. 
    Finally, we conduct comparative analysis of the various approaches in terms of the networks' mean average precision, mean average recall, and the speed of inference\footnote[1]{
        The TorchSig toolkit is available at \href{https://github.com/torchdsp/torchsig}{https://github.com/torchdsp/torchsig} with additional documentation available at \href{https://torchsig.com}{https://torchsig.com}.
        The WBSig53 dataset can be generated using the TorchSig toolkit.
    }.
\end{abstract}

%% file: 01_introduction.tex
\section{Introduction}
\label{sec:introduction}

Radio frequency machine learning (RFML) has brought about innovative solutions to the problem of modulation classification \citep{o2016radio}, \citep{o2018radio}, \citep{boegner2022sig53}.
These efforts provide a foundation for training ML models to classify signals in an environment where the center frequency and bandwidth of 
the signal is known \textit{a-priori} but do not address the significantly more difficult context of a wideband environment.   
In a wideband environment, a signal must first be detected, localized in the time-frequency domain, and then finally classified. 
What makes this problem more difficult than the narrowband case is the presence of many other potentially interfering signals, 
receiver impairments such as dynamic range that are not as prominent in a narrowband environment, and the sheer amount of data to be processed. 
Attempting to detect and classify potentially thousands of signal bursts per second coming from numerous types of receivers in real-time remains 
an open engineering problem even when considering classical detection methods.
A key limiter of more research in this area is the lack of openly available datasets by which researchers can more objectively compare their results.
In this work, we introduce WBSig53, a benchmark dataset that includes many peculiarities of the wideband signal detection problem and initial results using 
state of the art (SoTA) deep learning models adapted to the RF domain.

\cref{sec:prior_work} of this paper discusses prior RFML work and datasets for signal detection and recognition. 
\cref{sec:dataset} introduces the WBSig53 dataset and the TorchSig RFML software toolkit.
\cref{sec:neural_networks} describes the approaches and neural networks used in the experiments conducted below.
\cref{sec:signal_detection} shares initial experimentation and performance of various neural networks on the WBSig53 dataset for the task of signal detection, while 
\cref{sec:signal_recognition} addresses the task of signal recognition.
The paper then concludes with \cref{sec:conclusion}.

%% file: 02_prior_work.tex
\section{Prior Work}
\label{sec:prior_work}

In \cite{wong2021rfml}, researchers provide an excellent overview of the complexities of the RFML ecosystem. 
In particular, they describe the major areas of overlap and voids in the broader RFML research space.  
Notably, they highlight the importance of \emph{data} towards learned behavior and describe the intricacies of open and custom RFML datasets.

In \cite{vagollari2021jointdet}, wideband spectrograms are used to perform detection, localization, and classification.  
Object detection is performed via YOLO adaptations using simulated datasets that cover analog and digital modulations.  
The classes are somewhat limited,
and the authors group PSK and QAM into the PAM class due to their (1) reliance on spectrograms as the network input and (2) spectrogram similarity.  
The researchers do include some channel variations, albeit in a limited fashion.

As part of the IEEE SPAWC2021 Challenge, 
researchers developed and released a wideband dataset \citep{west2021spawc21dataset} to spur competition and promote research in this area.  
The dataset was synthetically generated, covering $130$ unique band layouts and $14$ modulations with some emulated real-world channel effects.  
\cite{west2021widebandseg} describes the dataset and also presents and analyzes a U-Net approach to the problem \citep{ronneberger2015unet}.  

Another wideband dataset was discussed in \cite{nguyen2021wrist} 
covering emissions from real devices using five signal protocols (Bluetooth, Lightbridge, Wi-Fi, XPD, and Zigbee).  
The data was collected at $100$ MHz and expertly labeled, yielding $1.4$ TBytes of RFML data for the community. 
A generalized YOLO approach was also presented.

%% file: 03_dataset.tex
\section{WidebandSig53 Dataset}
\label{sec:dataset}

The WidebandSig53 (WBSig53) dataset is envisioned to facilitate the advancement of SoTA signal detection and recognition techniques in the RFML domain.
Given the vastly diverse nature of signal types, wireless environments, and transmitter/receiver impairments in the real world, 
it is impossible to fully capture all possible distributions within a tractable dataset.
However, the WBSig53 dataset incorporates many of the significant, representative challenges observed during the practical application of signal detection and recognition.
When a model performs well on the WBSig53 dataset, ideally, it should also be able to successfully transfer-learn to a real world solution.
This concept mirrors how computer vision object detection and segmentation models can be pre-trained on the COCO dataset \citep{lin2014coco} and serve as a solid foundation for future, specific applications.
WBSig53 fills a gap in this research area by providing the following:

\textbf{Open-Source Dataset Generation:} While several efforts highlighted in \cref{sec:prior_work} introduced wideband signals datasets,
they focus on specific applications and provide limited scale, flexibility, and/or reproducibility.
By introducing WBSig53 as a synthetically generated dataset,
we provide the community with a common benchmark for collaborative exploration in a large scale manner that is easily \emph{extensible}, \emph{reviewable}, and \emph{reproducible}.

\textbf{Large Signal Diversity:} Many works in the wideband signal detection and recognition space have a limited number of signal classes.
While this is likely intentional due to the complexity of signal recognition tasks, 
we provide 53 unique signal classes in the WBSig53 dataset to maximize future research potential.
We also provide target transforms within the TorchSig toolkit 
such that the difficult task of signal recognition over 53 classes can be easily mapped down to fewer classes by means of modulation family groupings or all the way down to a signal detection task with a single ``signal'' class.

\textbf{Impairment Diversity:} Past work in \cite{miller2019policy}, \cite{scholl2022rfaugs}, and \cite{boegner2022sig53} have all shown the benefits of synthetic impairments to RFML training data.
During the WBSig53 dataset generation, the complex-valued samples are impaired by applying a diversified subset of 11 emulated real world RF impairments.

The WBSig53 dataset consists of 550 thousand examples containing $\sim$2 million signals split into 4 distinct sub-datasets:
\emph{Clean Training} (250k examples),
\emph{Clean Validation} (25k examples),
\emph{Impaired Training} (250k examples), and
\emph{Impaired Validation} (25k examples).

The WBSig53 sub-datasets split training and validation examples, such that the validation sets can be used collaboratively and competitively across the RFML community.
This is similar to the computer vision community's use of the COCO dataset (and corresponding challenge) as a benchmark.
The WBSig53 dataset also provides training and validation datasets in both clean and impaired formats. 
Impaired examples are passed through realistic RF impairments as defined below, while the clean examples enable researchers to apply customized impairments tailored to their use case.

The default example size across all sub-datasets is $262,144$ complex-valued samples, representing I (in-phase or real) and Q (quadrature or imaginary) samples.
This size enables the use of a $512$-point FFT operation (with similar step size for no overlap) in the creation of square $512 \times 512$ complex-valued spectrograms.
This size spectrogram can be used with many modern ML techniques used in the vision domain while still encompassing a large amount of relevant signal data.

\begin{figure}[t]
    \setlength\belowcaptionskip{-1.0\baselineskip}
    \centering
    \begin{tabular}{lr}
        \subfloat[Clean Dataset]{\label{fig:clean-wbsig53-dataset}
        \includegraphics[width=.30\textwidth]{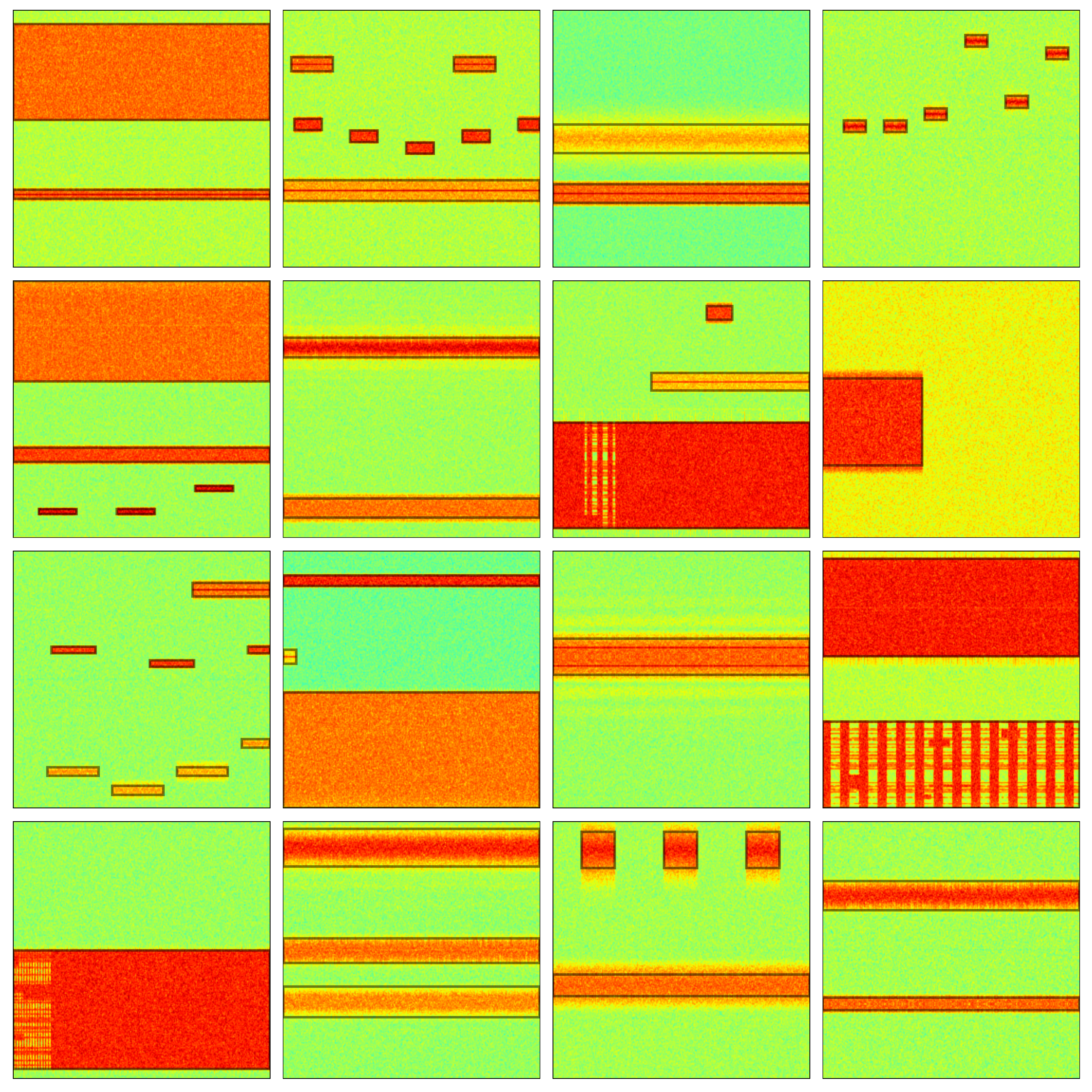}} &
        \subfloat[Impaired Dataset]{\label{fig:impaired-wbsig53-dataset}
        \includegraphics[width=.30\textwidth]{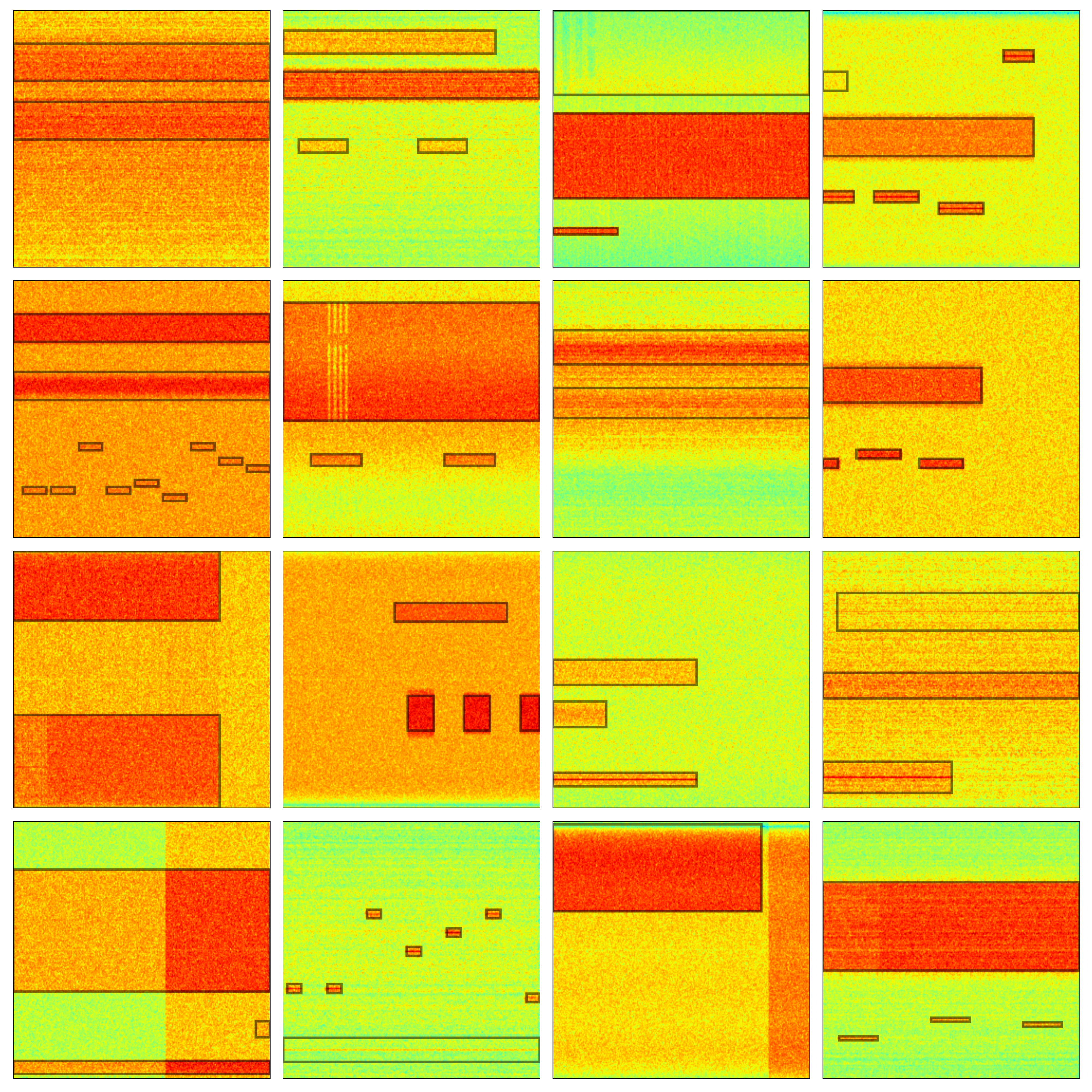}} \\
    \end{tabular}
    \caption{
        Spectrograms of examples from the clean and impaired WBSig53 datasets.
        } 
    \label{fig:wbsig53_dataset}
\end{figure}

\subsection{Clean WBSig53 Dataset}
\label{sec:clean-dataset}
The clean datasets randomly select $1$ to $6$ signal sources.
Each signal source is randomly given a signal-to-noise ratio (SNR) in the range of $20$ to $40$dB $E_s / N_0$ (energy per symbol to noise power spectral density).
The type of signal for each source is randomly selected from the full list of signal classes in \cref{tab:class-list}.
The modulation families within the list include: amplitude-shift keying (ASK), pulse-amplitude modulation (PAM),
phase-shift keying (PSK), quadrature amplitude modulation (QAM), frequency-shift keying (FSK), and orthogonal frequency-division multiplexing (OFDM).

\subsection{Impaired WBSig53 Dataset}
\label{sec:impaired-dataset}
The impaired datasets impose synthetic impairments emulating environmental effects and real-world system impairments.
Prior to any impairments, signal sources' SNRs change to be uniformly distributed between $0$ and $30$db $E_s / N_0$.

The following 5 impairments are applied to full data examples with parenthetically specified likelihoods\footnote[2]{
    In wideband multi-signal scenarios, there are typically different transmitter and wireless channel impairments for different signal sources.
    We currently limit the impairments to occur on the full data example, affecting all signal sources equally.
    We suspect this still encompasses a sufficiently challenging task 
    and models trained using this subset of realism should still provide an equally valuable baseline for transfer learning.
}:  \emph{Time Shift} (25\%), \emph{Frequency Shift} (25\%), \emph{Random Resampling} (25\%), \emph{Spectral Inversion} (50\%), and \emph{Additive White Gaussian Noise} (100\%).  

To introduce additional impairment diversity, we also use a modified RandAugment transform \citep{cubuk2020randaugment}.
Our RandAugment transform inputs 7 additional impairments and randomly selects 2 of them to apply for each data example.
These 7 additional impairments are:  \emph{Magnitude Rescaling}, \emph{RF Roll-Off}, \emph{Random Convolve}, \emph{Rayleigh Fading Channel}, \emph{Drop Samples}, \emph{Phase Shift}, and \emph{IQ Imbalance}.

Spectrograms of a subset of WBSig53 clean and impaired data can be seen in \cref{fig:wbsig53_dataset}.
More details and visualizations of these impairments and additional augmentations are shown in \cref{sec:appendix_dataset} and \cref{sec:appendix_tools}, respectively.

%% file: 04_neural_networks.tex
\section{Neural Networks}
\label{sec:neural_networks}

Signal presence detection and localization (i.e. time and/or frequency estimation) have been explored using signal processing-based techniques,
such as in \cite{spooner2007bandofinterest}, \cite{olivieri2005scalable}, and \cite{vartiainen2010two}.
A more comprehensive survey is discussed in \cite{yucek2009survey}.
While there exists opportunities to explore signal processing-based techniques using the WBSig53 dataset, 
our work is scoped to the exploration of neural networks' signal detection and recognition abilities.

We construct the tasks of signal detection and recognition with an object detection and a segmentation approach.
Within each approach, we experiment with one neural architecture that is convolutional-based and one that is transformer-based \citep{vaswani2017attention}.

\subsection{Object Detection}
Object detection algorithms are a class of ML techniques that directly infer bounding boxes and classes given an input image.
These algorithms lend themselves well to signal detection and recognition tasks because most signals (and all signals currently within the WBSig53 dataset) are rectangular in time and frequency.
Within this class of algorithms, we choose to explore YOLOv5 \citep{jocher2022yolov5} as a fast, convolutional-based approach
and DETR \citep{carion2020detr} as a more powerful, transformer-based approach.

With YOLOv5, we experiment with the YOLOv5-nano and YOLOv5-small scales.
We also introduce a further scaled down version, which we term YOLOv5-pico.
With DETR, we modify the architecture slightly by using EfficientNet \citep{tan2019efficientnet} backbones and an XCiT transformer \citep{el2021xcit}.
We make these changes for data efficiency and increased model inference speed.
For all DETR experiments, we use the smallest scale XCiT-nano.
For the backbones, we vary scale by using EfficientNet-B0, B2, and B4.
We term these models DETR-B0-Nano, DETR-B2-Nano, and DETR-B4-Nano, respectively.

\subsection{Segmentation}
\begin{figure}[t]
    \centering
    {\includegraphics[width=0.7\textwidth]{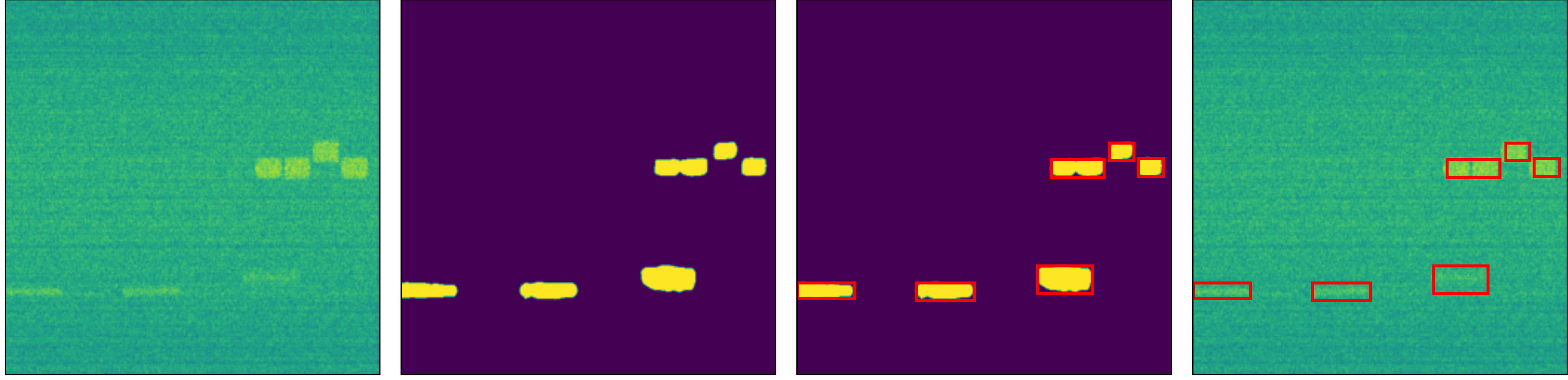}}
    \caption{Segmentation masks are converted to boxes for signal predictions.}
    \label{fig:mask-to-box}
\end{figure}

Segmentation algorithms are a class of ML techniques that infer object instance and/or class associations pixel-by-pixel in an input image.
Within segmentation algorithms, there are three common subcategories: semantic, instance, and panoptic segmentation.
Semantic segmentation infers pixel-by-pixel class labels, where overlapping classes are left ambiguous.
Instance segmentation infers pixel-by-pixel object instance labels, where overlapping objects of the same class are disambiguated;
however, there is no class information associated with each instance.
Panoptic segmentation combines semantic and instance segmentation, where all instances and class information for each instance are inferred.
In the WBSig53 dataset, there are no overlapping signals in both time and frequency,
so we focus on the task of semantic segmentation for both signal detection and recognition.
We impose a naive post-processing step that converts the masks to bounding boxes (\cref{fig:mask-to-box}).
This allows us to report signals in familiar time-frequency values
and easily compare segmentation models to object detection models.
Within this class of algorithms, we explore PSPNet \citep{zhao2016pspnet} as a convolutional-based approach, 
and Mask2Former \citep{cheng2021mask2former} as a more powerful, transformer-based approach.

With PSPNet, we modify the original architecture slightly by using EfficientNet as our encoders at three distinct scales: EfficientNet-B0, B2, and B4.
We term these models PSPNet-B0, PSPNet-B2, and PSPNet-B4, respectively.
Similarly with Mask2Former, we use EfficientNet backbones at the same scales.
We term these models Mask2Former-B0, Mask2Former-B2, and Mask2Former-B4, respectively.

%% file: 05_signal_detection.tex
\section{Signal Detection Experiments}
\label{sec:signal_detection} 

In this section, we present results using the previously introduced neural networks to perform signal detection on the WBSig53 impaired dataset. 
We define signal detection as the ability of an algorithm to detect the presence of any and all signals within an input sample,  
as well as discern the location in time and frequency of each signal. 
This is analogous to object detection in the vision domain, where instead of inferring $(x,y)$-coordinates of objects within an image,  we infer ${(t,f)}$-coordinates of signals within a data capture. 
 
To better leverage research from the vision domain,  
we frame our tasks as closely to object detection as possible by applying a complex-valued spectrogram transform to WBSig53's complex-valued IQ samples. 
This transform is applied with an FFT size of $512$, a segment length of $512$, no overlap, and a Blackman-Harris window,  
resulting in a critically sampled complex-valued spectrogram of size $(512,512)$. 
As a final step, the complex values are split into real and imaginary components and concatenated in the channel dimension,  
similar to how RGB pixels are handled in the vision domain. 
The resulting shape of our input samples is then $(2,512,512)$. 
With this shape, many vision-based networks can be leveraged with a simple modification to the channel dimension from $3$ (RGB) down to $2$ (real/imaginary). 
 
In addition to the spectrogram data transform,  
we also apply a target transform that maps the WBSig53 signal labels to each neural networks' expected target format. 
In our signal detection experiments, these target transforms map all $53$ signal classes contained within WBSig53 to a single ``signal'' class. 
In the object detector approaches (DETR and YOLOv5), the transformed targets are bounding box labels. 
In the vision domain, these bounding box labels are often represented as $(x_c, y_c, w, h)$, where $x_c$ represents the center $x$ position, 
$y_c$ represents the center $y$ position, $w$ represents the width of the box, and $h$ represents the height of the box. 
We follow a similar convention; however, our bounding box labels are represented as $(t_c, f_c, d, B)$, 
where $t_c$ represents the center time, $f_c$ represents the center frequency, $d$ represents the signal duration, and $B$ represents the signals' bandwidth. 
In the case of the segmentation approaches (PSPNet and Mask2Former), the transformed targets are semantic masks  
where all areas containing signals in the input sample are labeled with $1$ and all areas outside these signals are labeled with $0$, 
forming a pixel-by-pixel label of all signal instances as masks within a $(512,512)$-sized matrix. 

\begin{table}[t]
    \setlength\abovecaptionskip{-0.7\baselineskip}
    \caption[fontsize=10pt]{
        Signal detection results for various networks using the WBSig53 impaired dataset.
        YOLOv5-pico is the fastest, while DETR-B4-Nano is the best performer.
        Throughput is measured in samples per second (SPS) on a single Nvidia V100 GPU.
        }
    \label{tab:wbsig53-det-results}
    \vskip 0.1in
        \begin{center}
            \centering
            \small
            \begin{tabular}{l|lllllllll}
                \toprule[1.5pt]
                Model & Params & SPS & mAP & AP$_{50}$ & AP$_{75}$ & AP$_{S}$ & AP$_{M}$ & AP$_{L}$ & mAR \\ \midrule
                \rowcolor{LightCyan}YOLOv5-pico & 0.32 M & 1335.21 M & 73.03 & 87.00 & 78.81 & 67.28 & 77.56 & 68.64 & 75.17 \\
                YOLOv5-nano & 1.8 M & 796.35 M & 73.82 & 86.99 & 79.97 & 68.82 & 78.29 & 69.29 & 75.92 \\
                YOLOv5-small & 7.0 M & 432.61 M & 73.64 & 86.98 & 79.94 & 68.86 & 78.09 & 68.92 & 75.78 \\ \midrule
                DETR-B0-Nano & 8.2 M & 161.21 M & 85.49 & 97.93 & 92.24 & 73.67 & 84.62 & 90.92 & 89.62 \\
                DETR-B2-Nano & 11.9 M & 119.17 M & 85.83 & 97.93 & 92.22 & 74.47 & 85.12 & 90.98 & 90.06 \\
                \rowcolor{LightCyan}DETR-B4-Nano & 21.9 M & 74.70 M & 86.98 & 98.92 & 93.35 & 75.99 & 86.01 & 91.54 & 90.90 \\ \midrule
                PSPNet-B0 & 4.1 M & 215.54 M & 68.52 & 92.65 & 73.54 & 39.32 & 62.66 & 86.34 & 70.89 \\
                PSPNet-B2 & 7.8 M & 155.07 M & 72.30 & 93.76 & 77.37 & 44.42 & 68.01 & 88.18 & 74.44 \\
                PSPNet-B4 & 17.6 M & 97.03 M & 73.59 & 93.83 & 79.42 & 46.79 & 69.68 & 88.81 & 75.79 \\ \midrule
                Mask2Former-B0 & 22.2 M & 19.28 M & 71.97 & 84.48 & 78.89 & 62.12 & 65.75 & 84.13 & 78.86 \\
                Mask2Former-B2 & 26.0 M & 18.47 M & 77.29 & 90.77 & 84.98 & 61.76 & 73.30 & 92.40 & 82.45 \\
                Mask2Former-B4 & 36.3 M & 16.87 M & 81.05 & 94.31 & 89.25 & 64.16 & 78.64 & 91.15 & 85.35 \\
            \bottomrule[1.5pt]
            \end{tabular}
        \end{center}
    \vskip -0.1in
\end{table}

For evaluation, we adopt the Mean Average Precision (mAP) metric. 
mAP is a widely-used metric within object detection tasks in the vision domain due to its ability to quantify performance across classification and localization, 
while jointly measuring precision and recall across varying detection criteria, namely a discrete set of intersection over union (IoU) thresholds. 
In addition to the mAP score, we also report AP\textsubscript{50}, AP\textsubscript{75}, AP\textsubscript{S}, AP\textsubscript{M}, and AP\textsubscript{L}, 
as defined in the TorchMetrics toolkit \citep{skafte2020torchmetrics}.
We also inspect the Mean Average Recall (mAR) score.
mAR measures an algorithm's ability to detect all instances within an input sample,
where a target object is considered detected if a predicted box's IoU score meets a given threshold.
While the mAR score is often omitted in vision domain evaluations, 
we include it because of its similarities with a signal detection rate that is more commonly used in the signals domain.

\begin{figure}[b]
    \setlength\belowcaptionskip{-1.0\baselineskip}
    \centering
    \begin{tabular}{lr}
        \subfloat[
            mAP vs speed
        ]{\label{fig:wbsig53-det-results}
        \includegraphics[width=.42\textwidth]{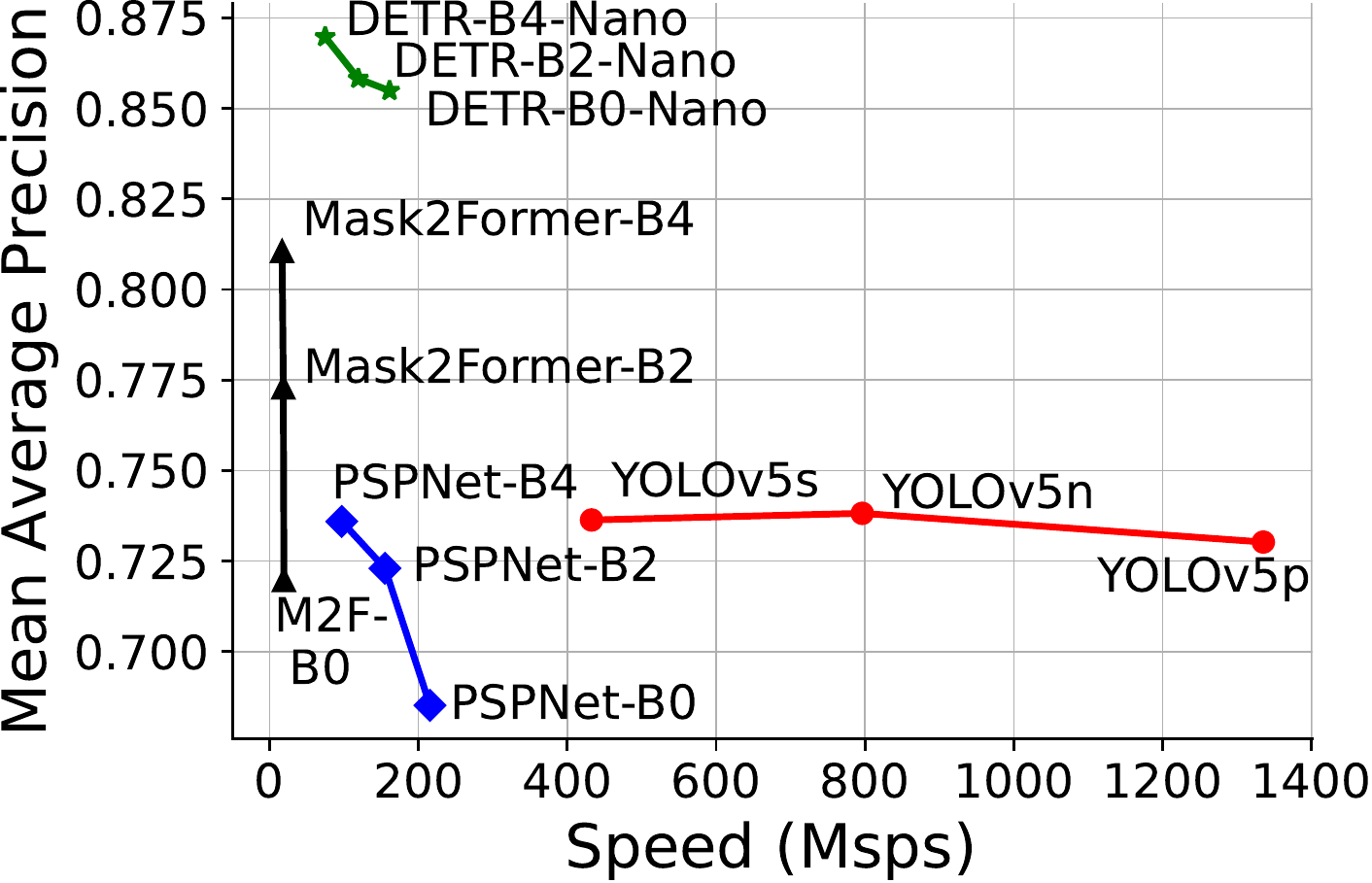}} &
        \subfloat[
            mAR vs SNR
        ]{\label{fig:detection-mars-snrs}
        \includegraphics[width=.42\textwidth]{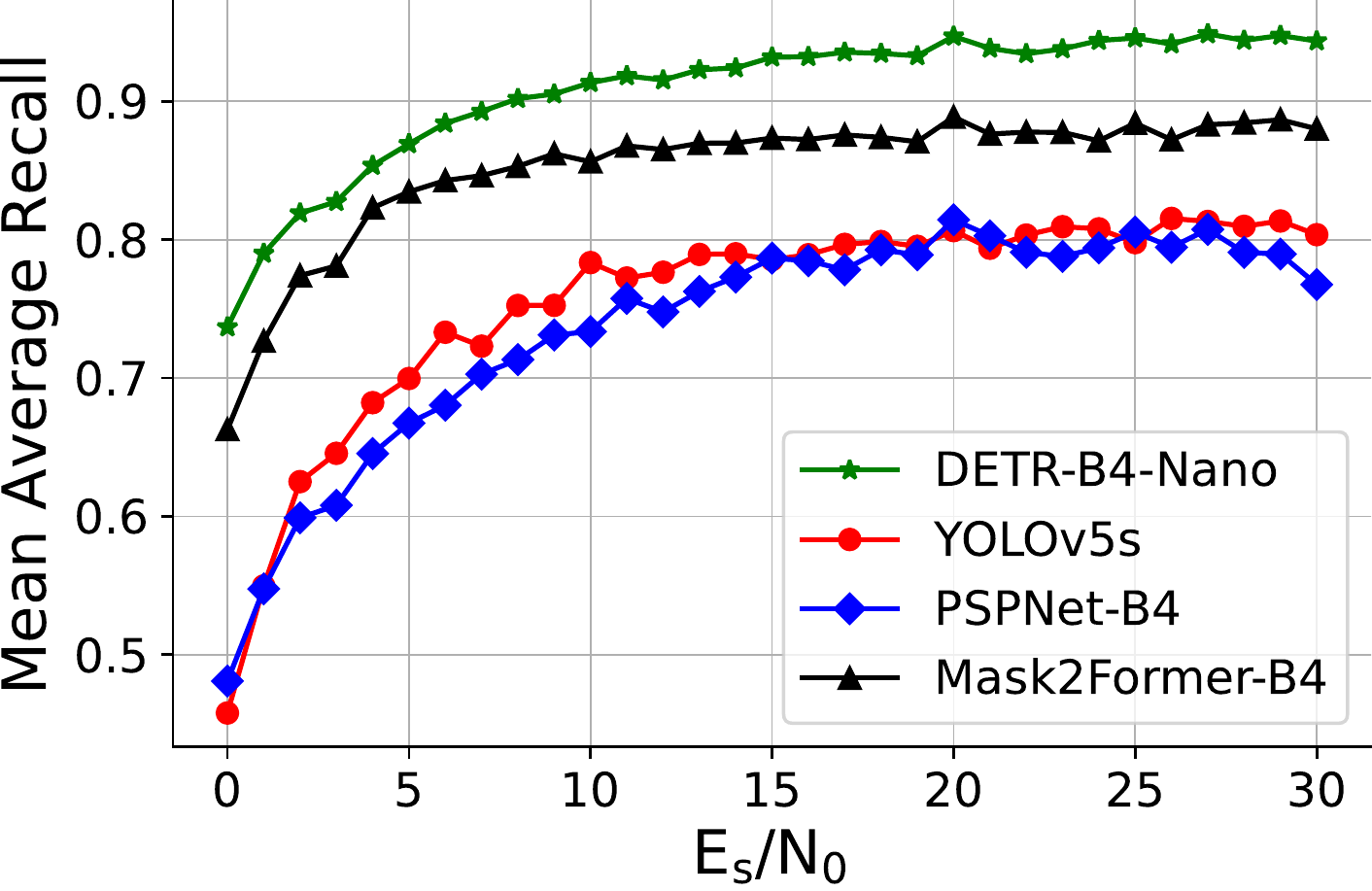}} \\
    \end{tabular}
    \caption{
        Signal detection results.
        \protect\subref{fig:wbsig53-det-results} mean average precision vs speed shows that
        DETR is the best performer while YOLOv5 is the fastest network.
        \protect\subref{fig:detection-mars-snrs}
        mean average recall versus SNR shows that
        DETR performs best across all SNRs.
    }
    \label{fig:detection-results}
\end{figure}

All networks are pulled from their original authors' PyTorch implementations and then modified as discussed previously \citep{paszke2017pytorch}.
Despite the benefits of data augmentations,
we limit our experiments to the static data samples found directly in the WBSig53 impaired dataset.
Within each network architecture, we train the various scales with identical training parameters; 
however, we vary parameters across architectures.
All networks are trained on a single Nvidia V100 GPU, lasting from days to weeks depending on network speeds.

YOLOv5 models are trained for $128$ epochs with a batch size of $32$.
We use the Adam optimizer \citep{kingma2014adam} with a weight decay of $1e^{-4}$ and initialize the learning rate to $2e^{-3}$ while using a cosine annealing scheduler \citep{loshchilov2016sgdr}.
DETR models are trained for $1$M steps with a batch size of $32$.
We use the AdamW optimizer \citep{loshchilov2017weightdecay} with a weight decay of $0.04$ and initialize the learning rate to $5e^{-4}$ while using a cosine annealing scheduler with $20$k warm-up steps.
PSPNet models are trained for $128$ epochs with a batch size of $32$.
We use the Adam optimizer with a weight decay of $1e^{-4}$ and initialize the learning rate to $2e^{-3}$ while using a cosine annealing scheduler.
Mask2Former models are trained for $1$M steps with a batch size of $16$.
We use the AdamW optimzer with a weight decay of $0.04$ and initialize the learning rate to $5e^{-4}$ while using a cosine annealing scheduler with $20$k warm-up steps.

During training, we monitor the validation loss and save the lowest score for evaluation.
For metric calculations, we use TorchMetrics for all models' inferred results on the WBSig53 impaired validation set.
YOLOv5 and DETR models output bounding boxes with probabilities directly,
and as such, their outputs undergo very minor processing prior to the metric computation.
Mask2Former outputs masks with probabilities, so simply converting the masks to boxes with their associated probabilities is sufficient processing prior to computing the metrics.
PSPNet only outputs masks, so in addition to the mask to box conversion, a probability is assigned as the average of every pixel-level probability within the box region.

\begin{figure}[t]
    \centering
    \setlength\belowcaptionskip{-1.0\baselineskip}
    {\includegraphics[width=0.70\textwidth]{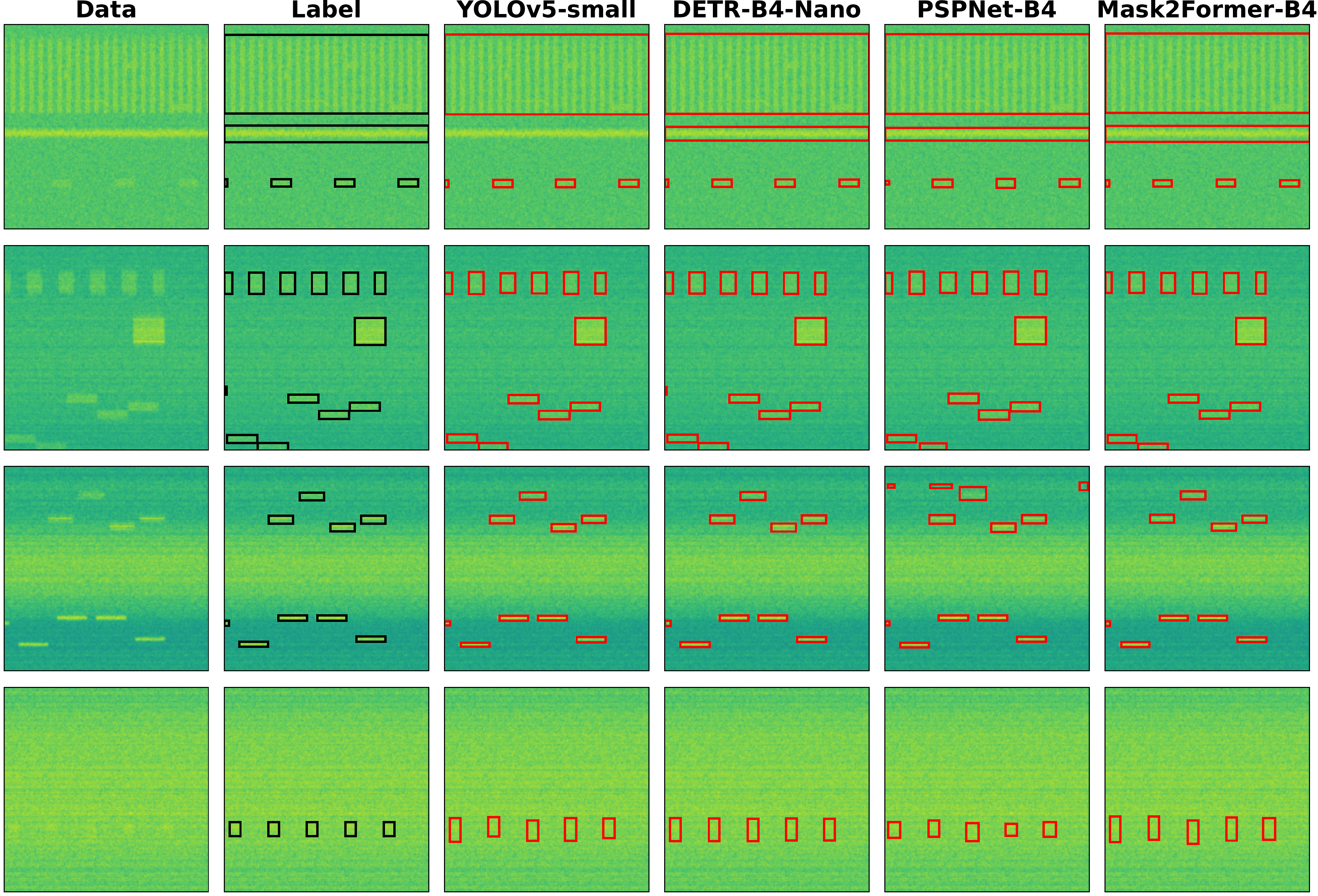}}
    \caption{
        Signal detection model comparisons. 
        DETR-B4-Nano is the only network to detect all signals due to a very small signal at the start of the second data sample.
        YOLOv5-small misses an addition signal in the first data example, despite the signal having fairly high SNR.
        PSPNet-B4 contains a few false positives in the third data example.
        Mask2Former-B4 performs well, only missing the small signal at the start of the second data example.}
    \label{fig:detection-model-comparisons}
\end{figure}

Under our training settings, DETR reports the best performance in terms of both mAP and mAR.
YOLOv5-pico achieves modest performance at significant speeds relative the other networks.
Interestly, PSPNet reports the lowest mAP score; however, its AP\textsubscript{50} score is greater than YOLOv5.
This may indicate PSPNet is better at coarse detection than YOLOv5, but its localization suffers;
this could be due to the naive mask to box conversion process, 
where errors at the pixel-level bias PSPNet's output detections to be larger, thus hurting its localization.
It is also interesting that Mask2Former fails to achieve the best performance given its large parameter advantage over the other networks.
But, the increased parameter complexity comes with additional hyperparameters, likely indicating that its performance has significant room for improvement via hyperarameter tuning.

We next analyze each networks' largest scale's performance across SNRs in \cref{fig:detection-mars-snrs}.
With this metric, we analyze the networks' abilities to detect signals across various SNRs.
Interestingly, this metric displays mAR differences across all SNR values across all networks.
In a traditional signal detection algorithm, there is typically high performance at high SNR with a drop at lower SNRs,
where the location of the drop is the main shift in performance,
i.e. a better detector would see its drop in performance occur at lower SNRs than a lower quality detector's drop at higher SNRs.
These findings suggest that even at high SNRs, our lower quality detectors still only detect at a rate of roughly $80$\%.
This difference could be due to our mAR metric's joint detection and localization measure, 
whereas the traditional detectors prioritize presence detection without locality.

In a final analysis of the signal detectors' performances, 
we retrieve four random data examples from the WBSig53 dataset 
and plot their spectrograms, labels, and inferred results across all architectures' largest scales (\cref{fig:detection-model-comparisons}).
Here, we see DETR-B4-Nano detecting all target signals with only minor bandwidth errors.
YOLOv5-small has a missed detection in the top data example, despite the signal having fairly high SNR, and it misses a very small signal at the beginning of the second example.
PSPNet-B4 has multiple false positives in the third data example, and it also misses the very small signal at the start of the third example.
Mask2Former performs well; however, it also misses the very small signal at the start of the third example.

%% file: 06_signal_recognition.tex
\section{Signal Recognition Experiments}
\label{sec:signal_recognition}

\begin{table}[t]
    \setlength\abovecaptionskip{-0.7\baselineskip}
    \caption[fontsize=10pt]{
        Signal recognition results on modulation family classes using the WBSig53 impaired dataset.
        YOLOv5-pico is the fastest network while DETR-B4-Nano is the best performer.
        }
    \label{tab:wbsig53-rec-results}
    \vskip 0.1in
        \begin{center}
            \centering
            \small
            \begin{tabular}{l|lllllllll}
                \toprule[1.5pt]
                Model & Params & SPS & mAP & AP$_{50}$ & AP$_{75}$ & AP$_{S}$ & AP$_{M}$ & AP$_{L}$ & mAR \\ \midrule
                \rowcolor{LightCyan}YOLOv5-pico & 0.32 M & 1335.21 M & 39.46 & 50.08 & 42.70 & 22.28 & 49.18 & 44.45 & 48.74 \\
                YOLOv5-nano & 1.8 M & 796.35 M & 57.02 & 72.51 & 63.24 & 37.61 & 55.13 & 52.39 & 62.51 \\
                YOLOv5-small & 7.0 M & 432.61 M & 58.50 & 72.37 & 64.29 & 38.35 & 56.68 & 54.27 & 64.35 \\ \midrule
                DETR-B0-Nano & 8.2 M & 161.21 M & 76.52 & 86.15 & 82.45 & 54.30 & 74.15 & 84.39 & 83.62 \\
                DETR-B2-Nano & 11.9 M & 119.17 M & 78.95 & 88.92 & 85.65 & 57.50 & 76.77 & 86.89 & 84.74 \\
                \rowcolor{LightCyan}DETR-B4-Nano & 21.9 M & 74.70 M & 80.65 & 88.64 & 85.41 & 59.24 & 78.48 & 88.48 & 86.03 \\ \midrule
                PSPNet-B0 & 4.1 M & 215.54 M & 27.30 & 39.07 & 28.29 & 07.07 & 21.47 & 41.85 & 44.50 \\
                PSPNet-B2 & 7.8 M & 155.07 M & 36.67 & 49.40 & 38.26 & 11.75 & 29.26 & 44.65 & 47.89 \\
                PSPNet-B4 & 17.6 M & 97.03 M & 51.84 & 67.23 & 56.54 & 16.85 & 47.59 & 68.53 & 62.61 \\ \midrule
                Mask2Former-B0 & 22.2 M & 19.28 M & 14.78 & 17.68 & 15.23 & 04.20 & 11.35 & 15.85 & 38.19 \\
                Mask2Former-B2 & 26.0 M & 18.47 M & 22.01 & 25.80 & 23.29 & 09.20 & 16.96 & 25.27 & 45.18 \\
                Mask2Former-B4 & 36.3 M & 16.87 M & 27.03 & 32.13 & 29.22 & 08.23 & 22.85 & 33.38 & 52.81 \\
            \bottomrule[1.5pt]
            \end{tabular}
        \end{center}
    \vskip -0.1in
    \end{table}

\begin{figure}[b]
    \setlength\belowcaptionskip{-1.0\baselineskip}
    \centering
    \begin{tabular}{lr}
        \subfloat[
            mAP vs speed
        ]{\label{fig:wbsig53-rec-results}
        \includegraphics[width=.42\textwidth]{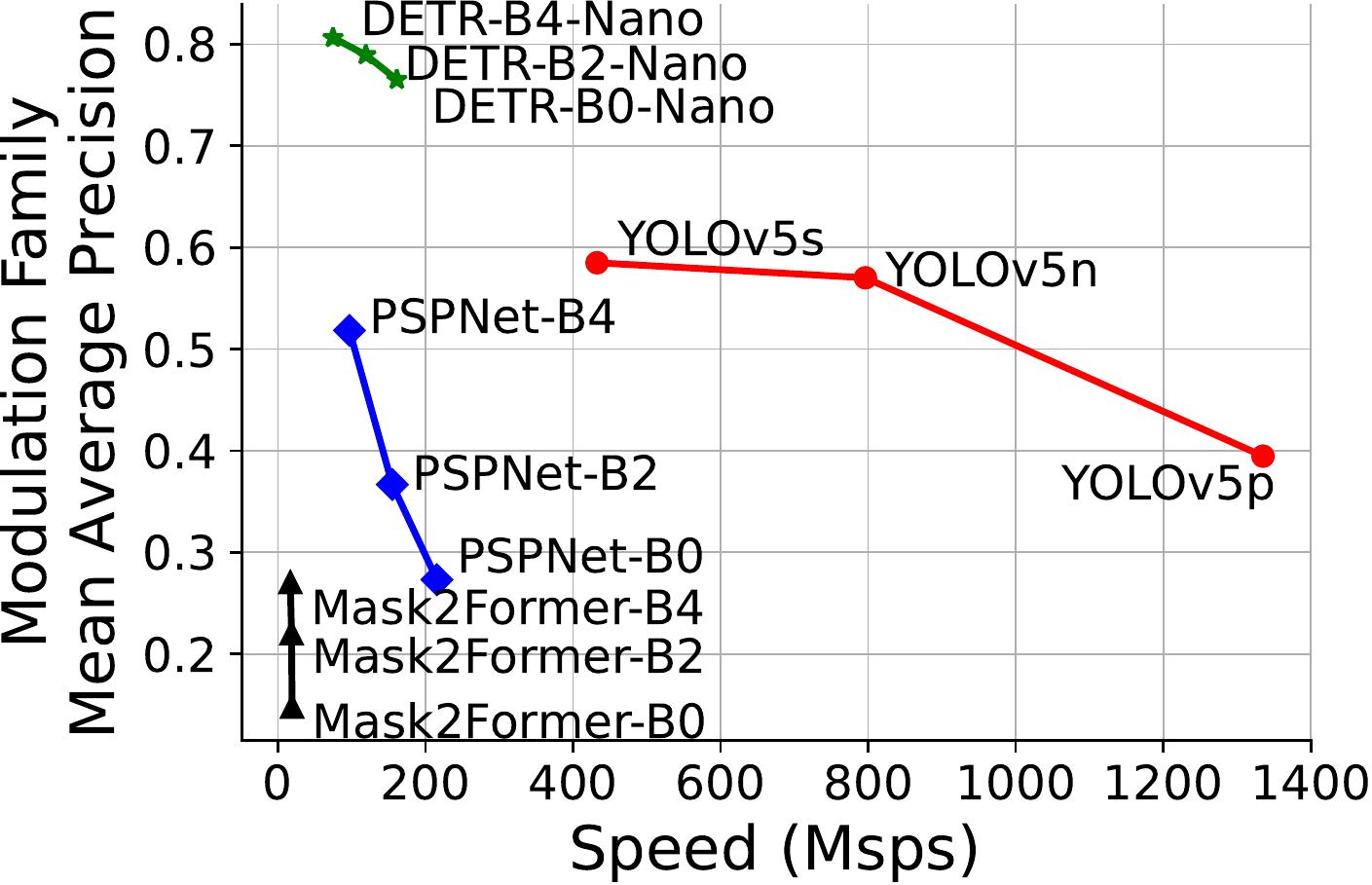}} &
        \subfloat[
            mAR vs SNRs
        ]{\label{fig:recognition-mars-snrs}
        \includegraphics[width=.42\textwidth]{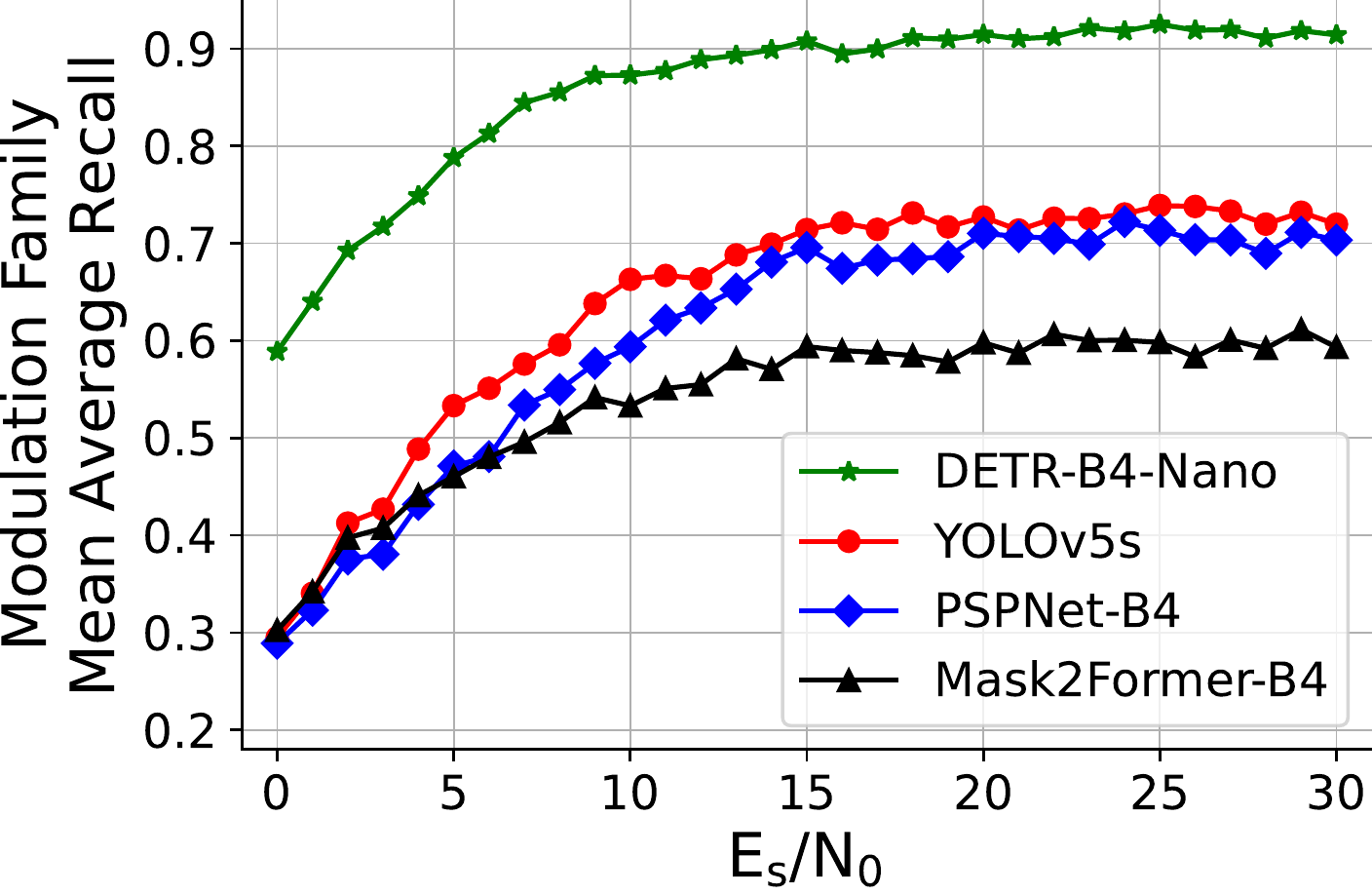}} \\
    \end{tabular}
    \caption{
        Signal recognition results across modulation families.
        \protect\subref{fig:wbsig53-rec-results} Mean average precision vs speed shows that DETR is the best performer while YOLOv5 is the fastest network.
        \protect\subref{fig:recognition-mars-snrs} Mean average recall versus SNR shows that DETR performs best across all SNRs.
        } 
    \label{fig:recognition-results}
\end{figure}

We now present signal recognition results on the WBSig53 impaired dataset. 
We define signal recognition as an algorithm's ability to discern the presence, location (in time and frequency), and class of any and all signals within an input example.
Here, we simplify the task to perform recognition over modulation family classes, rather than the granular signal class contained within the WBSig53 dataset.
With recognition over modulation families, we reduce the valid classes to the $6$ modulation families: ASK, FSK, OFDM, PAM, PSK, and QAM.
Signal recognition is analogous to the vision domain's object recognition task, where instead of inferring $(x,y)$-coordinates and the class of objects within an image,  
our algorithms infer ${(t,f)}$-coordinates and the class of signals within a data capture. 

We follow the same data processing pipeline as the detection task, 
and we update our models to output $6$ modulation family classes, rather than the single ``signal'' class for detection.
Here, AP and AR metrics now average across all classes in addition to all IoU thresholds.
Once again, we train all networks using a single V100 GPU over the span of days to weeks, depending on model speed.
We train the models with the static WBSig53 impaired training set without the use of data augmentations.
We use the same training parameters and schedules as the detection experiments, except for the PSPNet models.
The PSPNet models are trained for $1M$ steps with a batch size of $32$.
We use the AdamW optimizer with a weight decay of $0.01$ and
initialize the learning rate to $1e^{-3}$ while using a cosine annealing scheduler with $4k$ warm-up steps.

During training, we monitor the validation loss and save the best value for evaluation.
We evaluate the best model using TorchMetrics and record the results in \cref{tab:wbsig53-rec-results}.
Additionally, we display the mAP versus speed for each network in \cref{fig:wbsig53-rec-results}.

DETR again reports the highest performance in terms of both mAP and mAR.
Interestingly, we see our YOLOv5-pico model takes a larger hit in performance compared to its larger scale, YOLOv5-nano.
We also see a sizeable decrease in performance with the Mask2Former experiments.
Once again, this is likely due to Mask2Former's large size and complexity, 
and we suspect performance will improve with additional training and hyperparameter tuning.

We next analyze each networks' performance across SNRs in \cref{fig:recognition-mars-snrs}.
Once again, we see DETR-B4-Nano perform the best across all SNRs.
YOLOv5-small slightly outperforms PSPNet-B4.
Mask2Former-B4 shows the weakest performance across all SNRs.

\begin{figure}[t]
    \setlength\belowcaptionskip{-1.0\baselineskip}
    \centering
    {\includegraphics[width=0.70\textwidth]{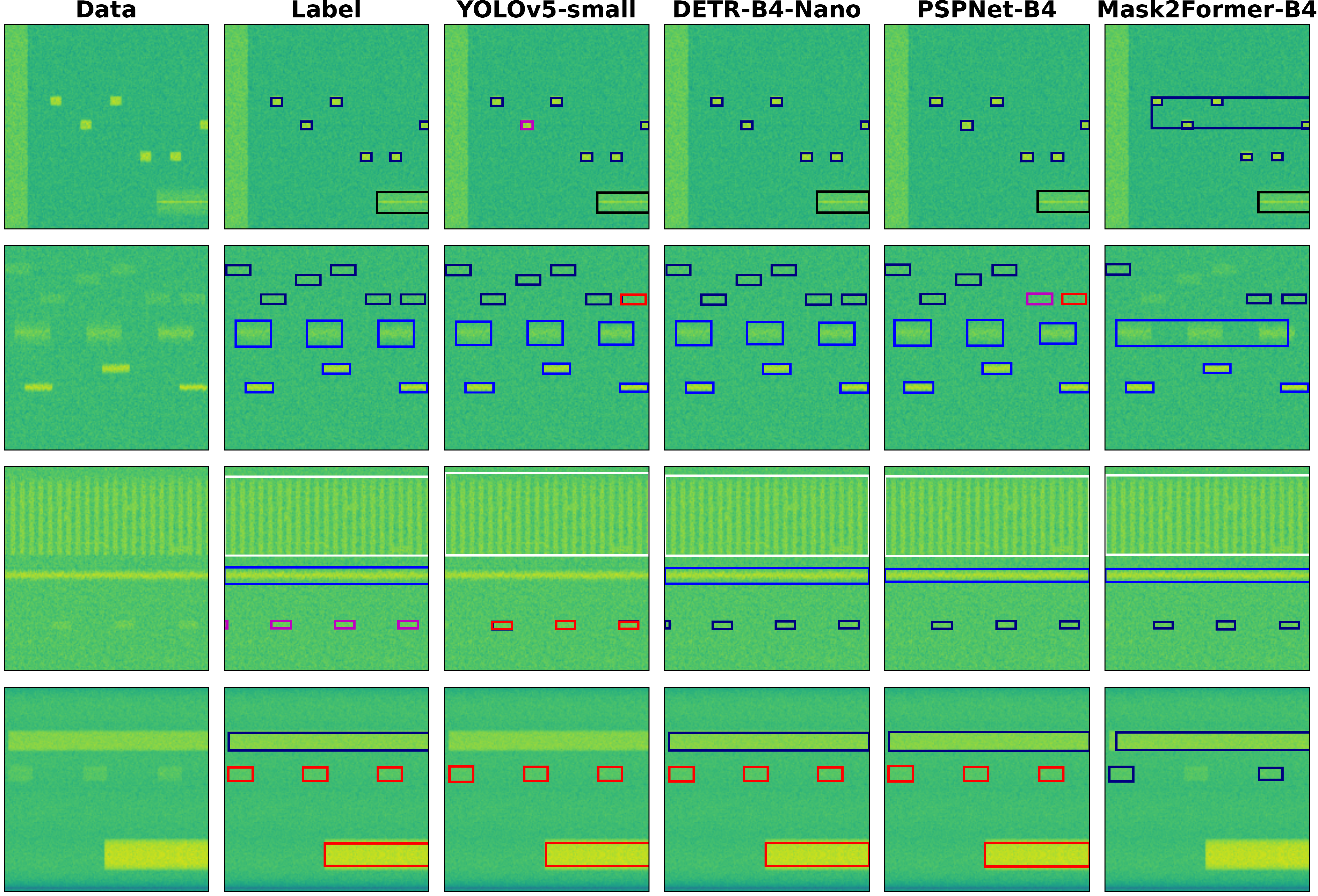}}
    \caption{
        Signal recognition for modulation family classes model comparisons. 
        We plot the modulation family label and prediction displayed as the color of the bounding boxes.
        DETR is the only network to detect all signals, but it misclassifies PSK signals as QAM in the third example.
        YOLOv5 suffers from missed detections of high SNR signals, and makes several misclassifications throughout.
        PSPNet detects all but one small signal at the start of the third example; however, it also suffers from misclassifications of several signals.
        Mask2Former struggles to make consistent, clean detections and classifications, likely due to non-optimal training parameters.
        }
    \label{fig:recognition-model-comparisons}
\end{figure}

We retrieve four random data examples from the WBSig53 dataset 
and plot their spectrograms, labels, and inferred results across all architectures' largest scales (\cref{fig:recognition-model-comparisons}).
We color the bounding boxes to represent the modulation families, 
where ASK is red, FSK is blue, OFDM is white, PAM is black, PSK is magenta, and QAM is navy.
DETR-B4-Nano is the only network to detect every signal, 
but it misclassifies a sequence of bursty PSK signals as QAM bursts in the third example.
YOLOv5-small suffers from more misclassifications, and it fails to detect several signals despite high SNR.
PSPNet-B4 detects all but a small signal at the start of the third data example; 
however, it suffers from misclassifying multiple signals throughout.
Mask2Former-B4 shows relatively poor performance with missed detections, poor localizations, and misclassifications.
We suspect Mask2Former's large complexity requires additional training optimizations in order to regain its performance advantage.

%% file: 07_conclusion.tex
\section{Conclusion}
\label{sec:conclusion}

In this work, we highlight a gap in RFML research in signal detection and signal recognition algorithms due to a lack of datasets and tools to conduct this work.
We introduce the WidebandSig53 dataset, which aims to fill this gap by providing an open-source benchmark enabling future RFML research.
We share reproducible experimental results modifying SoTA scalable vision domain networks: YOLOv5, DETR, PSPNet, and Mask2Former.
The application and thorough analysis of these networks marks one of the earliest applications of modern convolutional neural networks and transformers to the problems of large scale signal detection and signal recognition.
We also extend the open-source toolkit, TorchSig, to be an easily usable tool for the generation, augmentation, and utilization of wideband signals datasets and models.
While our work reports performances of multiple networks, we expect that these performances can improve with additional research, 
and we expect that other network architectures may outperform the limited subset explored in our work.
We invite others to build upon our research by leveraging our work and open-source tools to demonstrate better performance than what we report here.
We also invite others to investigate models capable of performing signal recognition across the full suite of fine-grain classes within the WBSig53 dataset.
We hope the accessibility of this dataset and toolkit allows for broad collaborative advances in the field of wideband RFML research.

%% file: 09_acknowledgements.tex
\subsubsection*{Acknowledgments}
This research was funded by the Laboratory for Telecommunication Sciences.

%% file: 10_data_appendix.tex
\subsection{Dataset Appendix}
\label{sec:appendix_dataset}

\begin{table}[ht]
  \setlength\abovecaptionskip{-0.7\baselineskip}
  \caption[fontsize=9pt]{Full WBSig53 Class List}
  \label{tab:class-list}
  \centering
  \vskip 0.1in
  \begin{center}
  \begin{small}
  \begin{sc}
  \begin{tabular}{l|l|l}
    \toprule[1.5pt]
    Class Name & Modulation Family & Class Index  \\ 
    \midrule
    4ASK & ASK Family & 3 \\
    8ASK & ASK Family & 6 \\
    16ASK & ASK Family & 10  \\
    32ASK & ASK Family & 15  \\
    64ASK & ASK Family & 19  \\
    OOK (On-Off Keying) & PAM Family & 0  \\
    4PAM & PAM Family & 2 \\
    8PAM & PAM Family & 5 \\
    16PAM & PAM Family & 9 \\
    32PAM & PAM Family & 14 \\
    64PAM & PAM Family & 18 \\
    BPSK (Binary Phase-Shift Keying) & PSK Family & 1 \\
    QPSK (Quadrature Phase-Shift Keying) & PSK Family & 4 \\
    8PSK & PSK Family & 7 \\
    16PSK & PSK Family & 11 \\
    32PSK & PSK Family & 16 \\
    64PSK & PSK Family & 20 \\
    16QAM & QAM Family & 8 \\
    32QAM & QAM Family & 12 \\
    32QAM\_Cross & QAM Family & 13 \\
    64QAM & QAM Family & 17 \\
    128QAM\_Cross & QAM Family & 21 \\
    256AM & QAM Family & 22 \\
    512AM\_Cross & QAM Family & 23 \\
    1024AM & QAM Family & 24 \\
    2FSK & FSK Family & 25 \\
    2GFSK & FSK Family & 26 \\
    2MSK & FSK Family & 27 \\
    2GMSK & FSK Family & 28 \\
    4FSK & FSK Family & 29 \\
    4GFSK & FSK Family & 30 \\
    4MSK & FSK Family & 31 \\
    4GMSK & FSK Family & 32 \\
    8FSK & FSK Family & 33 \\
    8GFSK & FSK Family & 34 \\
    8MSK & FSK Family & 35 \\
    8GMSK & FSK Family & 36 \\
    16FSK & FSK Family & 37 \\
    16GFSK & FSK Family & 38 \\
    16MSK & FSK Family & 39 \\
    16GMSK & FSK Family & 40 \\
    OFDM-64 & OFDM Family & 41 \\
    OFDM-72 & OFDM Family & 42 \\
    OFDM-128 & OFDM Family & 43 \\
    OFDM-180 & OFDM Family & 44 \\
    OFDM-256 & OFDM Family & 45 \\
    OFDM-300 & OFDM Family & 46 \\
    OFDM-512 & OFDM Family & 47 \\
    OFDM-600 & OFDM Family & 48 \\
    OFDM-900 & OFDM Family & 49 \\
    OFDM-1024 & OFDM Family & 50 \\
    OFDM-1200 & OFDM Family & 51 \\
    OFDM-2048 & OFDM Family & 52 \\
    \bottomrule[1.5pt]
  \end{tabular}
  \end{sc}
  \end{small}
  \end{center}
\vskip -0.1in
\end{table}

\clearpage
\textbf{Dataset Parameterization.} The clean datasets randomly select a number of signal sources in the range of $1$ to $6$.
Each signal source is randomly given an SNR in the range of $20$dB to $40$dB $E_s / N_0$ (energy per symbol to noise power spectral density).
The signal source is randomly selected from the full list of signal classes seen in \cref{tab:class-list}.
The modulation families within the full class list include: amplitude-shift keying (ASK), pulse-amplitude modulation (PAM),
phase-shift keying (PSK), quadrature amplitude modulation (QAM), frequency-shift keying (FSK), and orthogonal frequency-division multiplexing (OFDM).

OFDM signals are the only multi-carrier modulation family, 
and as such, we bias OFDM bandwidths to be larger than the single-carrier waveforms in examples that contain multiple signals of differing modulation families.
While an OFDM signal may have a bursty substructure, these signals are still considered to be a single signal.
Conversely, the non-OFDM signals may be bursty or frequency-hopping, where each burst is considered a different signal.
The non-OFDM signals are bursty or frequency-hopping with a probability of $0.2$ for either behavior.
Since non-OFDM signals have smaller bandwidths and a single source may represent numerous bursts, the number of signals of these families are larger.
To compensate this imbalance, the probability of signal source selection for OFDM signals is twice as likely to be selected as signals belonging to other modulation families.

When a non-OFDM signal is bursty, its burst duration is randomly selected to be within the range of $0.05$ to $0.2$ of the full sample duration.
Additionally, its silence period is randomly selected to be in the range of $1$ to $3$ times its burst duration.
When the signal is frequency-hopping, its number of frequency channels are randomly selected to be in the range $2$ to $16$ channels, centered around its original center frequency.

All signals' center frequencies are randomly selected to be in the range $-0.4$ to $0.4$ of the normalized bandwidth of the full example.
Bandwidths range from $0.0125$ to $0.7$ of the full normalized bandwidth of the example; however, as previously stated, OFDM signals bias towards the larger bandwidths.
The start and stop samples of each signal source are also randomized.
A signal source may start or stop during each example with a probability of $0.2$.
If a signal source starts at the beginning of the example and stops during the example, 
the stop point is randomly determined to be in the range $0.05$ to $0.95$ of the full example duration.
Conversely, if a signal source stops at the end of the example and starts during the example,
the start point is randomly determined to be in the range $0.0$ to $0.95$ of the full example duration.
Note that all signals in each example are intentionally separated in time and/or frequency such that no signals are overlapping.

\begin{figure*}[!h]
  \centering
  \includegraphics[width=0.42\textwidth]{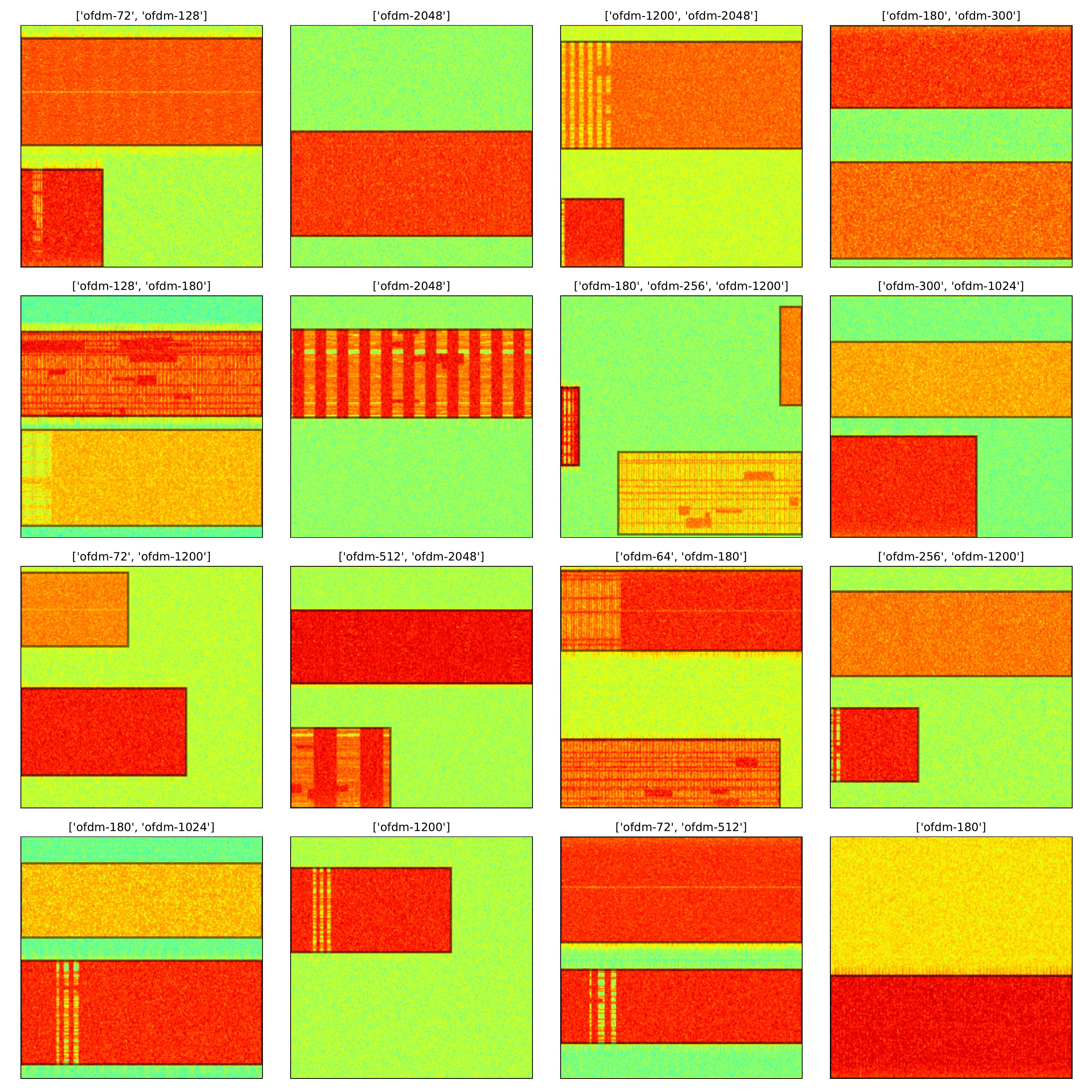}
  \caption{OFDM Randomizations}
  \label{fig:ofdm}
\end{figure*}

\clearpage
\textbf{OFDM Realism.} In the narrowband Sig53 dataset, 
OFDM signals are generated with randomized sidelobe suppression methods and randomized inclusion or omission of the DC subcarrier.
In the WBSig53 dataset, we also introduce additional randomized time and frequency effects with bursty symbols, pilot carriers, and emulated resource blocks.
These effects emulate more realistic OFDM communications signals.
\cref{fig:ofdm} shows several spectrograms of randomized OFDM signals, where we can see the newly introduced effects appearing randomly throughout the data examples.

\begin{figure*}[!h]
  \centering
  \subfloat[Original Data]{\includegraphics[width=0.48\textwidth]{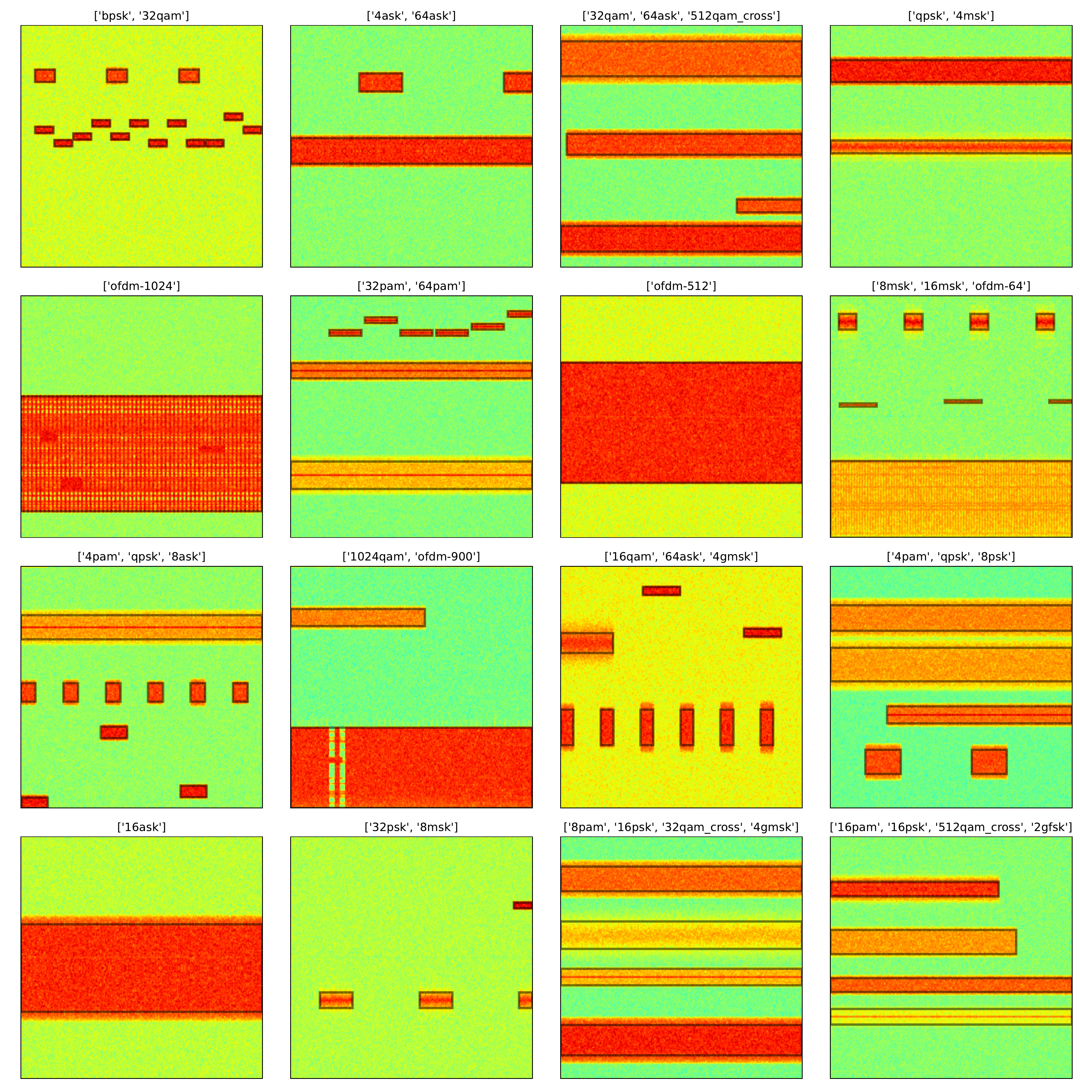}}
  \subfloat[Impaired Data]{\includegraphics[width=0.48\textwidth]{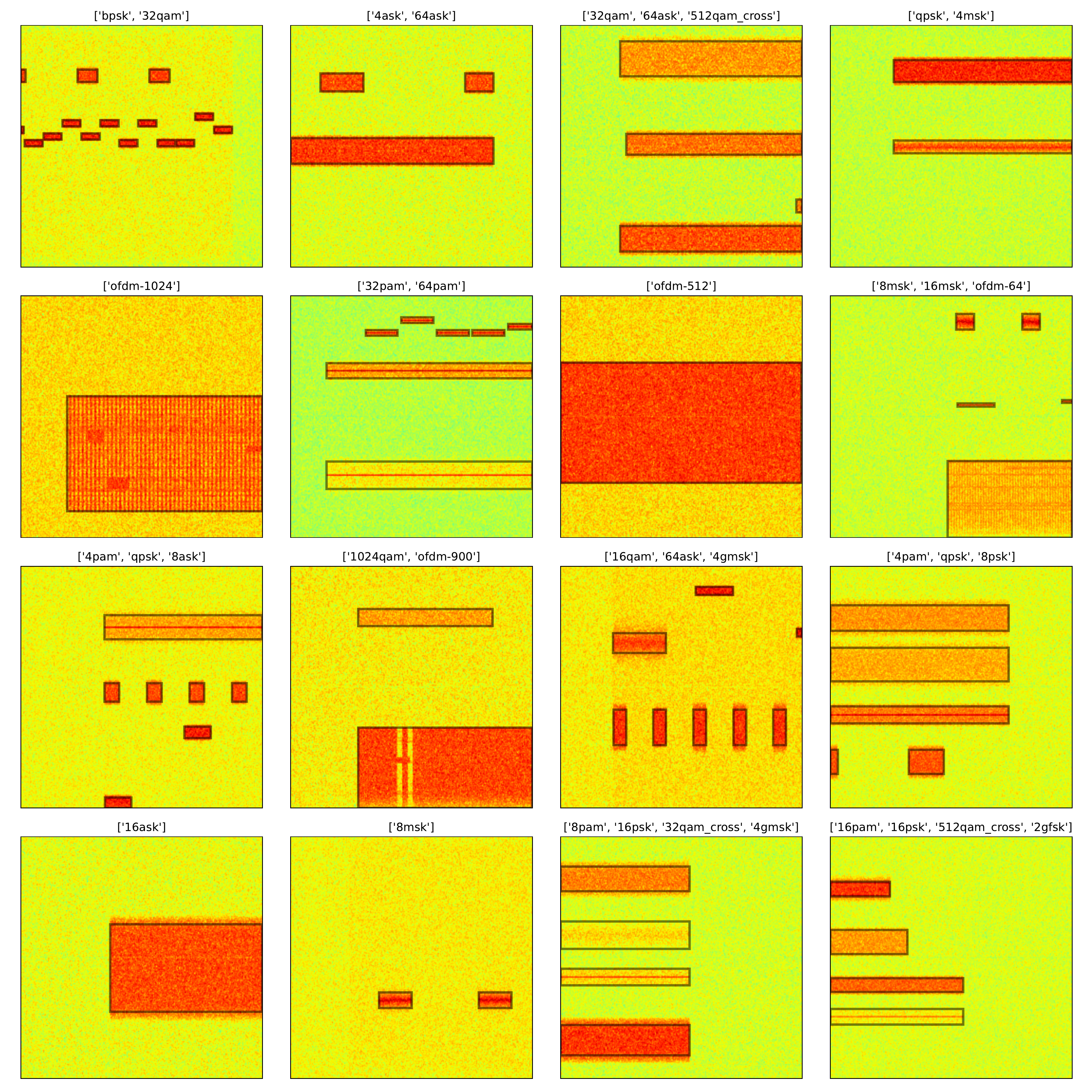}}
  \caption{Time Shift Impairment}
  \label{fig:time_shift}
\end{figure*}

\textbf{Time Shift Impairment.} The time shift impairment applies a randomized shift in time and pads the data back to the desired IQ sample length.
\cref{fig:time_shift} shows this impairment's effects when followed by adding noise to account for the zero-padding.
Note that the labels are adjusted accordingly.

\begin{figure*}[!h]
  \centering
  \subfloat[Original Data]{\includegraphics[width=0.40\textwidth]{images/wbsig53_clean.pdf}}
  \subfloat[Impaired Data]{\includegraphics[width=0.40\textwidth]{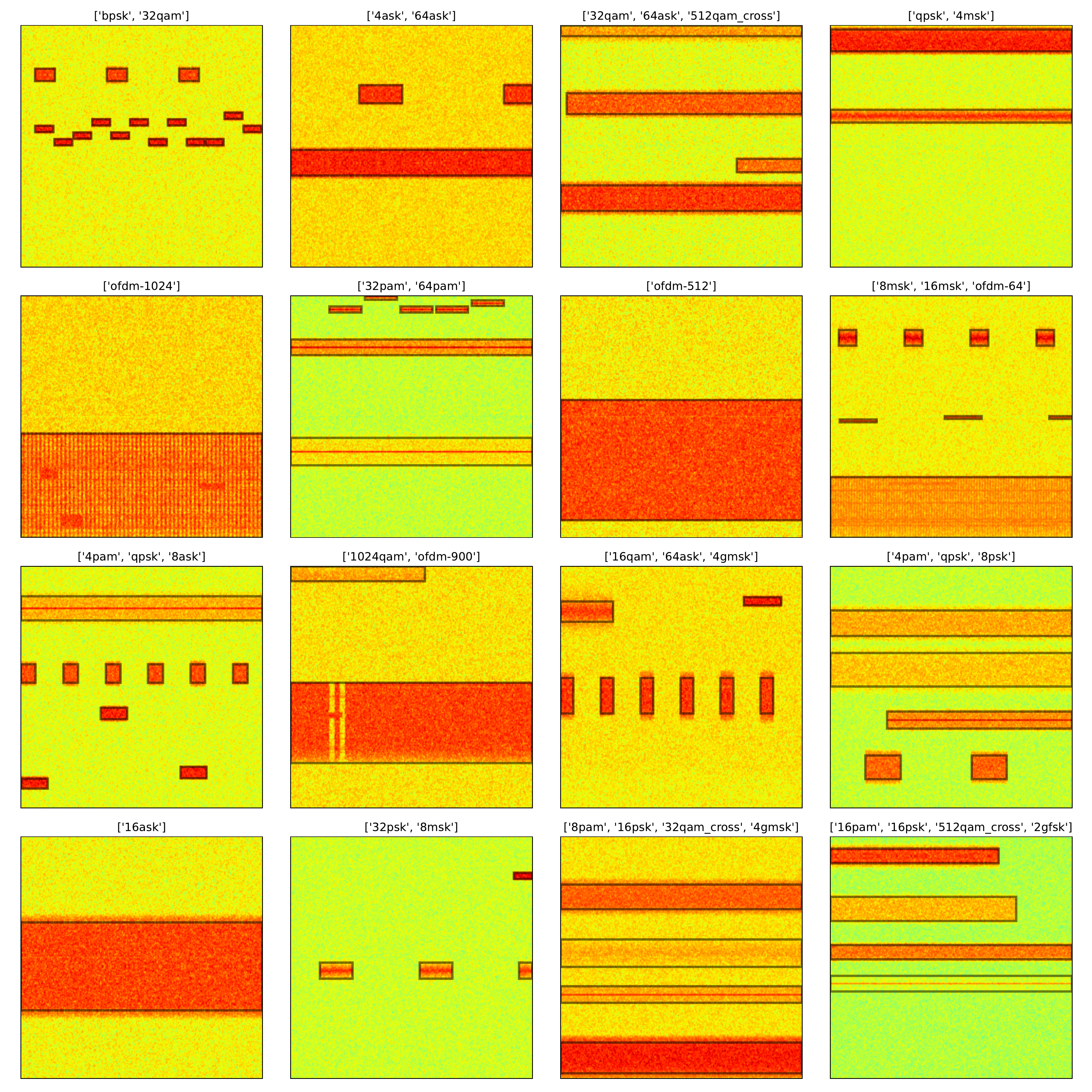}}
  \caption{Frequency Shift Impairment}
  \label{fig:freq_shift}
\end{figure*}

\clearpage
\textbf{Frequency Shift Impairment.} The frequency shift impairment applies a randomized shift in frequency.
This transform inspects the labels prior to performing the frequency shift in order to ensure aliasing is properly avoided.
If a signal is to be shifted past the boundaries, the impairment upsamples the input data, 
applies the frequency shift, applies a filter, and then downsamples the data to original rate.
The labels are adjusted to follow the signal's position post-shift.
\cref{fig:freq_shift} shows this impairment's effects.
Note that the labels are adjusted accordingly and no aliasing occurs with signals at the edges.

\begin{figure*}[!h]
  \centering
  \subfloat[Original Data]{\includegraphics[width=0.40\textwidth]{images/wbsig53_clean.pdf}}
  \subfloat[Impaired Data]{\includegraphics[width=0.40\textwidth]{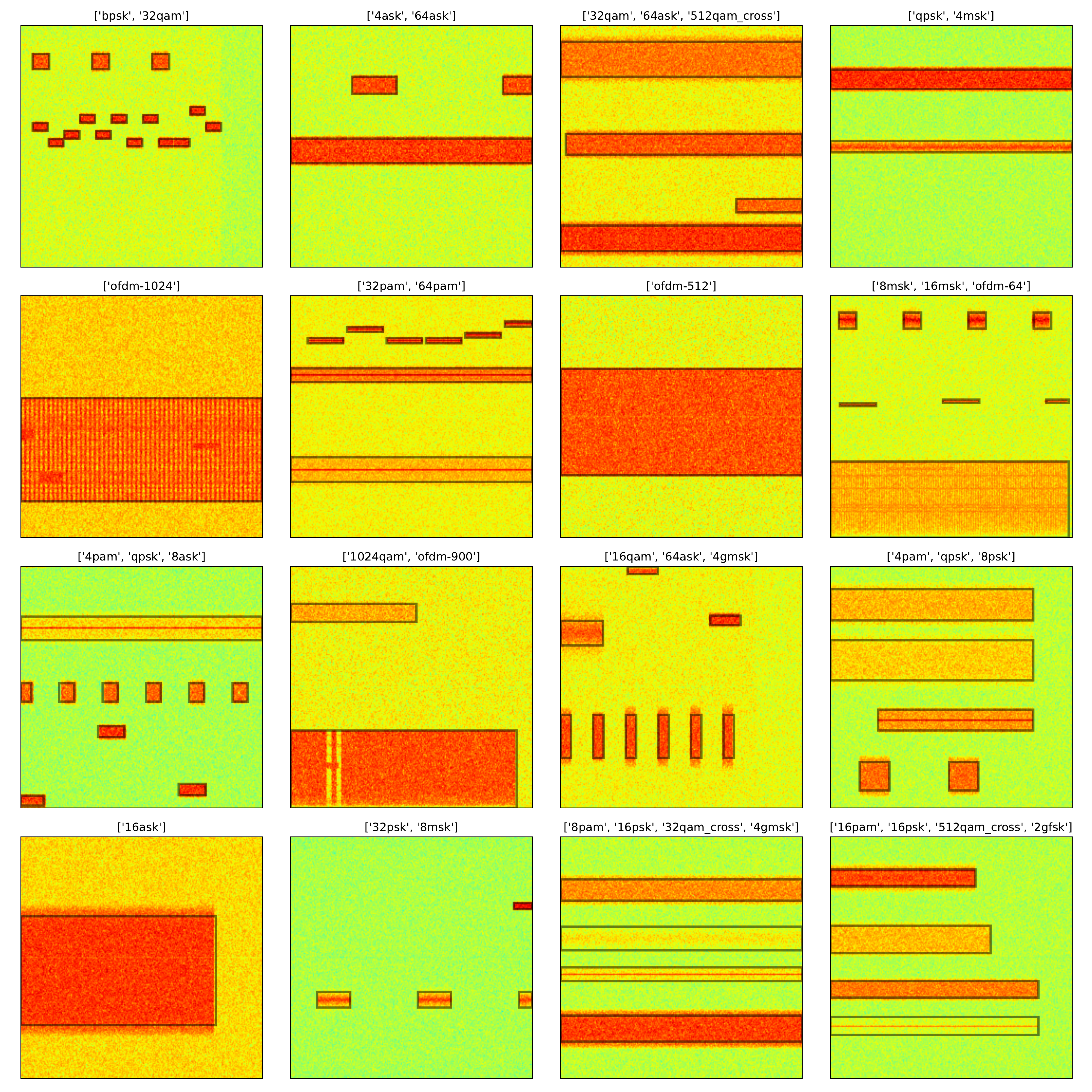}}
  \caption{Random Resample Impairment}
  \label{fig:random_resample}
\end{figure*}

\textbf{Random Resample Impairment.} The random resample impairment applies a randomized resampling to the input data.
Like the frequency shift impairment, this transform inspects the input labels prior to performing the resampling in order to properly handle/avoid aliasing.
Once again, the labels are adjusted to appropriately represent the impaired data's signals.
If downsampling occurs, the input data is zero-padded back to the original data size.
If upsampling occurs, the input data is truncated back to the original data size.
\cref{fig:random_resample} shows the random resample impairment effects.

\begin{figure*}[!h]
  \centering
  \subfloat[Original Data]{\includegraphics[width=0.40\textwidth]{images/wbsig53_clean.pdf}}
  \subfloat[Impaired Data]{\includegraphics[width=0.40\textwidth]{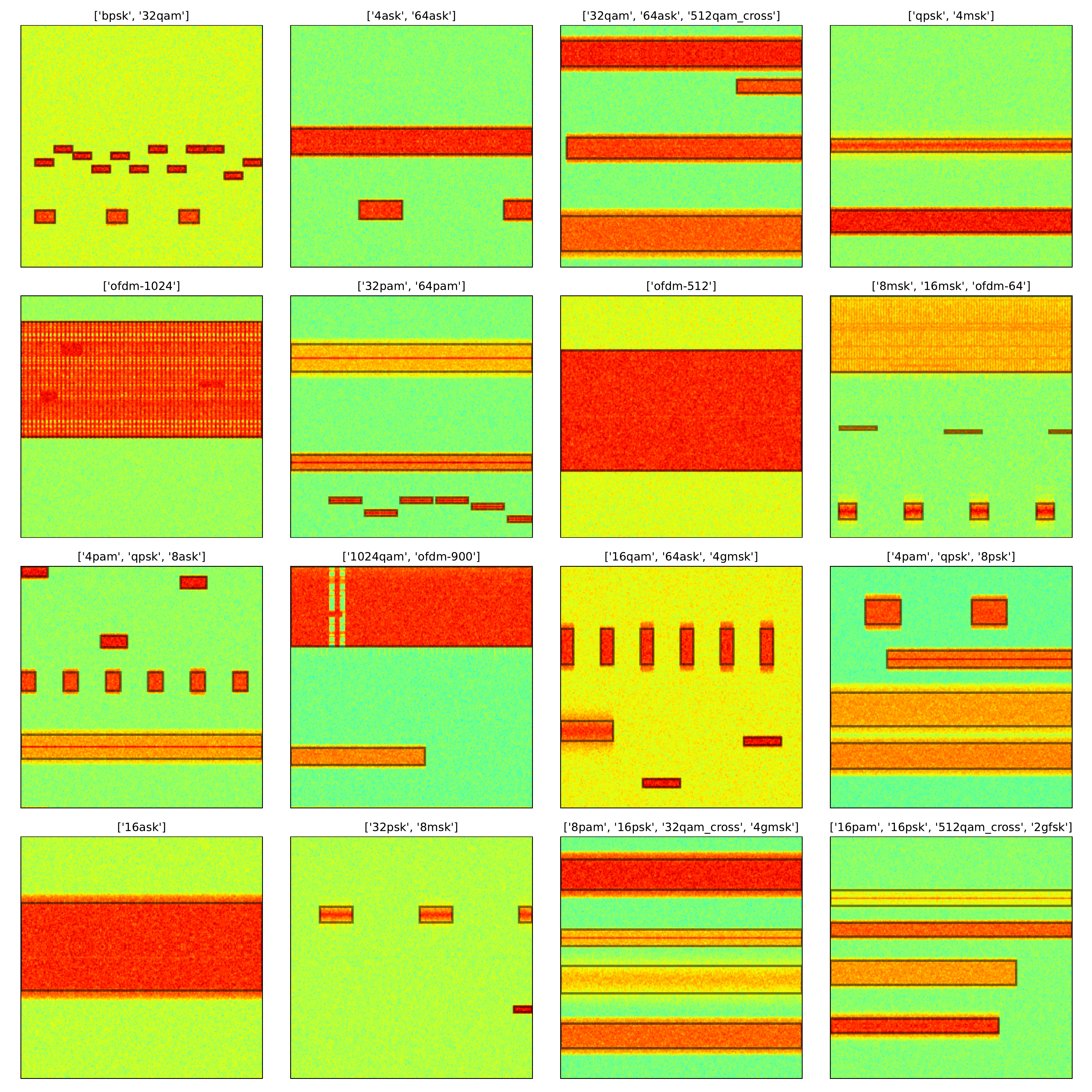}}
  \caption{Spectral Inversion Impairment}
  \label{fig:spec_inversion}
\end{figure*}

\clearpage
\textbf{Spectral Inversion Impairment.} The spectral inversion impairment inverts the frequency components of the input data by multiplying the imaginary values by $-1$.
Once again, the labels are adjusted to track the inverted signals' locations throughout the data.
\cref{fig:spec_inversion} shows the spectral inversion effects.

\begin{figure*}[!h]
  \centering
  \subfloat[Original Data]{\includegraphics[width=0.40\textwidth]{images/wbsig53_clean.pdf}}
  \subfloat[Impaired Data]{\includegraphics[width=0.40\textwidth]{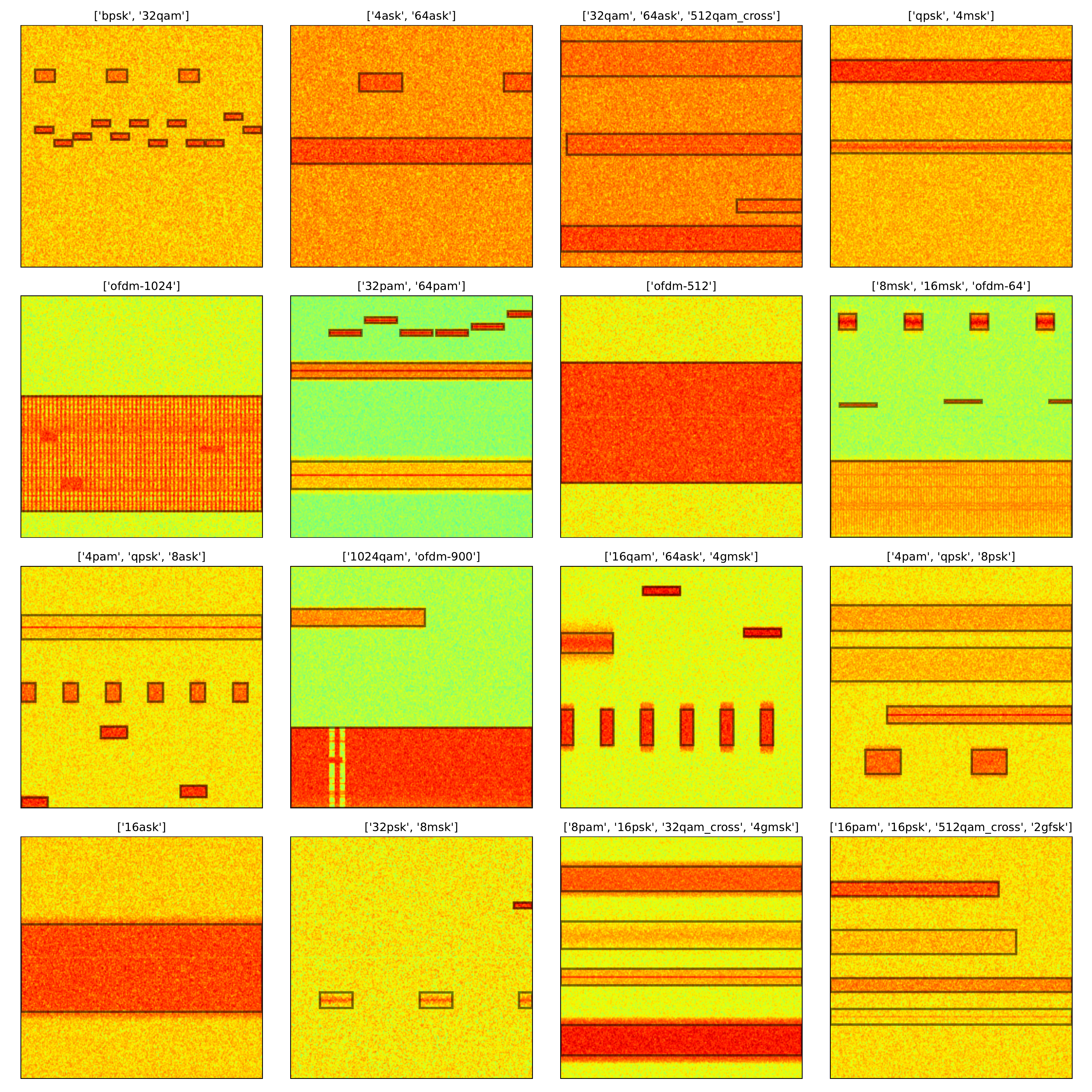}}
  \caption{Additive White Gaussian Noise Impairment}
  \label{fig:add_noise}
\end{figure*}

\textbf{Additive White Gaussian Noise Impairment.} The additive white Gaussian noise (AWGN) impairment adds noise to the full input data.
The magnitude of the AWGN is parameterized to pull from a specified random distribution.
\cref{fig:add_noise} shows varying degrees of AWGN applied to the input data.

\clearpage
\textbf{RandAugment Impairment.} In addition to the above impairments, we emulate the RandAugment transform for applying additional impairments \citep{cubuk2020randaugment}.
RandAugment is given a list of transforms and selects a subset of them to apply to the input data.
For WBSig53, we pass in the impairments: magnitude rescaling, RF roll-off, random convolve, Rayleigh fading, 
drop samples, phase shift, and IQ imbalance (all detailed below).
We set RandAugment to randomly select two of these impairments to apply at any given time.
Additionally, we condition the magnitude rescaling impairment with a random application rate of $50\%$ even when RandAugment selects it.

\begin{figure*}[!h]
  \centering
  \subfloat[Original Data]{\includegraphics[width=0.38\textwidth]{images/wbsig53_clean.pdf}}
  \subfloat[Impaired Data]{\includegraphics[width=0.38\textwidth]{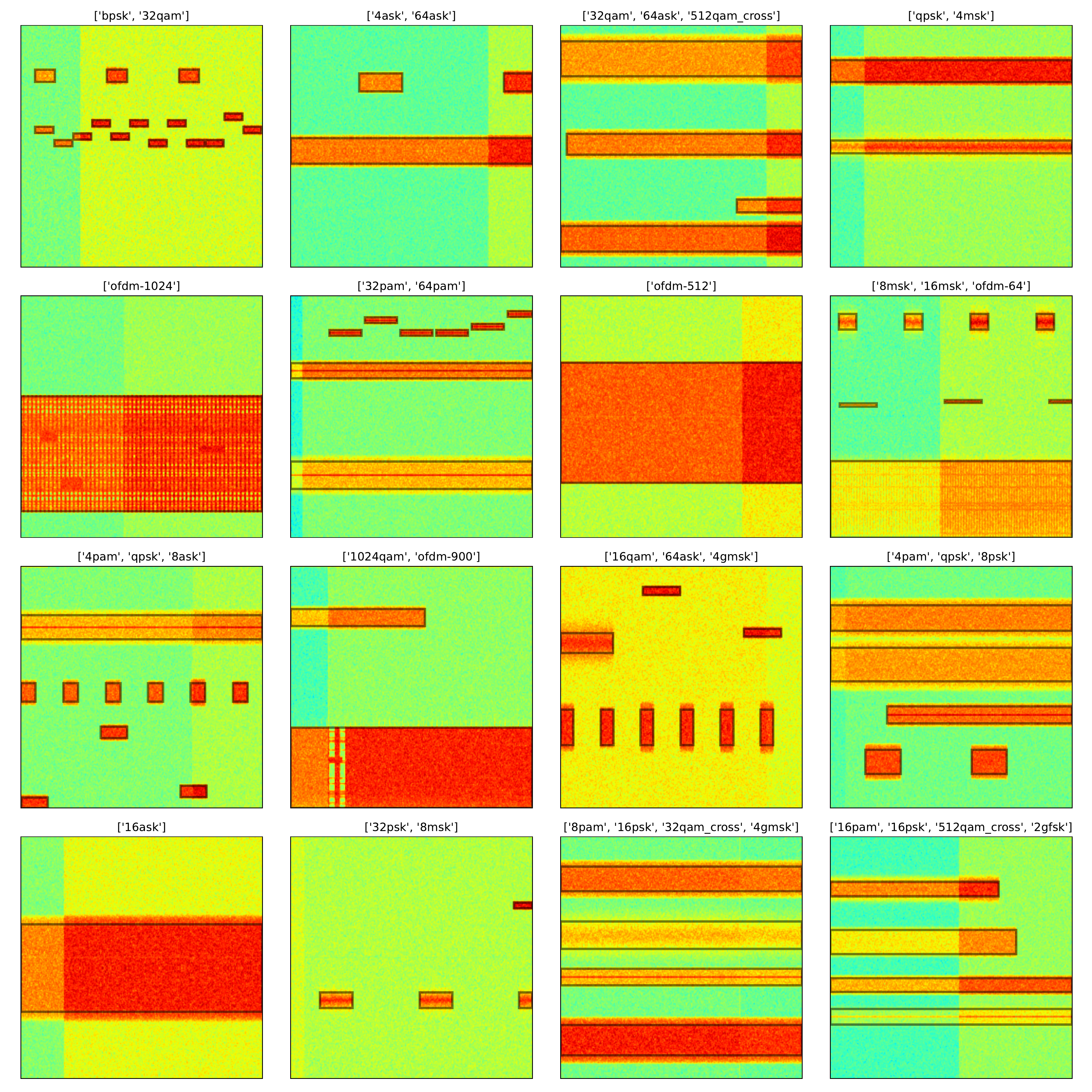}}
  \caption{Magnitude Rescaling Impairment}
  \label{fig:mag_rescale}
\end{figure*}

\textbf{Magnitude Rescaling Impairment.} The magnitude rescaling impairment randomly selects a time in the input data and then rescales the magnitude by a random rate.
This effect emulates an RF front end amplifier adjustment and can be seen in \cref{fig:mag_rescale}.

\begin{figure*}[!h]
  \centering
  \subfloat[Original Data]{\includegraphics[width=0.38\textwidth]{images/wbsig53_clean.pdf}}
  \subfloat[Impaired Data]{\includegraphics[width=0.38\textwidth]{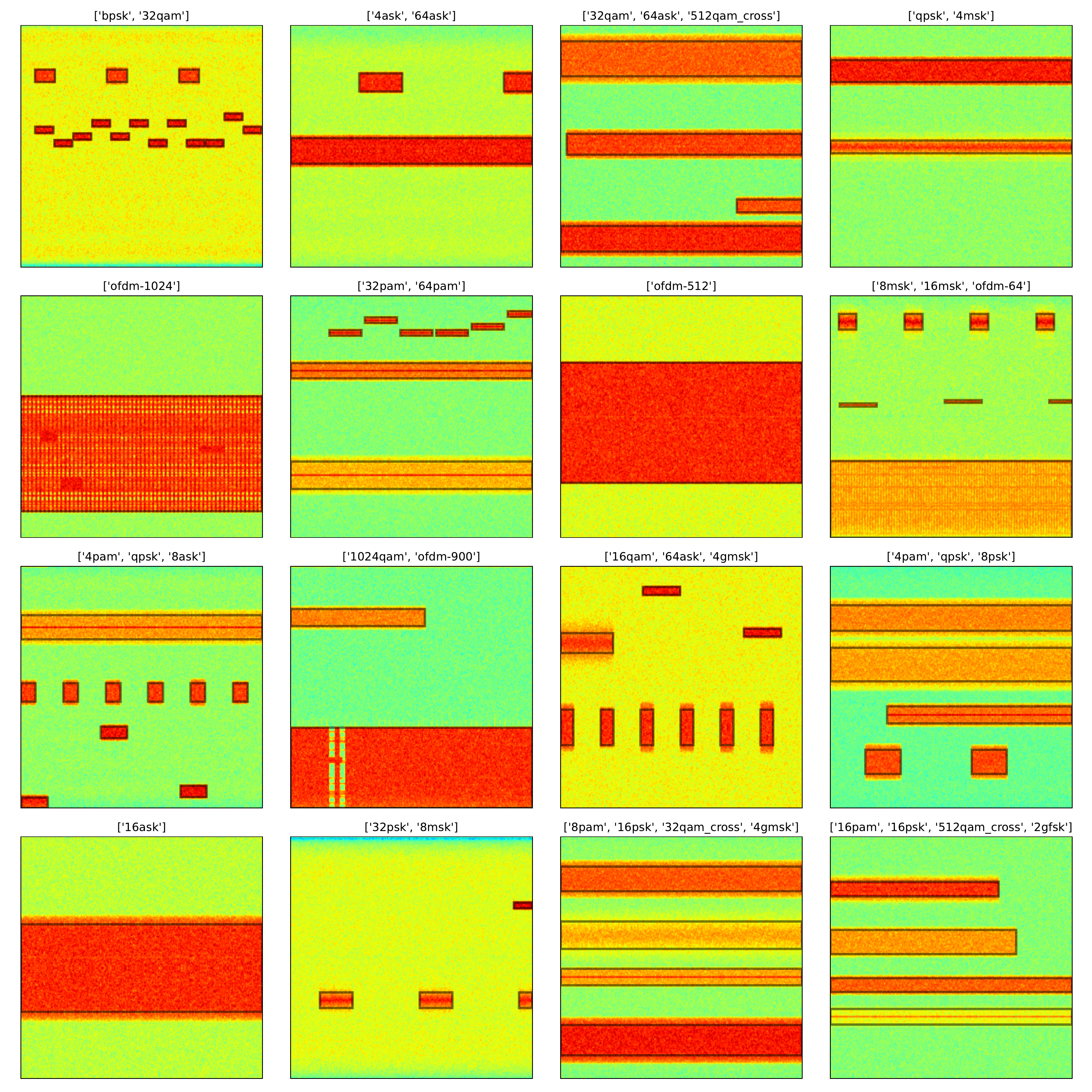}}
  \caption{RF Roll-Off Impairment}
  \label{fig:rolloff}
\end{figure*}

\textbf{RF Roll-Off Impairment.} The RF roll-off impairment randomly selects filter parameters in order to impose a roll-off effect at either the low frequency edge, the high frequency edge, or both.
This effect emulates an RF front end filter and can be seen in \cref{fig:rolloff}.

\begin{figure*}[!h]
  \centering
  \subfloat[Original Data]{\includegraphics[width=0.4\textwidth]{images/wbsig53_clean.pdf}}
  \subfloat[Impaired Data]{\includegraphics[width=0.4\textwidth]{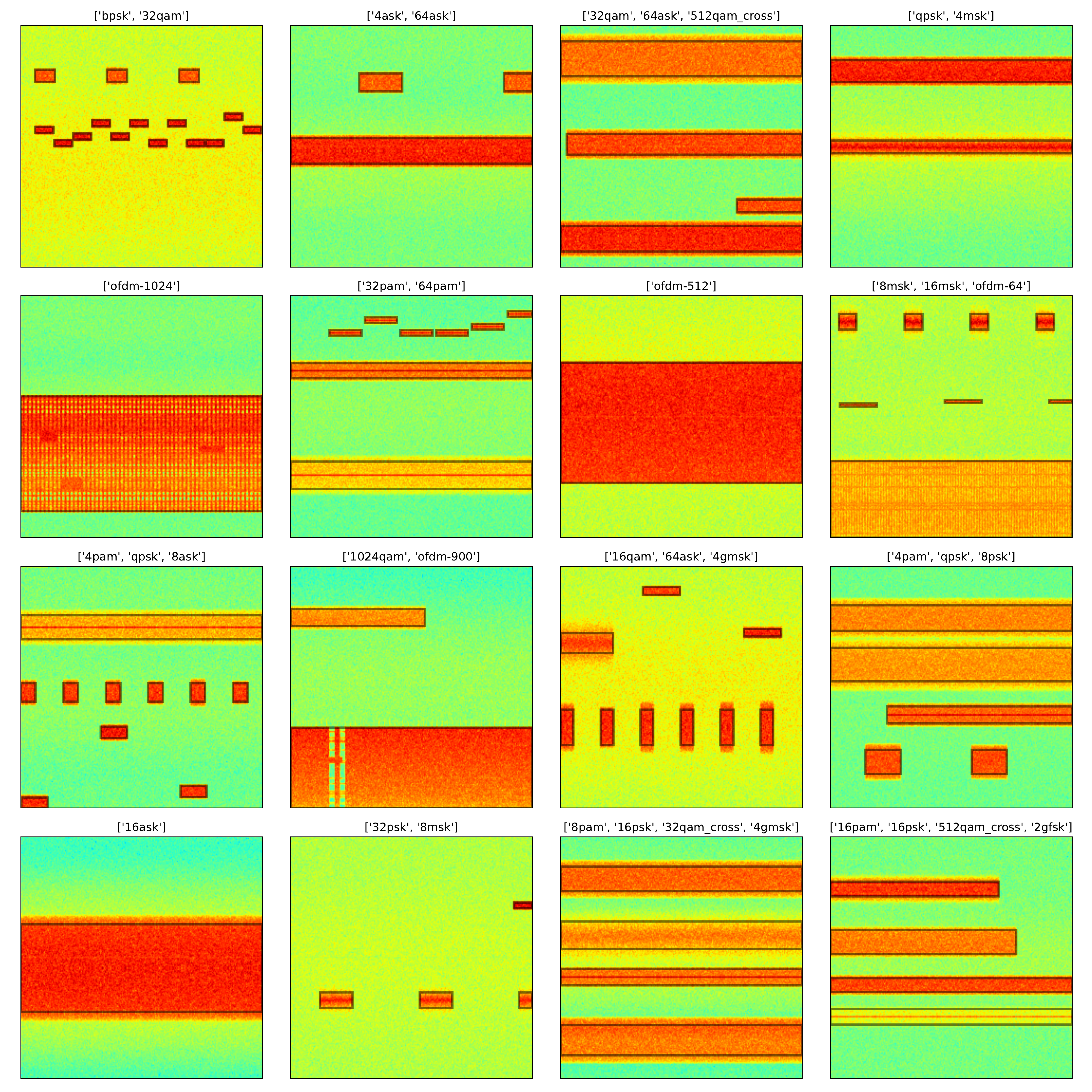}}
  \caption{Random Convolve Impairment}
  \label{fig:rand_convolve}
\end{figure*}

\clearpage
\textbf{Random Convolve Impairment.} The random convolve impairment inputs a random distribution of the number of taps in a filter, 
where each tap is assigned random values in range $0$ to $1$.
The randomly generated filter is then convolved with the input IQ data. 
An alpha value is also input to the transform, and it is used to dampen the effect of the randomly filtered data 
by weighting the newly filtered data and inversely weighting the original data and then summing the results. 
This random convolution is a relatively cheap form of applying a frequency-selective fading model (\cref{fig:rand_convolve}).

\begin{figure*}[!h]
  \centering
  \subfloat[Original Data]{\includegraphics[width=0.4\textwidth]{images/wbsig53_clean.pdf}}
  \subfloat[Impaired Data]{\includegraphics[width=0.4\textwidth]{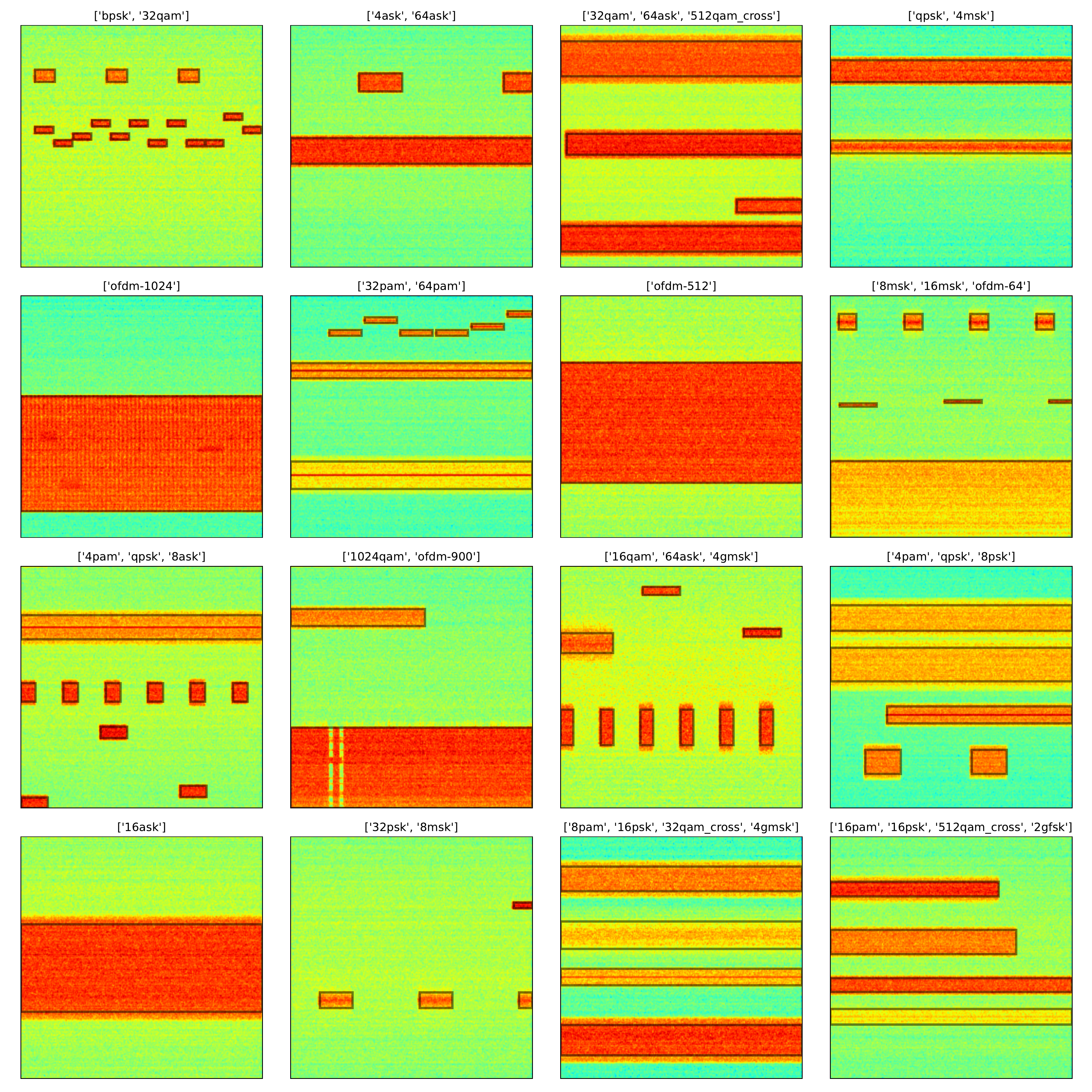}}
  \caption{Rayleigh Fading Impairment}
  \label{fig:rayleigh_fading}
\end{figure*}

\textbf{Rayleigh Fading Impairment.} A Rayleigh fading channel is modeled as a finite impulse response (FIR) filter with Gaussian distributed taps. 
The FIR filter is randomly generated under constraints and then convolved with the input data, simulating the effects of a frequency-selective, time-invariant Rayleigh fading channel. 
\cref{fig:rayleigh_fading} shows this Rayleigh fading channel applied to the full wideband input data;
however, the transform could also be applied directly to signal sources if desired.

\begin{figure*}[!h]
  \centering
  \subfloat[Original Data]{\includegraphics[width=0.40\textwidth]{images/wbsig53_clean.pdf}}
  \subfloat[Impaired Data]{\includegraphics[width=0.40\textwidth]{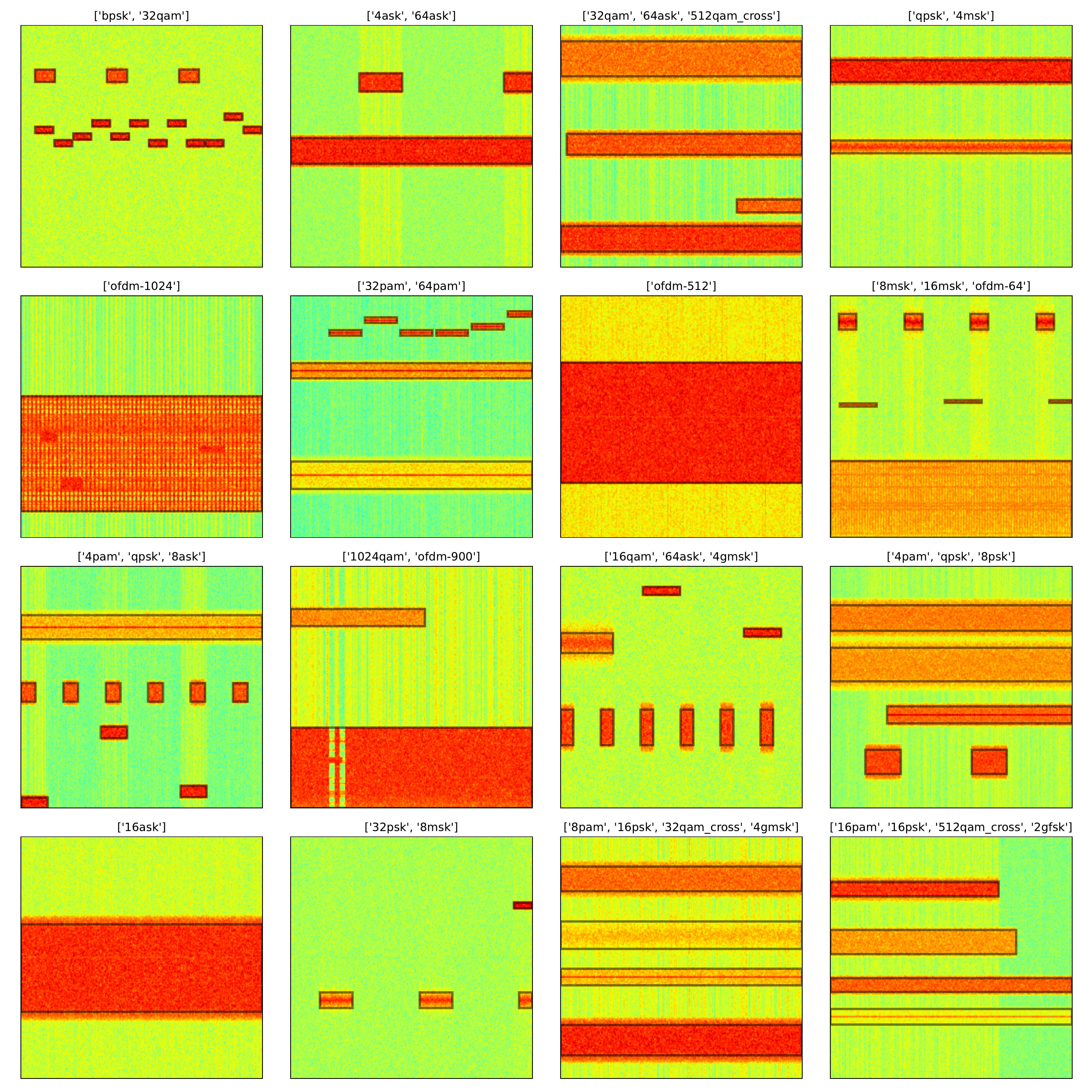}}
  \caption{Drop Samples Impairment}
  \label{fig:drop_samples}
\end{figure*}

\clearpage
\textbf{Drop Samples Impairment.} The drop samples augmentation randomly drops IQ samples from the input data using randomized values for the drop rate, the size of each dropped region, and the fill methods for how to replace the regions with dropped samples. 
The fill methods can be statically or randomly set to choose the following methods: front fill, back fill, mean, or zero; 
where front fill replaces each drop regions' samples with the last previous valid value, 
back fill replaces each drop regions' samples with the next valid value, 
mean replaces each drop regions' samples with the mean value of the full data example, 
and zero replaces each drop regions' samples with zeros. 
Note that the abrupt discontinuities cause high frequency spectral effects, as see in \cref{fig:drop_samples}.
This transform is loosely based on TSAug's DropOut transform \citep{wen2019tsaug}.

\begin{figure*}[!h]
  \centering
  \subfloat[Original Data]{\includegraphics[width=0.40\textwidth]{images/wbsig53_clean.pdf}}
  \subfloat[Impaired Data]{\includegraphics[width=0.40\textwidth]{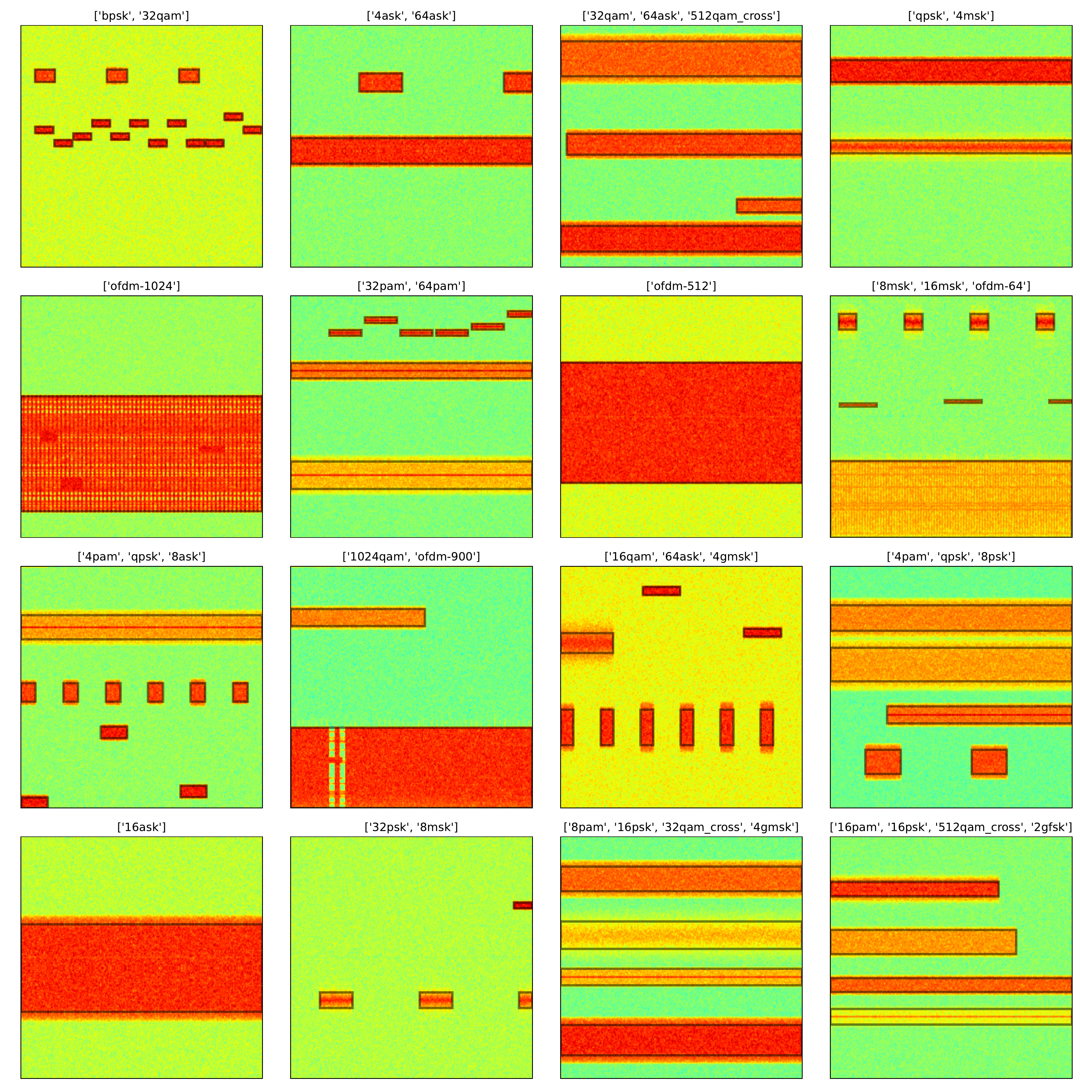}}
  \caption{Phase Shift Impairment}
  \label{fig:phase_shift}
\end{figure*}

\textbf{Phase Shift Impairment.} Phase shifts are applied by imposing a randomized rotation of IQ pairs about the complex origin.
Phase shift effects are best visualized with an IQ constellation plot;
however, for consistency, their spectral effects are shown in \cref{fig:phase_shift}.

\begin{figure*}[!h]
  \centering
  \subfloat[Original Data]{\includegraphics[width=0.48\textwidth]{images/wbsig53_clean.pdf}}
  \subfloat[Impaired Data]{\includegraphics[width=0.48\textwidth]{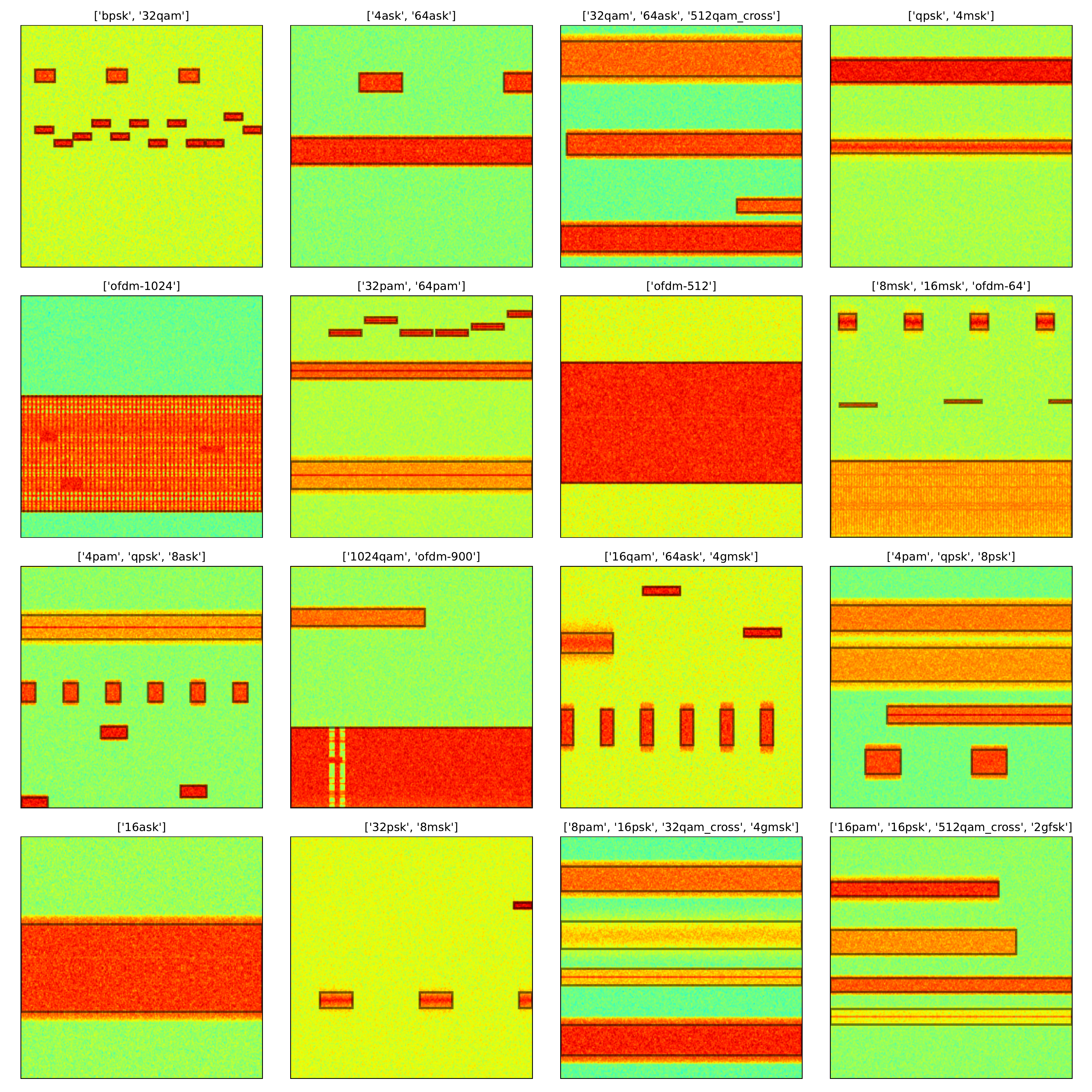}}
  \caption{IQ Imbalance Impairment}
  \label{fig:iq_imbalance}
\end{figure*}

\clearpage
\textbf{IQ Imbalance Impairment.} IQ imbalance is applied with three randomized parameters: amplitude imbalance, phase imbalance, and DC offset. 
Amplitude imbalance causes a growth or reduction of IQ pairs in a constellation,
phase imbalance causes a rotation of IQ pairs in a constellation, and
DC offset causes a shift in all IQ pairs.
These effects are best visualized with an IQ constellation plot; 
however, for consistency, their spectral effects are shown in \cref{fig:iq_imbalance}.

%% file: 11_tools_appendix.tex
\clearpage
\subsection{Tools Appendix}
\label{sec:appendix_tools}

In addition to the data impairments described in \cref{sec:appendix_dataset}, 
the TorchSig toolkit can also apply a number of augmentations to the WBSig53 data or any other dataset.
Here, we step through TorchSig's additional data augmentations and show their effects on wideband data.
We also extend TorchSig's list of augmentations to include augmentations that can be applied directly to the spectrogram data after the IQ data is transformed.

\subsubsection{Data Augmentations}

In addition to the impairments used in the generation of the WBSig53 dataset,
TorchSig provides support for other signal domain-specific data augmentations available for use during training.

\begin{figure*}[!h]
    \centering
    \subfloat[Original Data]{\includegraphics[width=0.40\textwidth]{images/wbsig53_clean.pdf}}
    \subfloat[Impaired Data]{\includegraphics[width=0.40\textwidth]{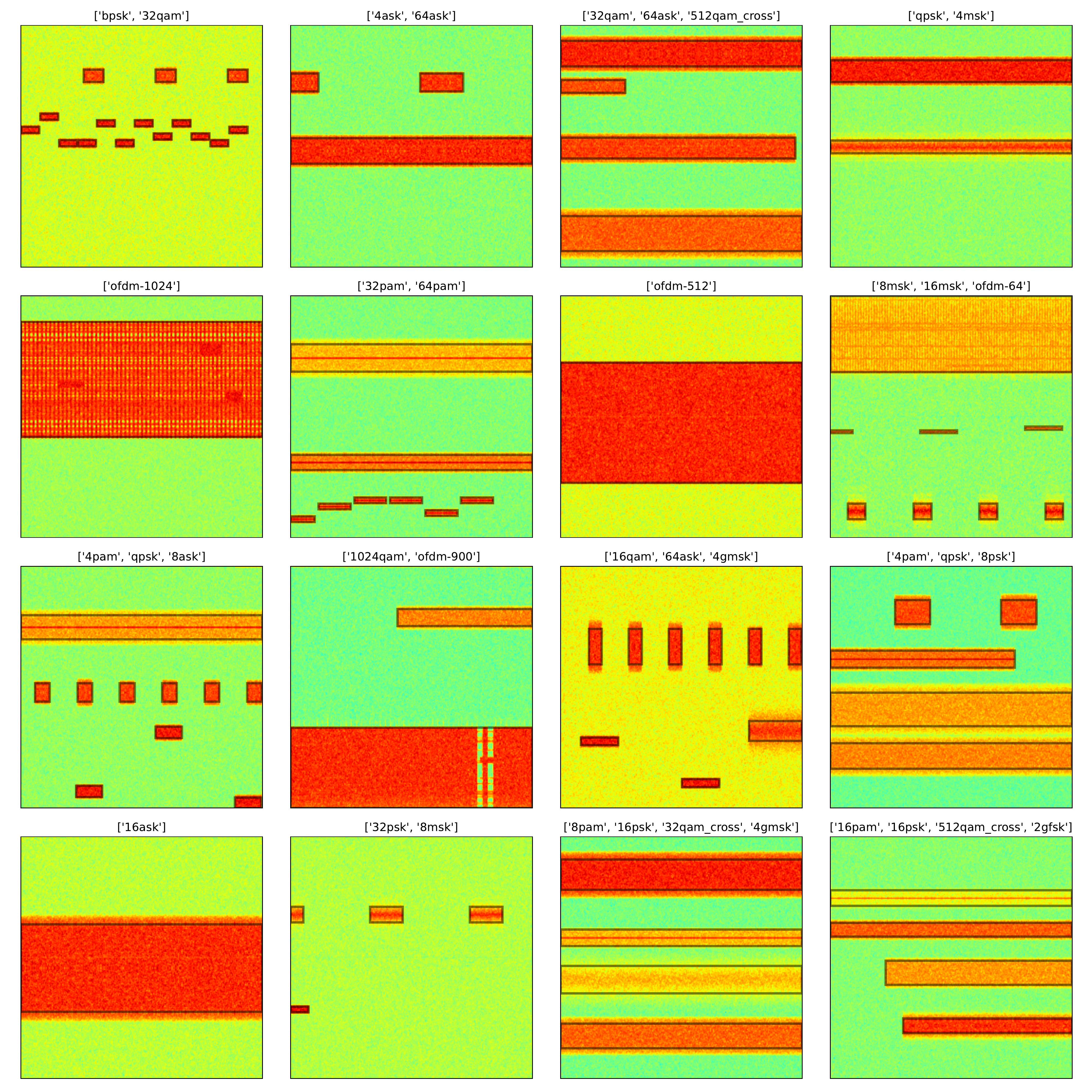}}
    \caption{Time Reversal Augmentation}
    \label{fig:time_reversal}
\end{figure*}

\textbf{Time Reversal.} The time reversal augmentation reverses the order of the IQ samples in the input. 
Since time reversal in the signal domain also results in a spectral inversion, 
the TorchSig time reversal augmentation additionally has the option to undo the spectral inversion effect if desired.
The time reversal augmentation is shown in \cref{fig:time_reversal}.
Note that the labels are also updated with the augmentation to properly track the updated time and frequency information.

\begin{figure*}[!h]
    \centering
    \subfloat[Original Data]{\includegraphics[width=0.40\textwidth]{images/wbsig53_clean.pdf}}
    \subfloat[Impaired Data]{\includegraphics[width=0.40\textwidth]{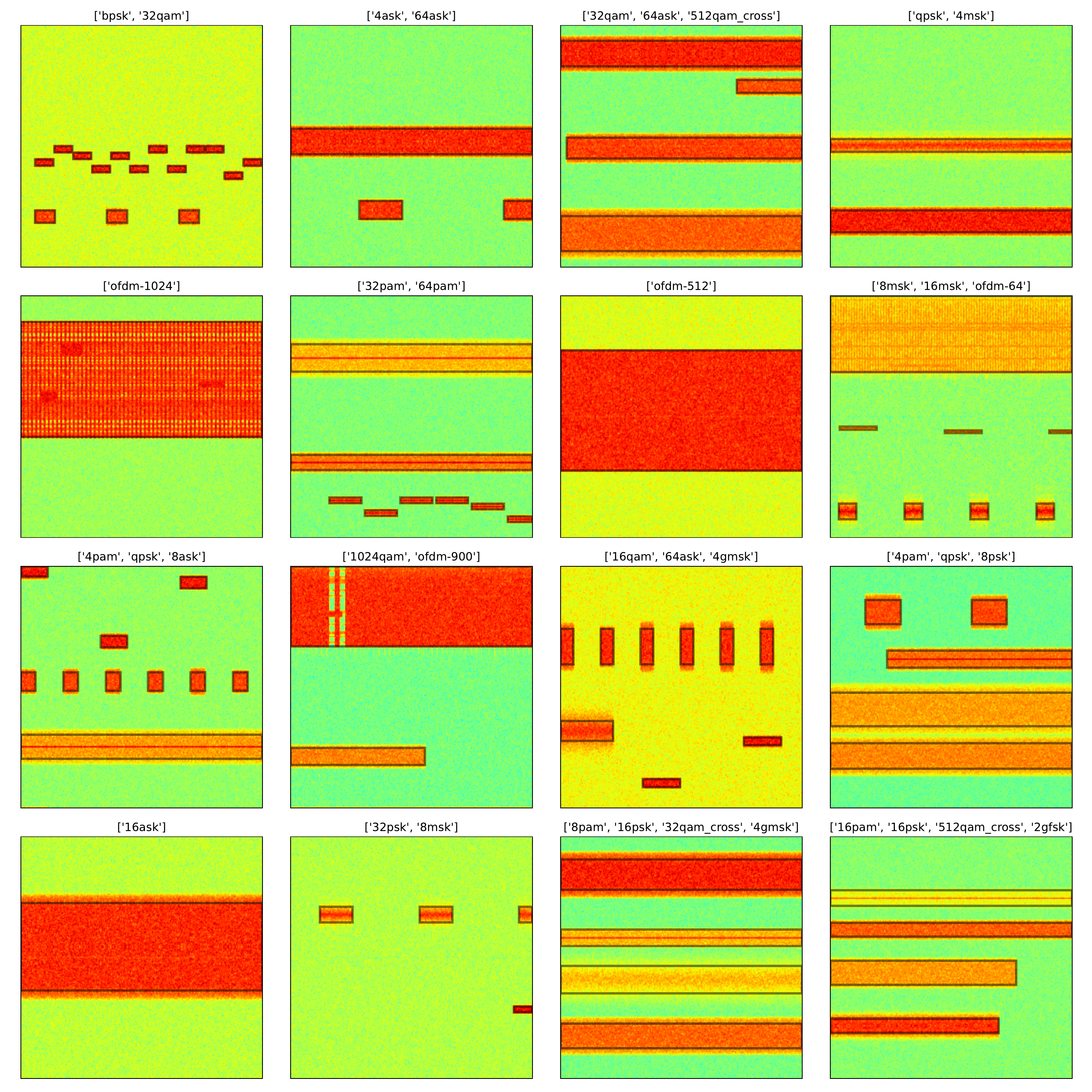}}
    \caption{Channel Swap Augmentation}
    \label{fig:channel_swap}
\end{figure*}

\clearpage
\textbf{Channel Swap.} The channel swap augmentation switches the real and imaginary componets of the input complex data. 
In the signal domain, this has the same effect as a spectral inversion followed by a static $\pi/2$ phase shift (\cref{fig:channel_swap}).

\begin{figure*}[!h]
    \centering
    \subfloat[Original Data]{\includegraphics[width=0.40\textwidth]{images/wbsig53_clean.pdf}}
    \subfloat[Impaired Data]{\includegraphics[width=0.40\textwidth]{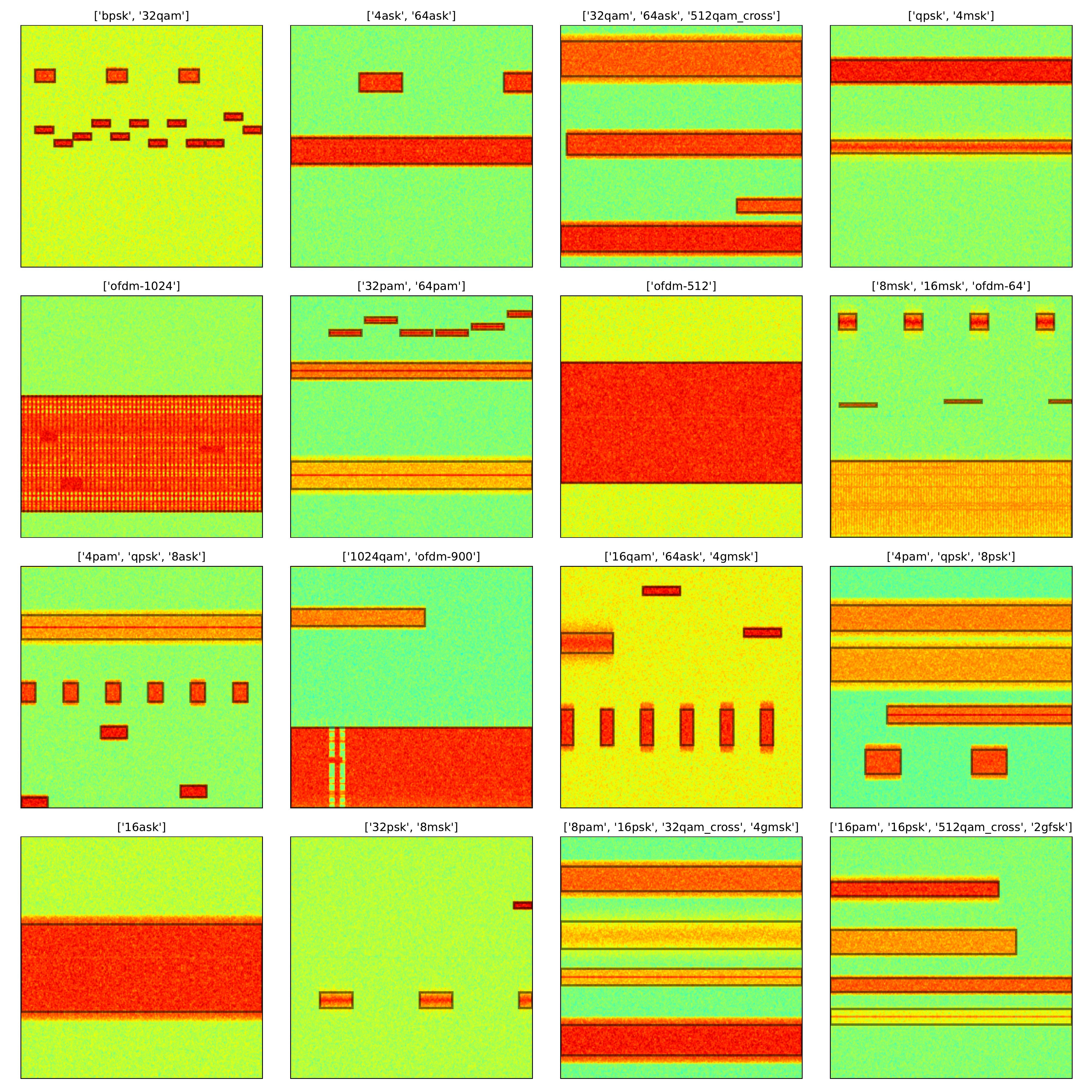}}
    \caption{Amplitude Reversal Augmentation}
    \label{fig:amplitude_reversal}
\end{figure*}

\textbf{Amplitude Reversal.} Amplitude reversal augments the input data by simply multiplying by $-1$ (\cref{fig:amplitude_reversal}).

\begin{figure*}[!h]
    \centering
    \subfloat[Original Data]{\includegraphics[width=0.40\textwidth]{images/wbsig53_clean.pdf}}
    \subfloat[Impaired Data]{\includegraphics[width=0.40\textwidth]{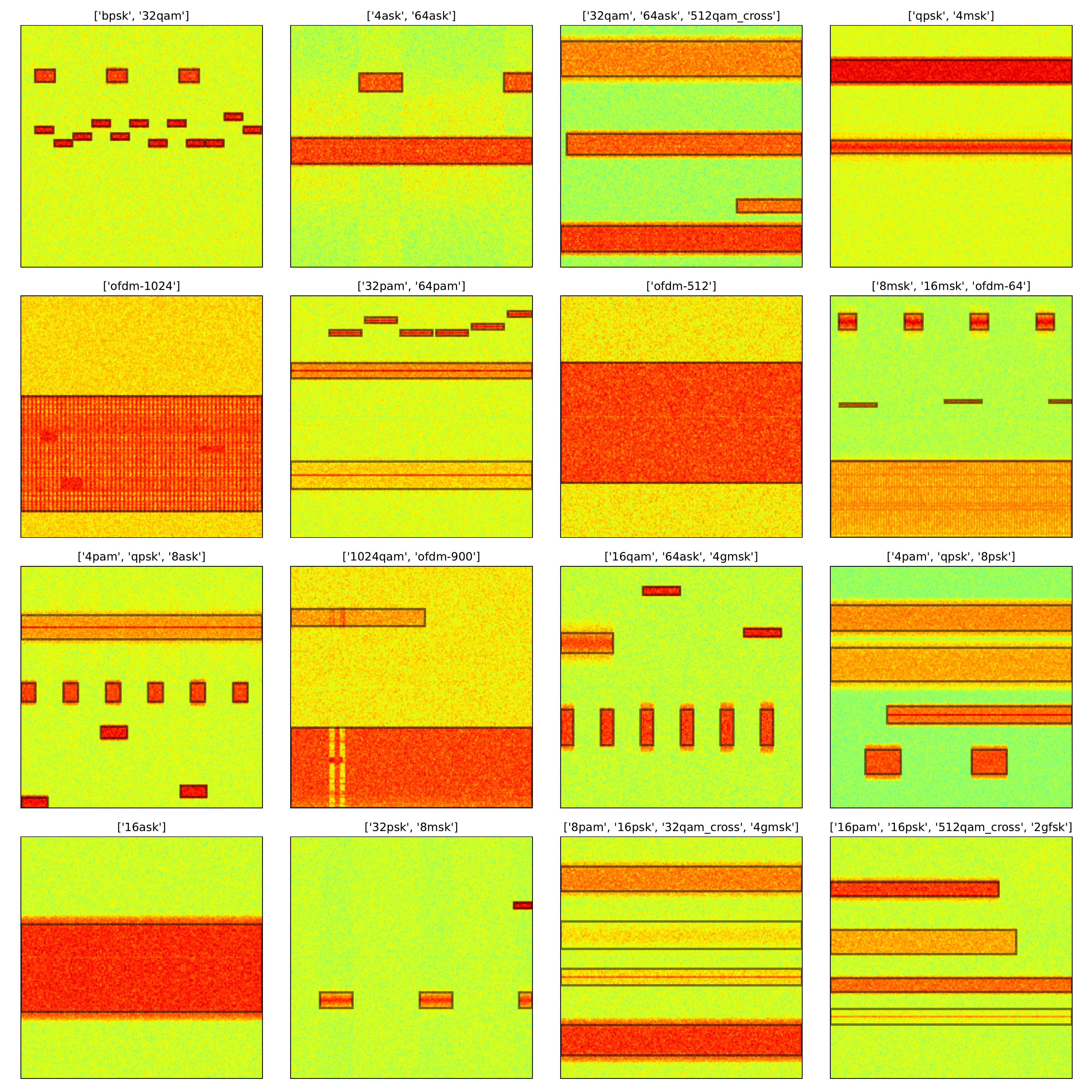}}
    \caption{Quantize Augmentation}
    \label{fig:quantize}
\end{figure*}

\clearpage
\textbf{Quantize.} Input data is quantized to a randomly selected number of levels with the quantize data transform, 
loosely emulating the bit-depth in an analog-to-digital converter (ADC) seen in digital RF systems. 
The quantization transform also allows for various rounding types between: 
flooring the observed values to the next-lowest valid quantized value, 
setting every value in a region to the middle value of the region, 
or rounding each value to the next largest valid quantized value (\cref{fig:quantize}).

\begin{figure*}[!h]
    \centering
    \subfloat[Original Data]{\includegraphics[width=0.40\textwidth]{images/wbsig53_clean.pdf}}
    \subfloat[Impaired Data]{\includegraphics[width=0.40\textwidth]{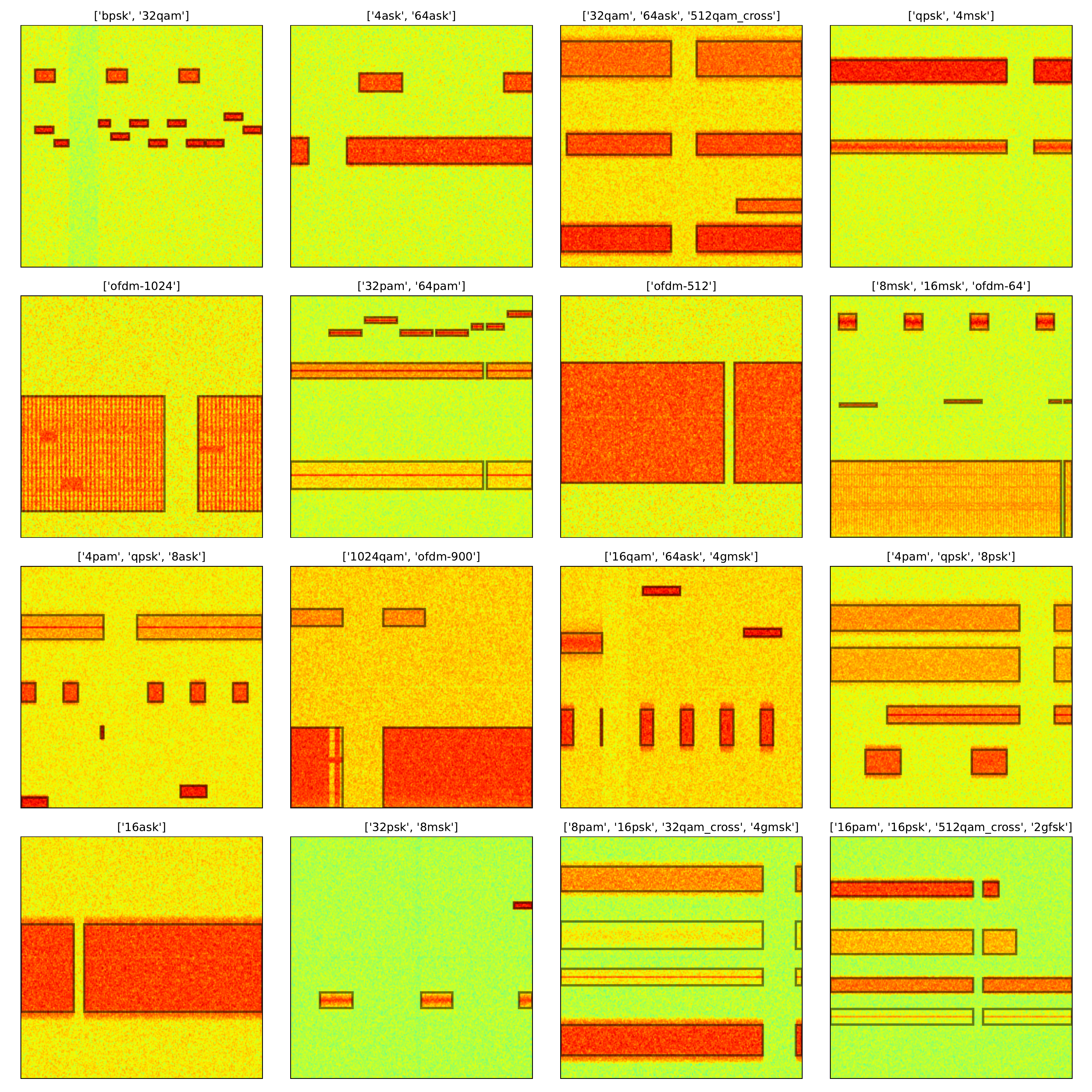}}
    \caption{CutOut Augmentation}
    \label{fig:cutout}
\end{figure*}

\textbf{CutOut.} The CutOut transform is a modified version of the computer vision domain's CutOut as seen in \cite{devries2017improved}.
Our version of CutOut inputs randomized cut durations and cut types to select how large a region in time should be cut out. 
The cut out region is then filled with either zeros, ones, low-SNR noise, average-SNR noise, or high-SNR noise.
In the context of wideband data, the CutOut augmentation appropriately omits the cut-out regions of signals.
If the cut-out region falls in the middle of a longer signal, the signal label is divided into separate signals of appropriate size.
The CutOut effects can be seen in \cref{fig:cutout}.

\begin{figure*}[!h]
    \centering
    \subfloat[Original Data]{\includegraphics[width=0.40\textwidth]{images/wbsig53_clean.pdf}}
    \subfloat[Impaired Data]{\includegraphics[width=0.40\textwidth]{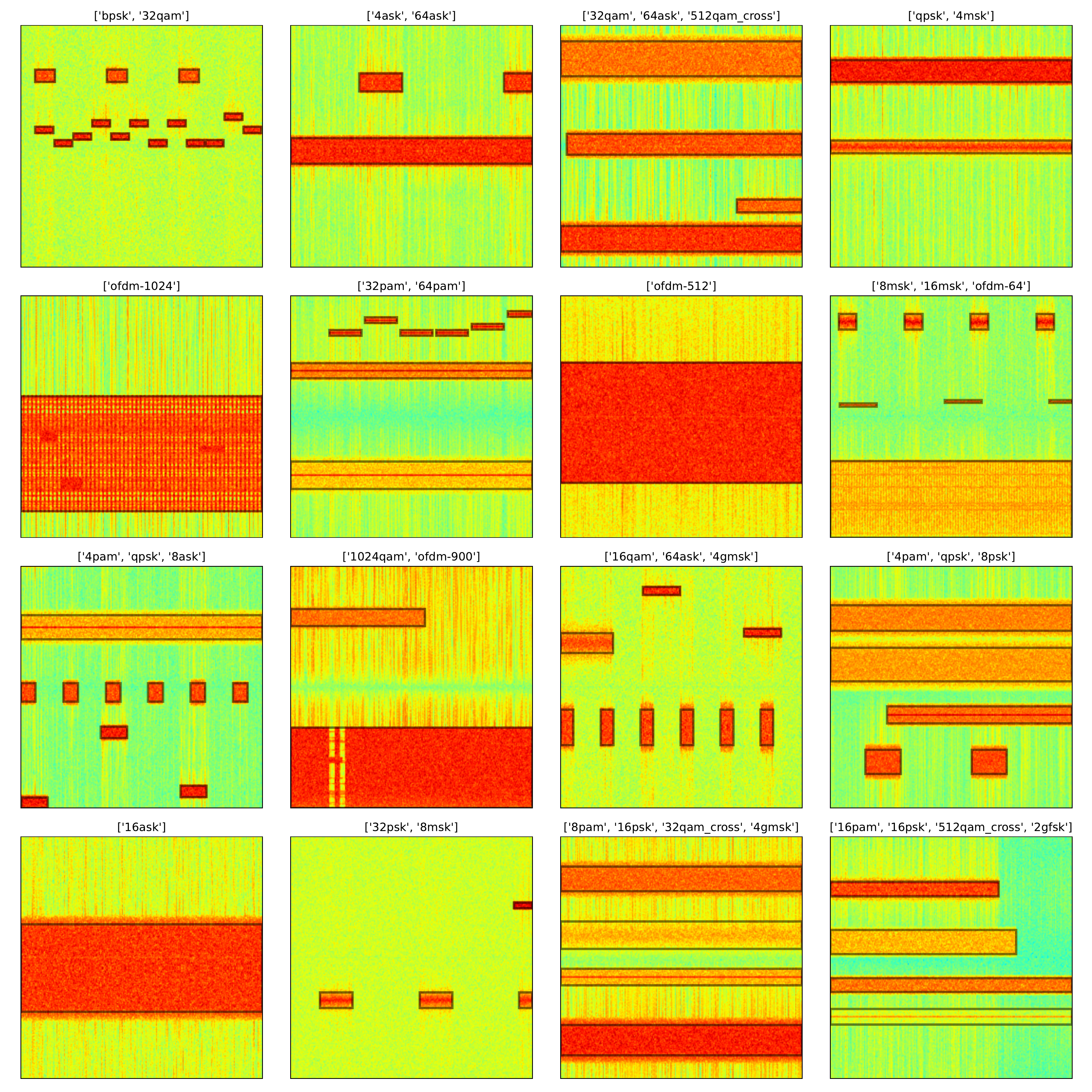}}
    \caption{PatchShuffle Augmentation}
    \label{fig:patch_shuffle}
\end{figure*}

\clearpage
\textbf{PatchShuffle.} The PatchShuffle transform is a modified version of the computer vision domain's PatchShuffle as seen in \cite{kang2017patchshuffle}. 
Our version of PatchShuffle operates solely in the time domain, 
randomly shuffling multiple local regions of IQ samples, 
using a randomized patch size input distribution and a randomized shuffle ratio to discern how many of the patches should undergo random local shuffling (\cref{fig:patch_shuffle}).

\begin{figure*}[!h]
    \centering
    \subfloat[Original Data]{\includegraphics[width=0.40\textwidth]{images/wbsig53_clean.pdf}}
    \subfloat[Impaired Data]{\includegraphics[width=0.40\textwidth]{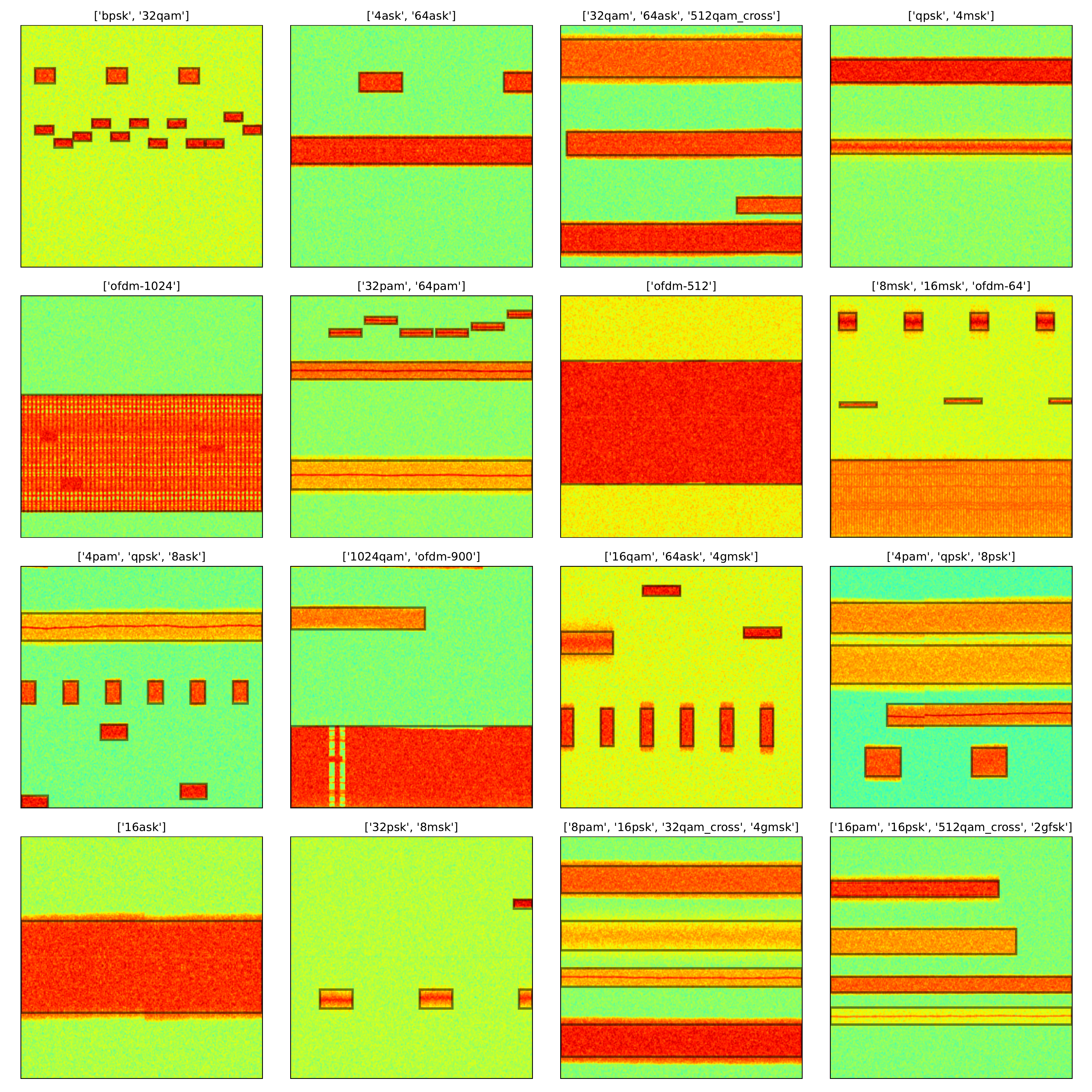}}
    \caption{Local Oscillator Drift Augmentation}
    \label{fig:lo_drift}
\end{figure*}

\textbf{Local Oscillator Drift.} The local oscillator (LO) drift transform emulates the imperfections of a receiver's LO. 
This transform models the LO drift by implementing a random walk in frequency with a drift rate and a max drift set as inputs, 
where when the max drift is reached, the frequency offset is reset to $0$ (\cref{fig:lo_drift}).

\begin{figure*}[!h]
    \centering
    \subfloat[Original Data]{\includegraphics[width=0.40\textwidth]{images/wbsig53_clean.pdf}}
    \subfloat[Impaired Data]{\includegraphics[width=0.40\textwidth]{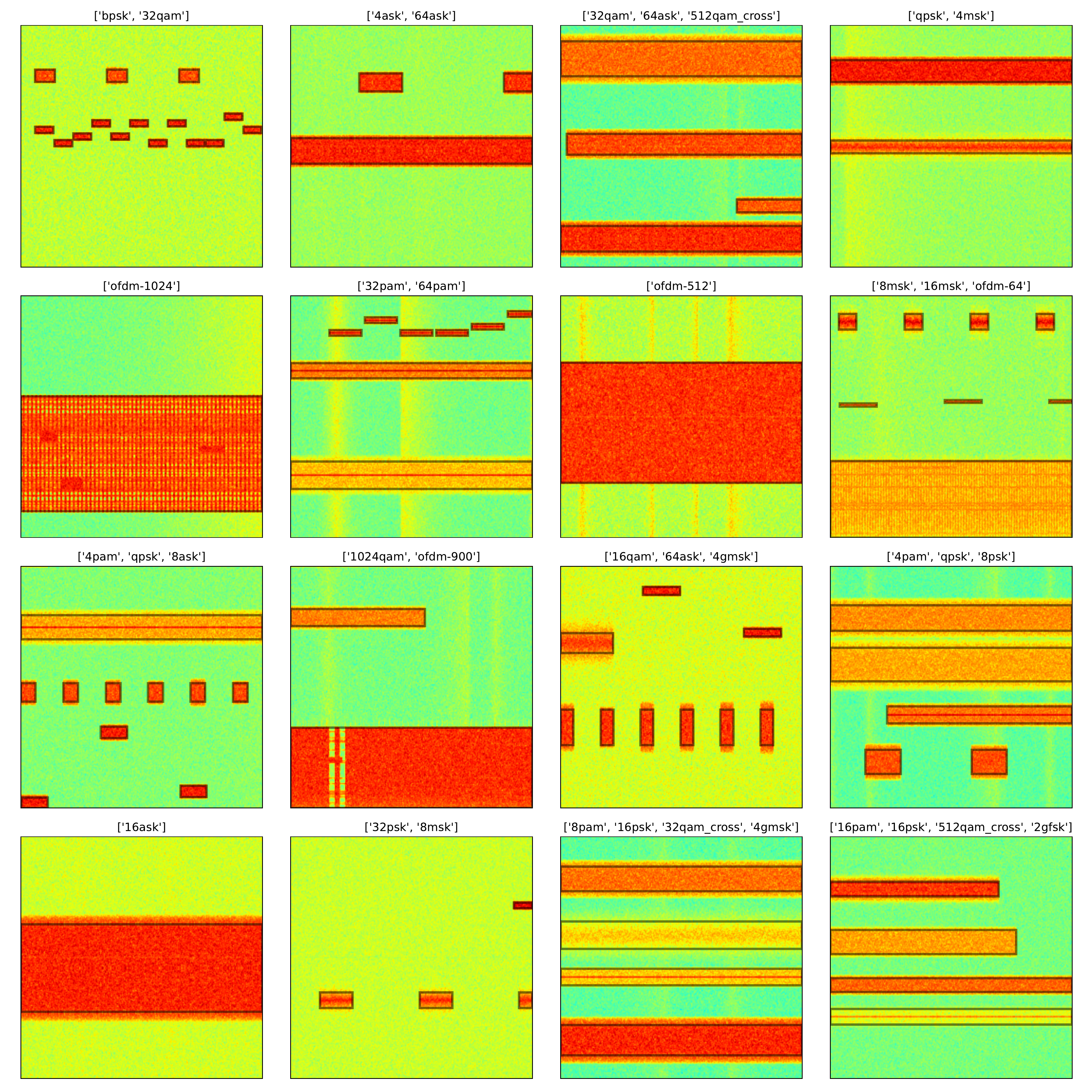}}
    \caption{Time-Varying Noise Augmentation}
    \label{fig:time_varying_noise}
\end{figure*}

\clearpage
\textbf{Time-Varying Noise.} The time-varying transform adds AWGN within a specified low to high SNR range with a specified number of inflection points at which point(s) the slope of the time-varying noise reverses direction (\cref{fig:time_varying_noise}).

\begin{figure*}[!h]
    \centering
    \subfloat[Original Data]{\includegraphics[width=0.40\textwidth]{images/wbsig53_clean.pdf}}
    \subfloat[Impaired Data]{\includegraphics[width=0.40\textwidth]{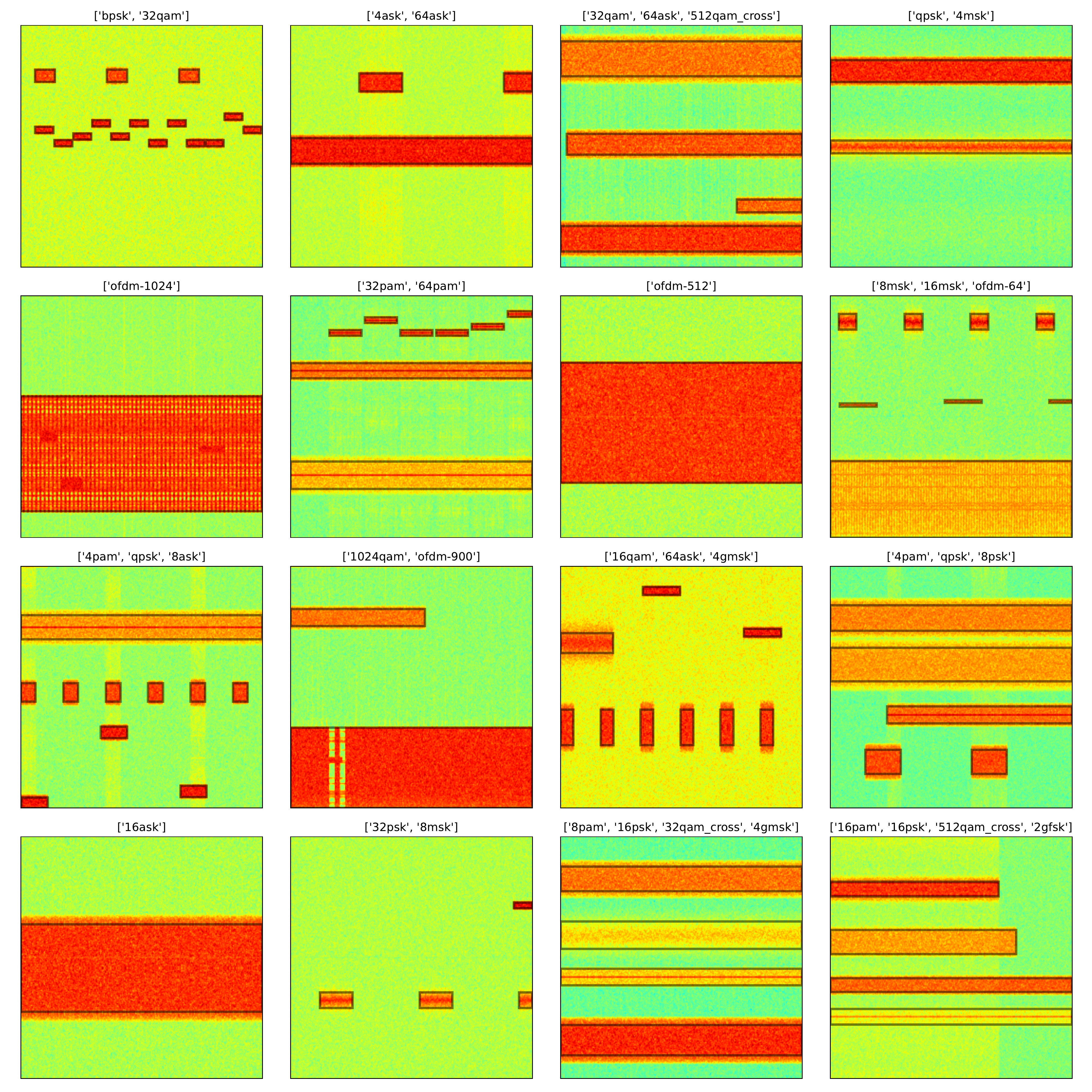}}
    \caption{Clip Augmentation}
    \label{fig:clip}
\end{figure*}

\textbf{Clip.} The clip transform inputs a percentage that it uses to calculate the max and min values allowable through the clipping transform, 
setting all values above and below these values to the max and min, respectively (\cref{fig:clip}).

\begin{figure*}[!h]
    \centering
    \subfloat[Original Data]{\includegraphics[width=0.40\textwidth]{images/wbsig53_clean.pdf}}
    \subfloat[Impaired Data]{\includegraphics[width=0.40\textwidth]{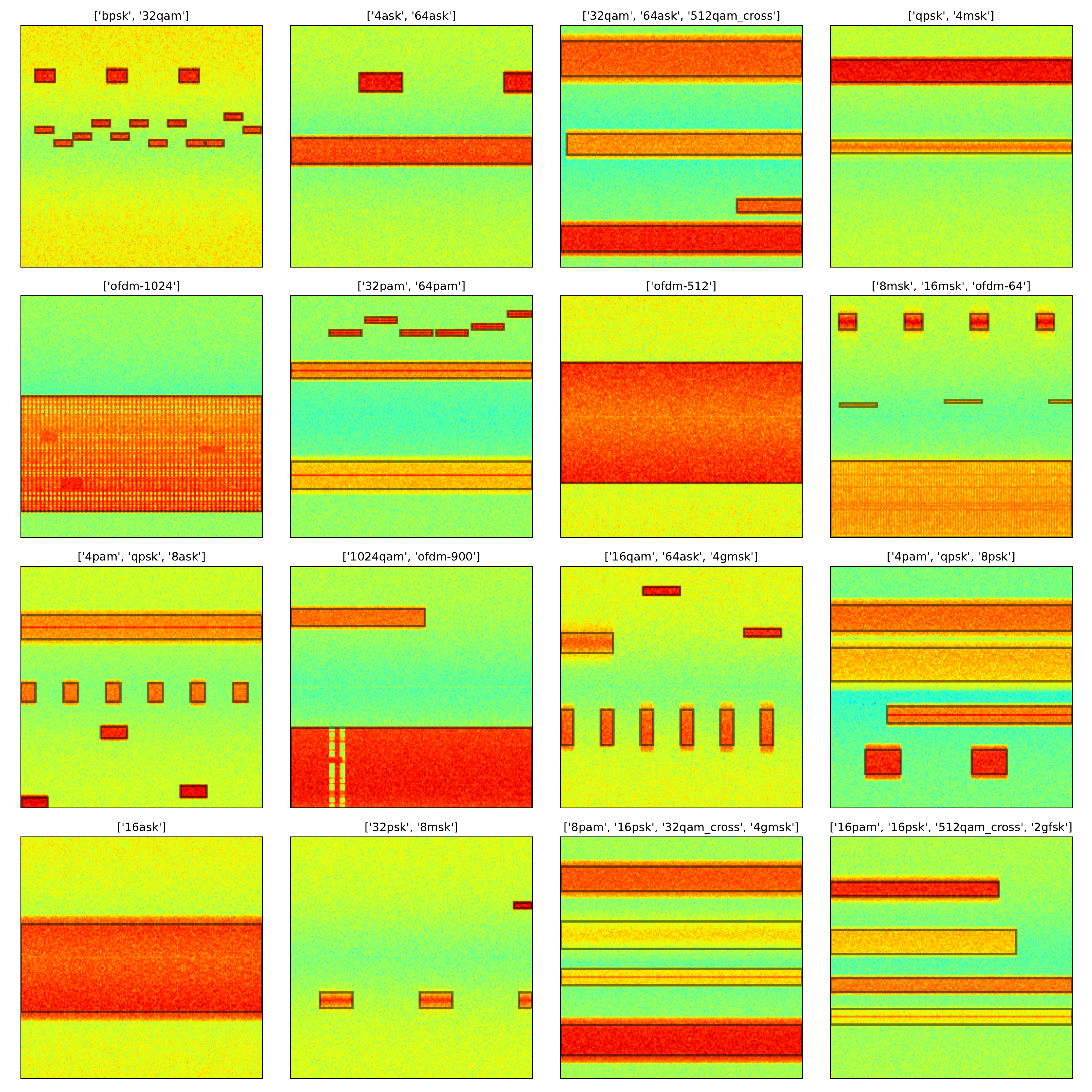}}
    \caption{Add Slope Augmentation}
    \label{fig:add_slope}
\end{figure*}

\clearpage
\textbf{Add Slope.} The add slope transform computes the slope between every IQ sample in the input with its preceding sample and adds the slope to its current IQ sample. 
This transform has the effect of amplifying higher frequency components more than the lower frequency components (\cref{fig:add_slope}).

\begin{figure*}[!h]
    \centering
    \subfloat[Original Data]{\includegraphics[width=0.40\textwidth]{images/wbsig53_clean.pdf}}
    \subfloat[Impaired Data]{\includegraphics[width=0.40\textwidth]{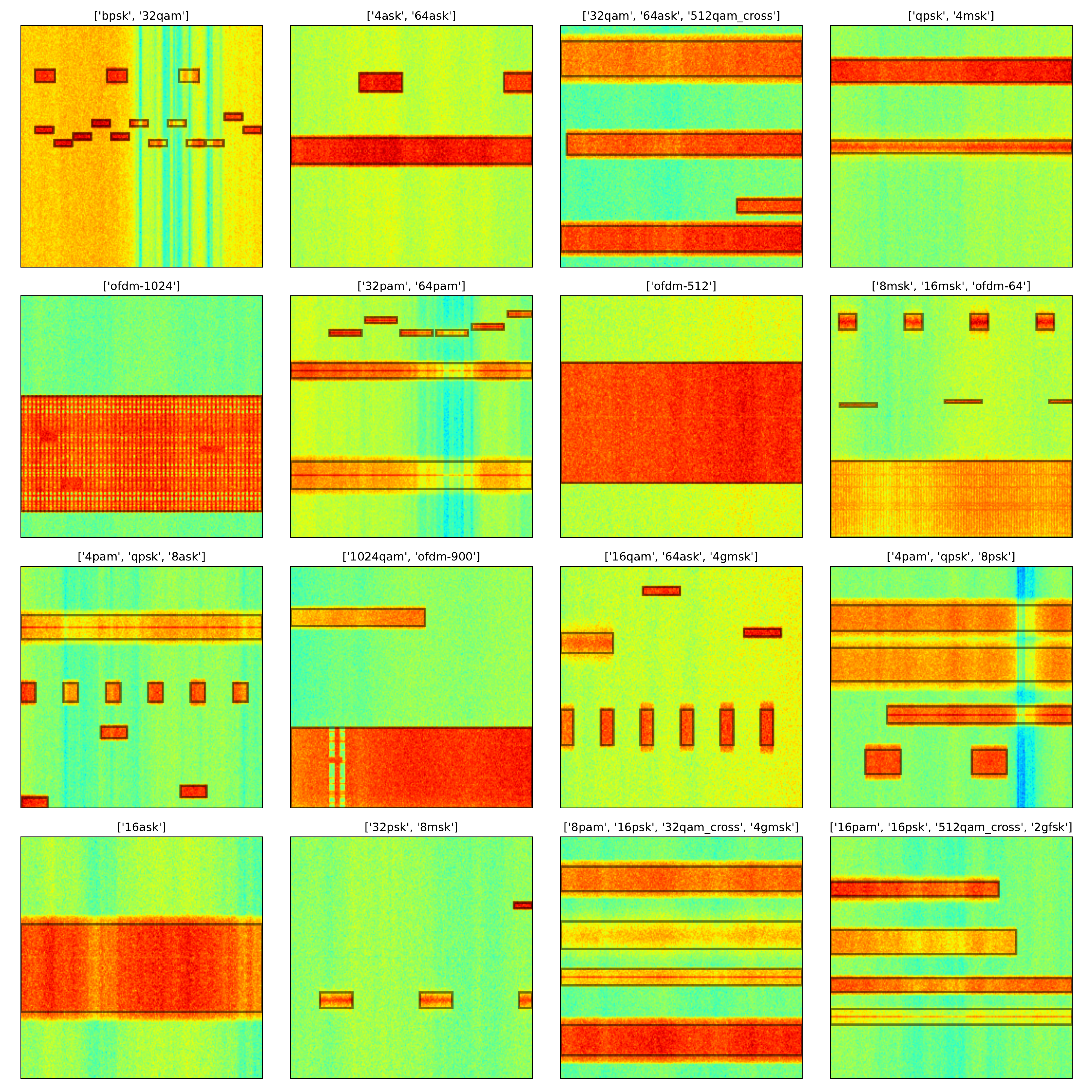}}
    \caption{Gain Drift Augmentation}
    \label{fig:gain_drift}
\end{figure*}

\textbf{Gain Drift.} A gain drift transform is also implemented to complement the LO drift transform's frequency effects with magnitude effects. 
The gain drift transform inputs similar max/min drift values and a drift rate, 
which are used in a random walk of adjusting the magnitudes of the input data IQ samples over time (\cref{fig:gain_drift}).

\begin{figure*}[!h]
    \centering
    \subfloat[Original Data]{\includegraphics[width=0.38\textwidth]{images/wbsig53_clean.pdf}}
    \subfloat[Impaired Data]{\includegraphics[width=0.38\textwidth]{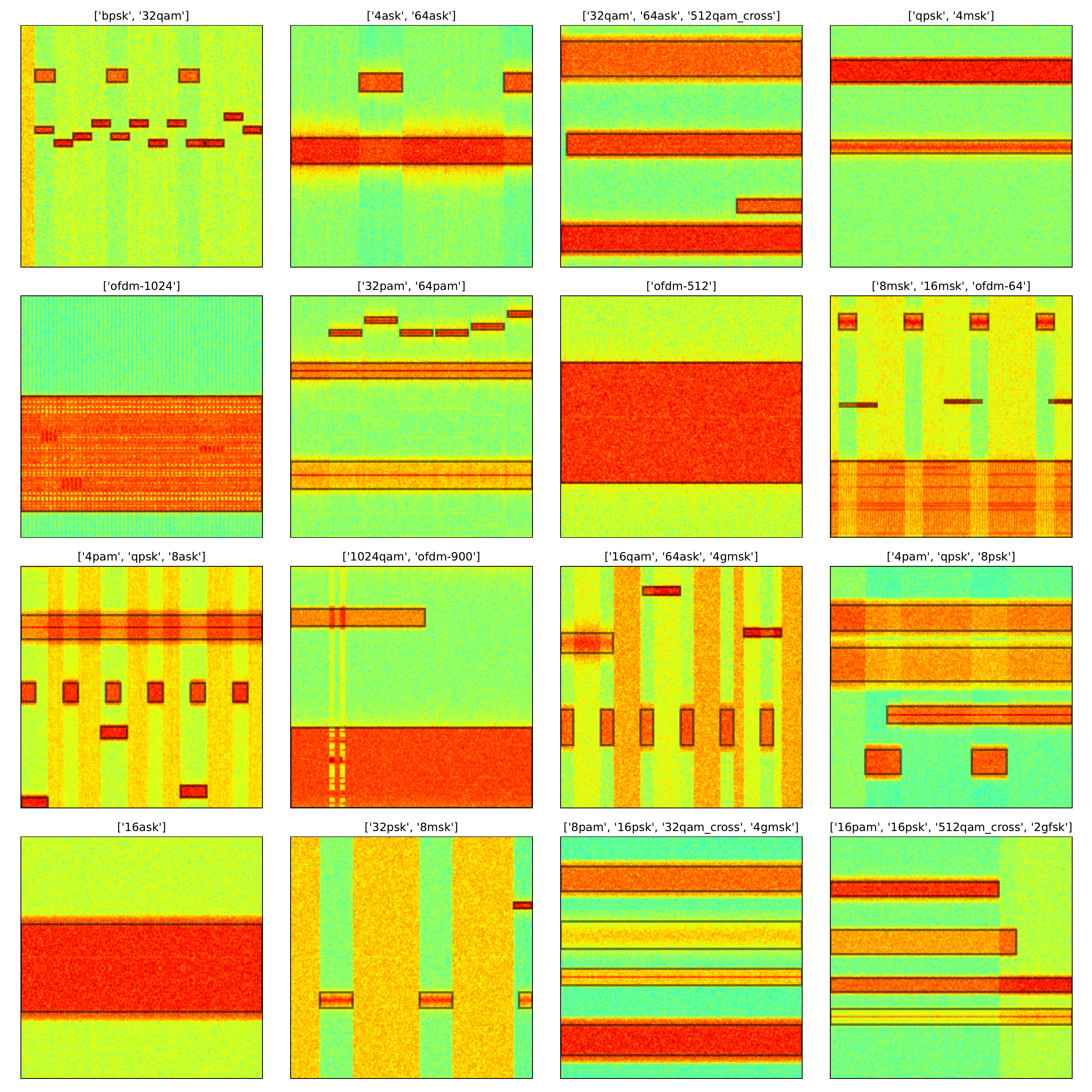}}
    \caption{Automatic Gain Control Augmentation}
    \label{fig:agc}
\end{figure*}

\clearpage
\textbf{Automatic Gain Control.} An automatic gain control (AGC) implementation is included in TorchSig as a data transform with an input scaling of default values for randomization across augmentation calls. 
The default values under random scaling can also be updated with arguments including: 
an initial gain value, 
an alpha for averaging the measure signal level, 
an alpha amount by which to adjust gain when in the tracking state, 
an alpha value by which to adjust gain when in the overflow state, 
an alpha value by which to adjust gain when in the acquire state, 
a reference level specifying the level of intended gain adjustment, 
a tracking range of allowable deviation before going into the acquire state, 
a low level which specifies when the AGC is disabled, and 
a high level which specifies when the AGC enters the overflow state (\cref{fig:agc}).

\begin{figure*}[!h]
    \centering
    \subfloat[Original Data]{\includegraphics[width=0.38\textwidth]{images/wbsig53_clean.pdf}}
    \subfloat[Impaired Data]{\includegraphics[width=0.38\textwidth]{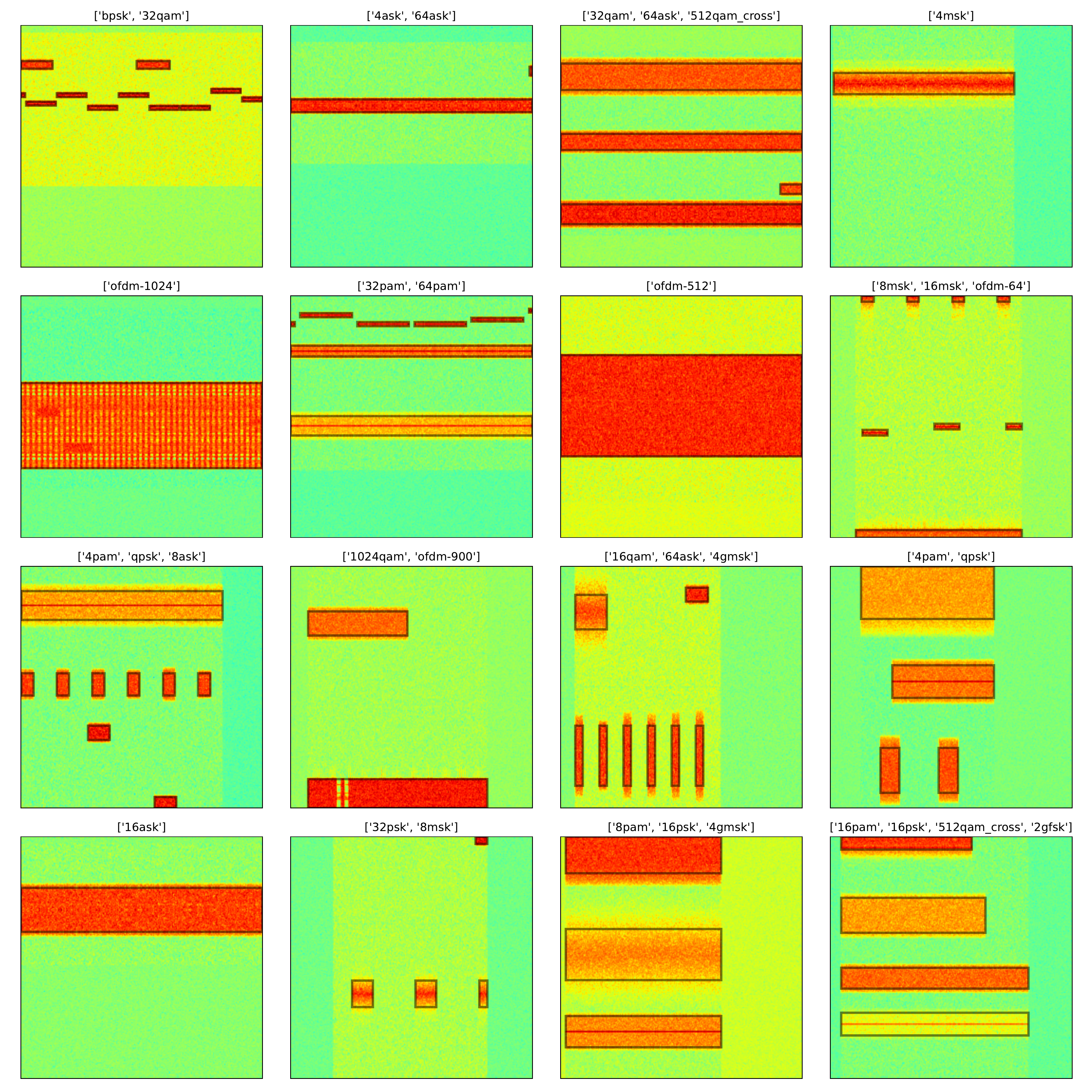}}
    \caption{Spectrogram Random Resize Crop Augmentation}
    \label{fig:spec_rrc}
\end{figure*}

\textbf{Spectrogram Random Resize Crop.} A common augmentation used in the vision domain is to randomly resize an input image and then crop the new image down to the expected input size.
Here, we emulate this technique by randomizing the parameters of the spectrogram operation.
The randomized paramters of the spectrogram operation result in variable sized spectrograms which we then crop down to the expected input size.
If the spectrogram is smaller than a desired dimension, we pad with emulated noise (\cref{fig:spec_rrc}).

\begin{figure*}[!h]
    \centering
    \subfloat[Original Data]{\includegraphics[width=0.40\textwidth]{images/wbsig53_clean.pdf}}
    \subfloat[Impaired Data]{\includegraphics[width=0.40\textwidth]{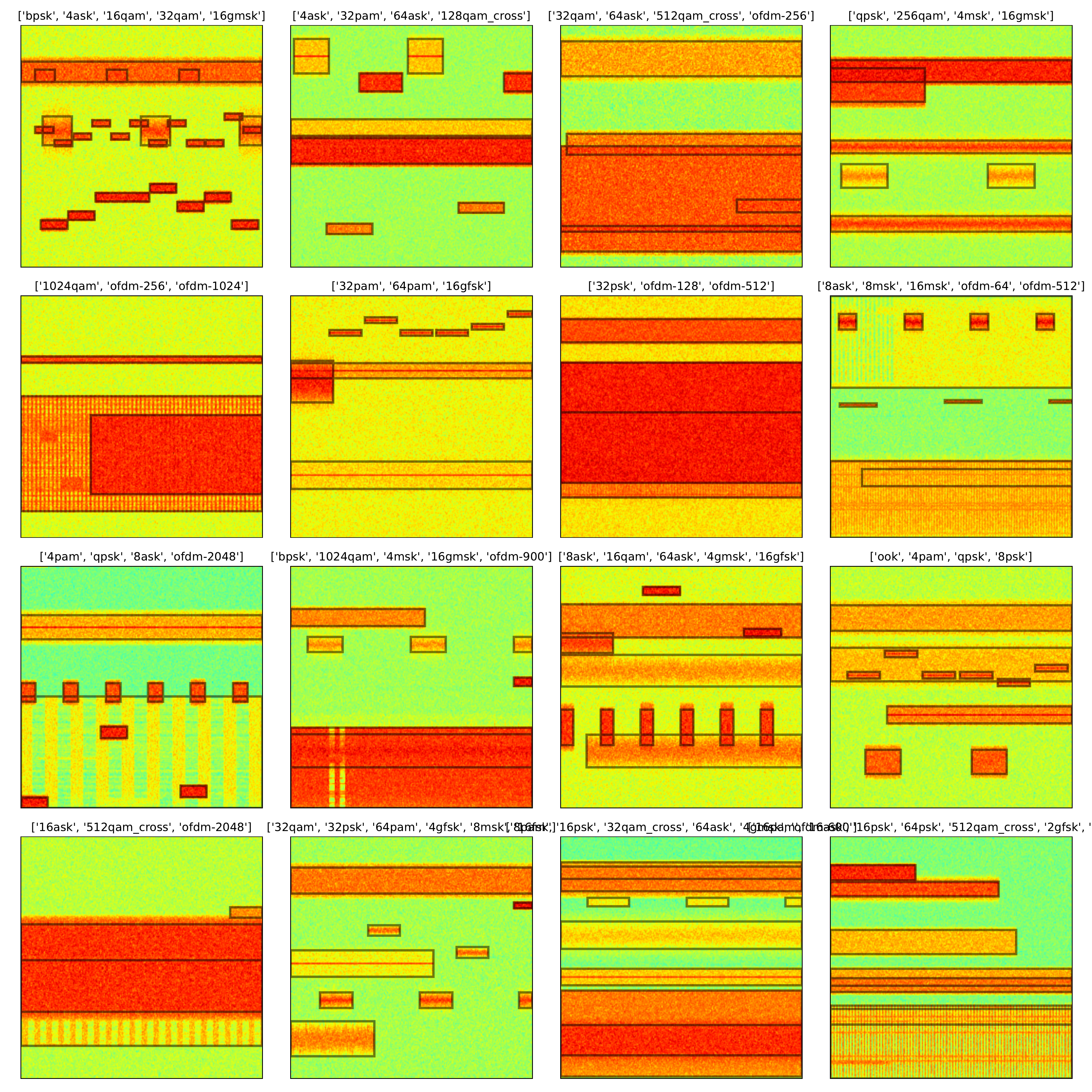}}
    \caption{MixUp Augmentation}
    \label{fig:mixup}
\end{figure*}

\clearpage
\textbf{MixUp.} The wideband MixUp transform implements a modified version of the vision domain's MixUp \citep{zhang2018mixup}.
Our MixUp transform inputs a secondary dataset from which additional data samples are drawn.
The new data sample is then mixed with the original data sample, resulting in additional signals being added to the data sample.
All labels from the new sample are added to the original label, such that a network can now be tasked with discovering additional, potentially-overlapping signals.
This effect is seen in \cref{fig:mixup}.

\begin{figure*}[!h]
    \centering
    \subfloat[Original Data]{\includegraphics[width=0.40\textwidth]{images/wbsig53_clean.pdf}}
    \subfloat[Impaired Data]{\includegraphics[width=0.40\textwidth]{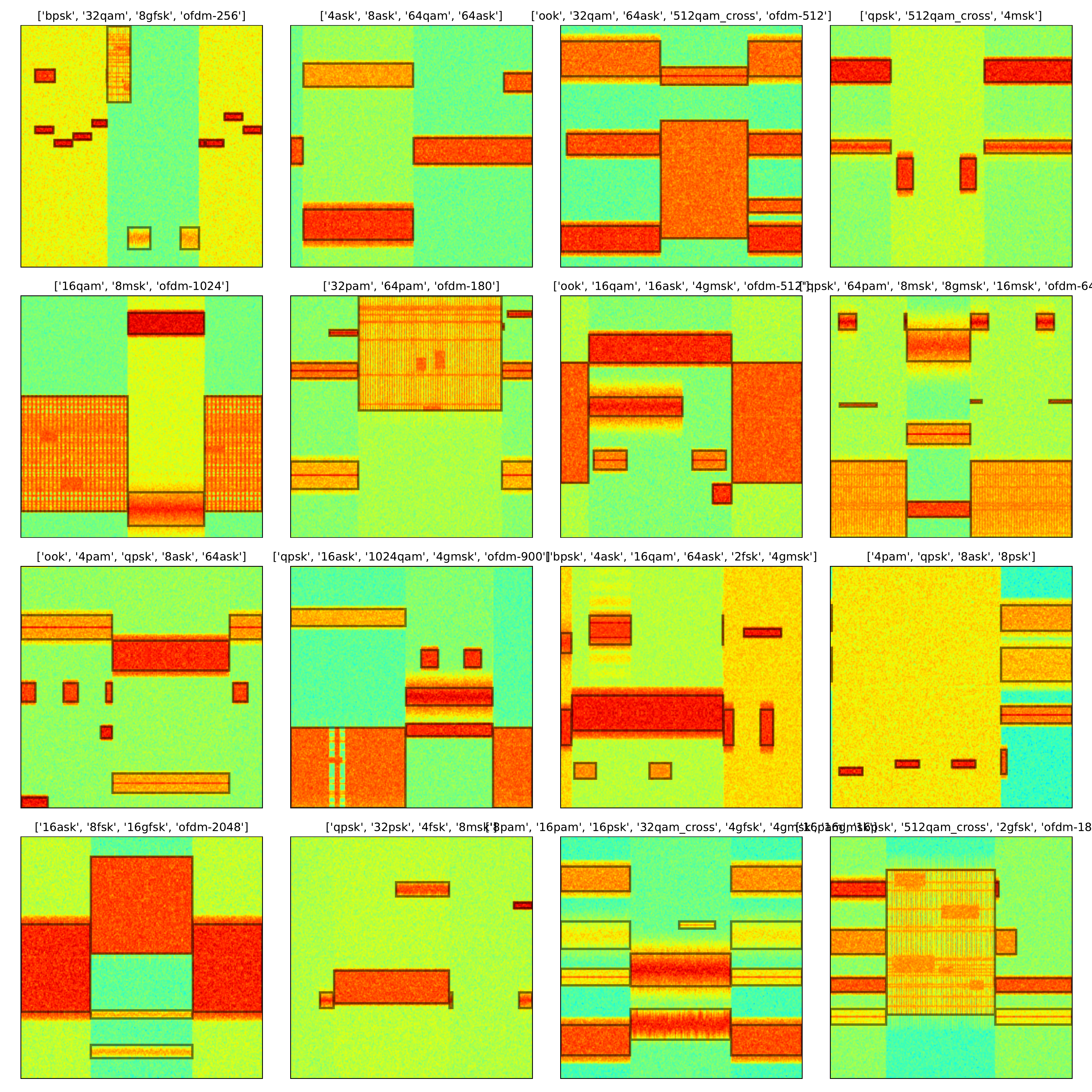}}
    \caption{CutMix Augmentation}
    \label{fig:cutmix}
\end{figure*}

\textbf{CutMix.} The wideband CutMix transform implements a modified version of the vision domain's CutMix \citep{yun2019cutmix}.
Like the MixUp transform, out CutMix transform inputs a secondary dataset from which additional data samples are drawn.
A random region of the original data sample is cut out and replaced with a region of appropriate size from the secondary data sample.
All labels are adjusted to appropriately capture all old and new signals as seen in \cref{fig:cutmix}.

\clearpage
\subsubsection{Spectrogram Data Augmentations}

TorchSig also now supports data augmentations that are applied after the raw IQ data has been converted to a spectrogram representation.
Details on these spectrogram data augmentations are below.

\begin{figure*}[!h]
    \centering
    \subfloat[Original Data]{\includegraphics[width=0.40\textwidth]{images/wbsig53_clean.pdf}}
    \subfloat[Impaired Data]{\includegraphics[width=0.40\textwidth]{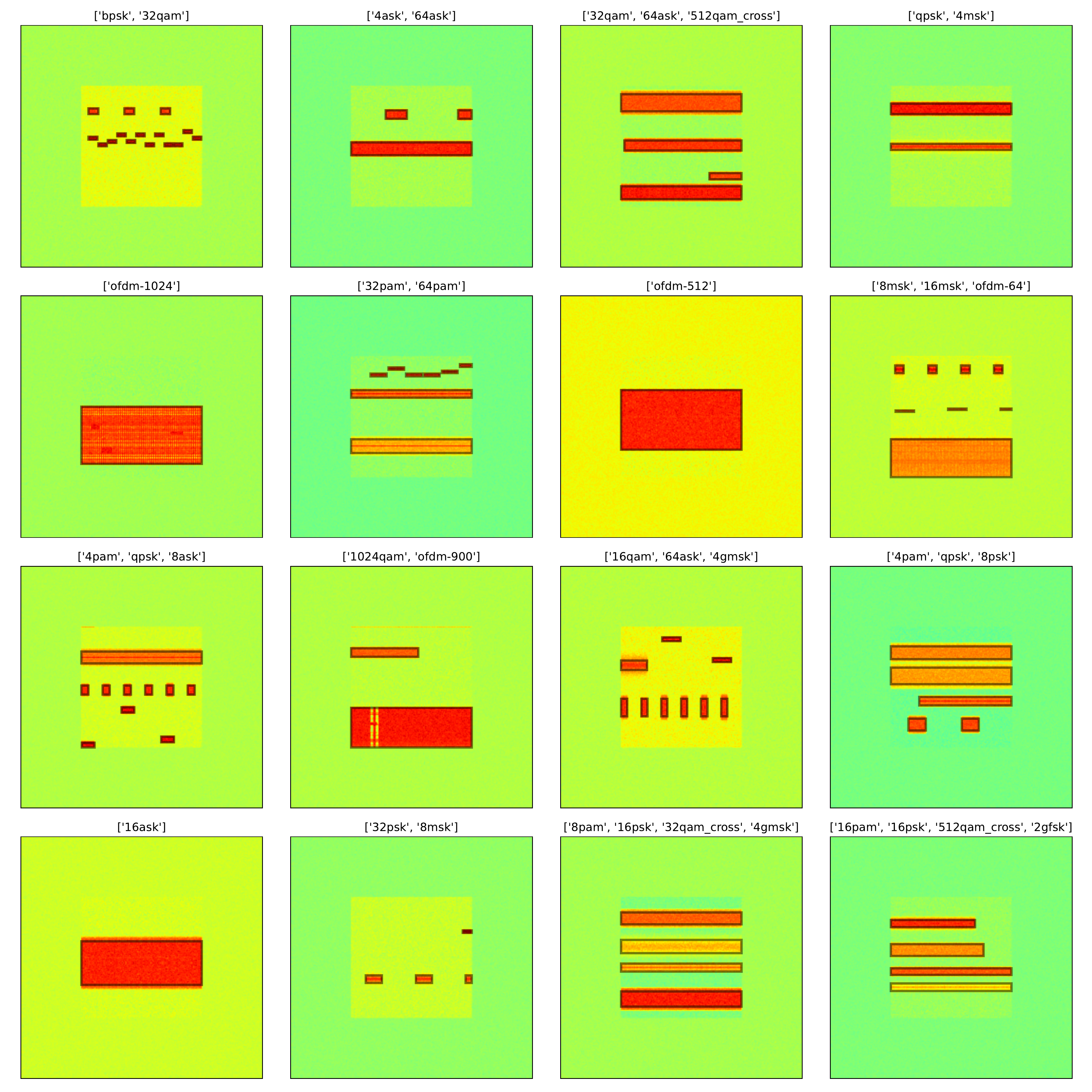}}
    \caption{Spectrogram Resize Transform}
    \label{fig:spec_resize}
\end{figure*}

\textbf{Spectrogram Resize.} The spectrogram resize transform inputs a spectrogram and resizes the data and labels to the specified dimensions.
If the input is larger than the desired size, the input is cropped.
If the input is too small to meet the desired size, the outer regions are padded with emulated noise as seen in \cref{fig:spec_resize}.

\begin{figure*}[!h]
    \centering
    \subfloat[Original Data]{\includegraphics[width=0.40\textwidth]{images/wbsig53_clean.pdf}}
    \subfloat[Impaired Data]{\includegraphics[width=0.40\textwidth]{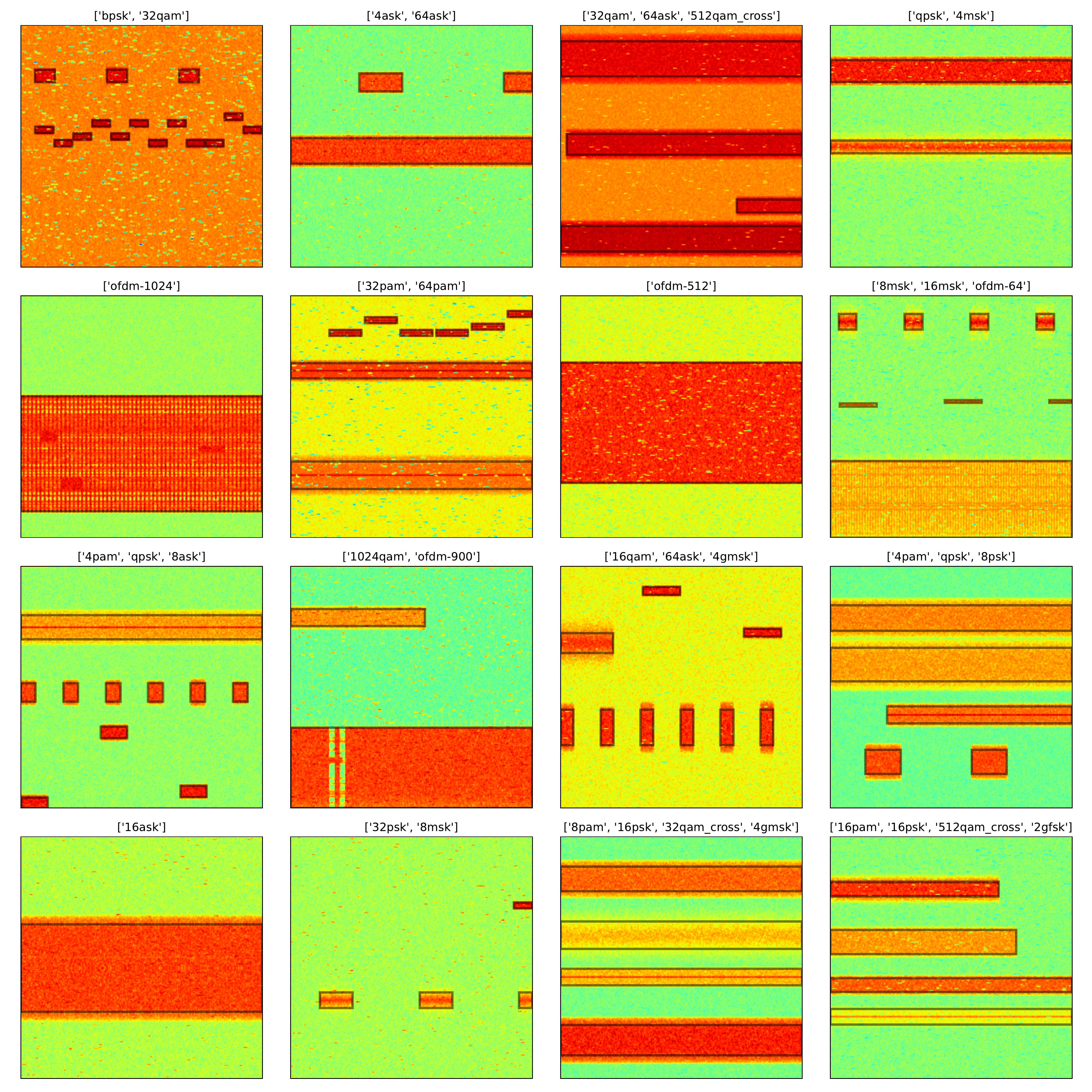}}
    \caption{Spectrogram Drop Samples Augmentation}
    \label{fig:spec_drop_samples}
\end{figure*}

\textbf{Spectrogram Drop Samples.} The spectrogram drop samples transform mirrors the IQ data's drop sample transform by randomly dropping samples throughout an input spectrogram.
Similar to the IQ data drop samples transform, the dropped samples are replaced by either the front fill, back fill, the mean value, zeros, low values, the minimum value, the maximum value, and/or ones, depending on the input options provided.
This effect can be seen in \cref{fig:spec_drop_samples}.

\begin{figure*}[!h]
    \centering
    \subfloat[Original Data]{\includegraphics[width=0.40\textwidth]{images/wbsig53_clean.pdf}}
    \subfloat[Impaired Data]{\includegraphics[width=0.40\textwidth]{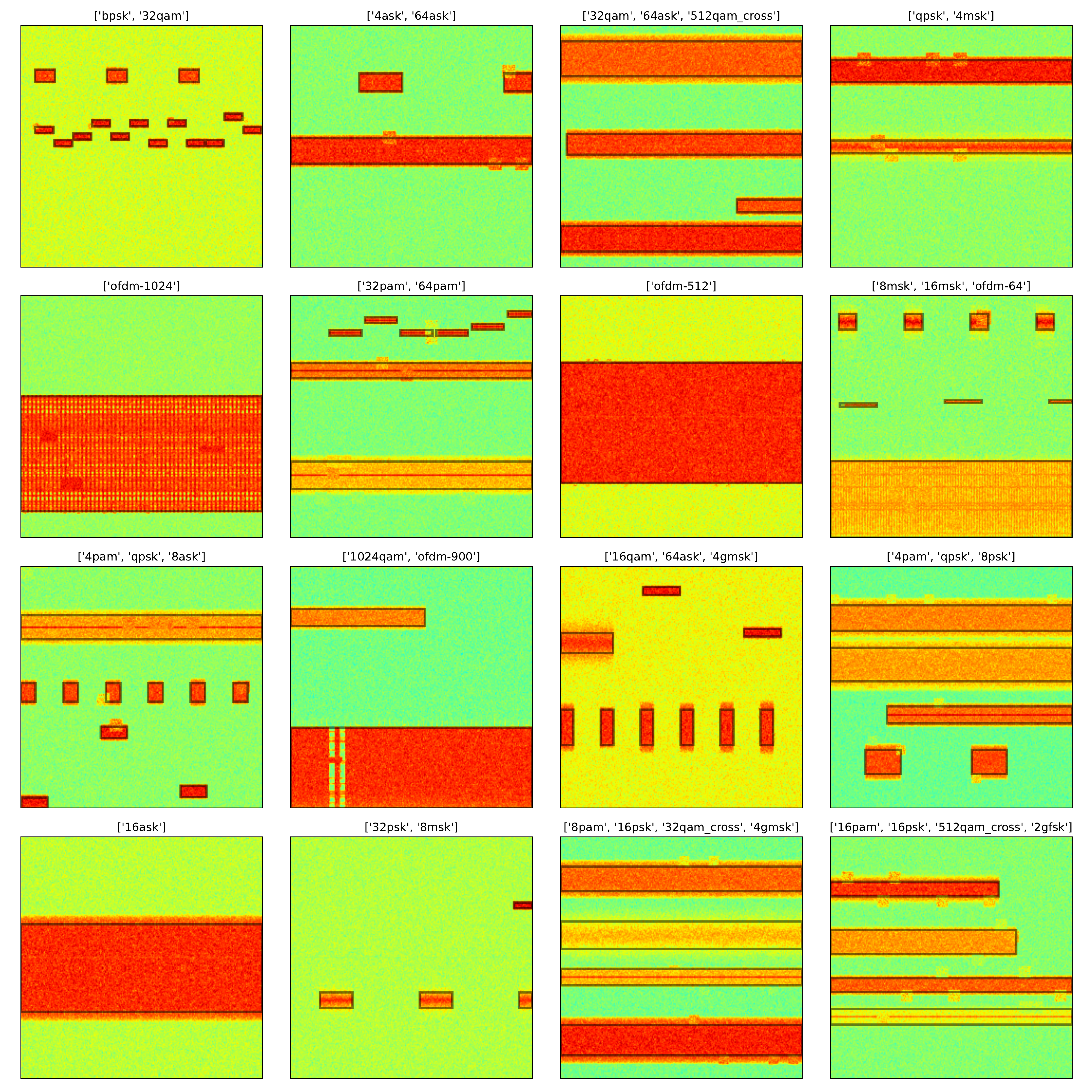}}
    \caption{Spectrogram PatchShuffle Augmentation}
    \label{fig:spec_patch_shuffle}
\end{figure*}

\clearpage
\textbf{Spectrogram PatchShuffle.} The spectrogram PatchShuffle transform mirrors the IQ data's PatchShuffle transform, 
but instead of shuffling local regions in time only, the spectrogram implementation performs the shuffling in time and frequency rectangular regions.
This effect is most easily seen at the edges of signals and can be viewed in \cref{fig:spec_patch_shuffle}.

\begin{figure*}[!h]
    \centering
    \subfloat[Original Data]{\includegraphics[width=0.40\textwidth]{images/wbsig53_clean.pdf}}
    \subfloat[Impaired Data]{\includegraphics[width=0.40\textwidth]{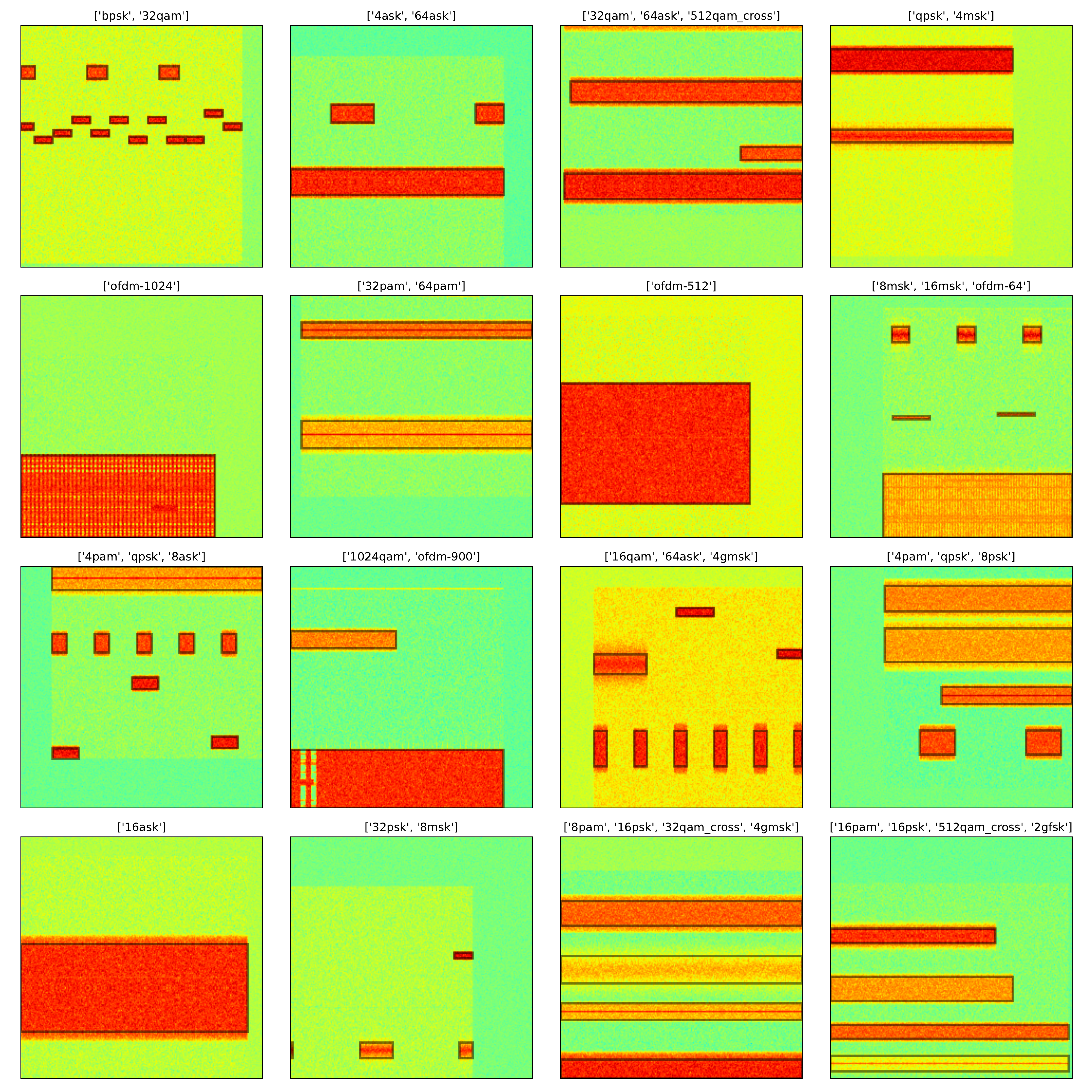}}
    \caption{Spectrogram Translation Augmentation}
    \label{fig:spec_translation}
\end{figure*}

\textbf{Spectrogram Translation.} The spectrogram translation augmentation applies a time and frequency shift of the spectrogram and fills new regions with emulated background noise (\cref{fig:spec_translation}).

\begin{figure*}[!h]
    \centering
    \subfloat[Original Data]{\includegraphics[width=0.40\textwidth]{images/wbsig53_clean.pdf}}
    \subfloat[Impaired Data]{\includegraphics[width=0.40\textwidth]{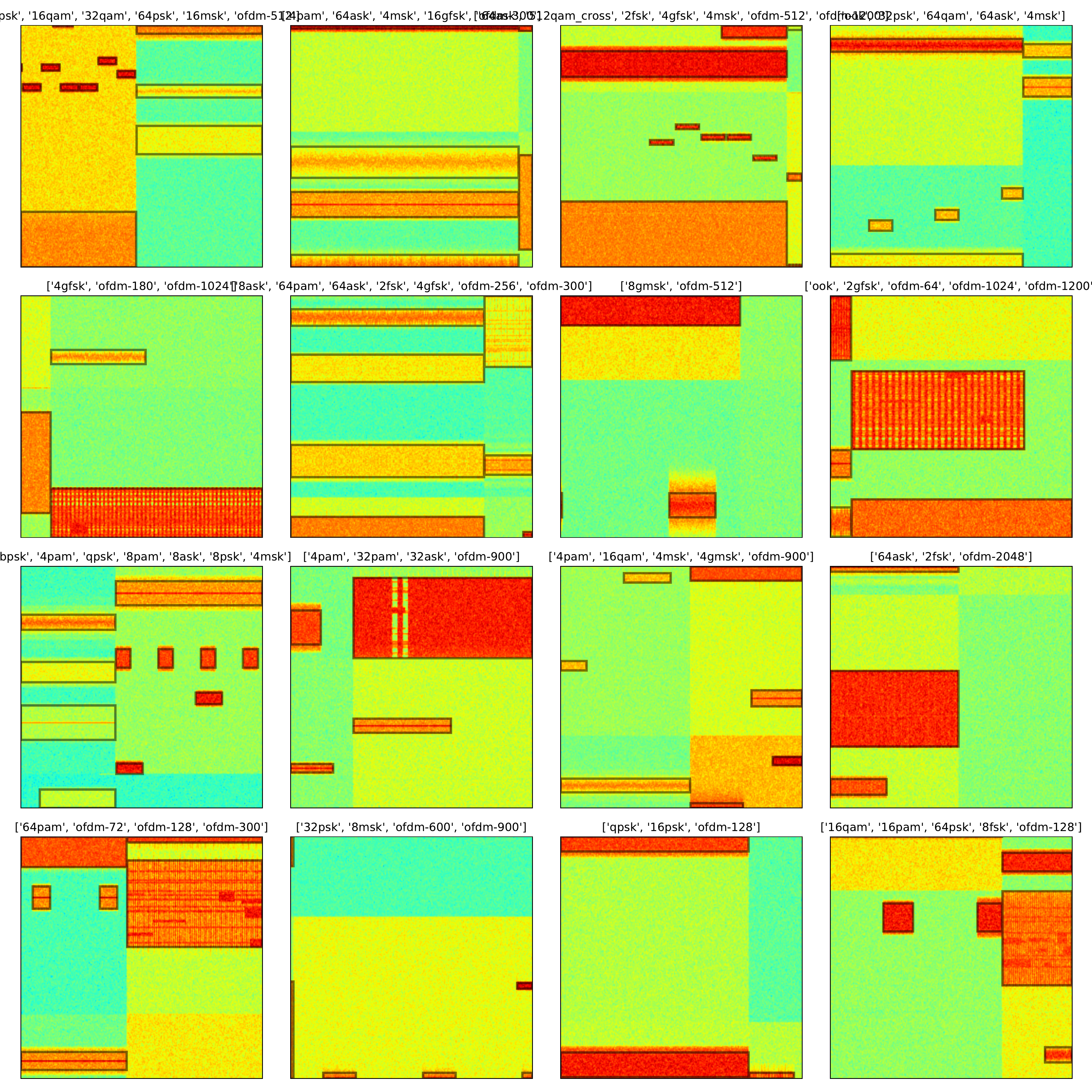}}
    \caption{Spectrogram Mosaic Crop Augmentation}
    \label{fig:spec_mosaic_crop}
\end{figure*}

\clearpage
\textbf{Spectrogram Mosaic Crop.} The spectrogram mosaic crop transform inputs a secondary dataset that is also transformed to an equivalent spectrogram representation, 
randomly samples three additional examples from the secondary dataset,
forms a $2x2$ grid of the four examples,
and then randomly crops a region the size of a single spectrogram.
The resulting spectrogram consists of signals from up to all four of the examples used in the generation of the grid.
\cref{fig:spec_mosaic_crop} shows examples of this augmentation.

\begin{figure*}[!h]
    \centering
    \subfloat[Original Data]{\includegraphics[width=0.40\textwidth]{images/wbsig53_clean.pdf}}
    \subfloat[Impaired Data]{\includegraphics[width=0.40\textwidth]{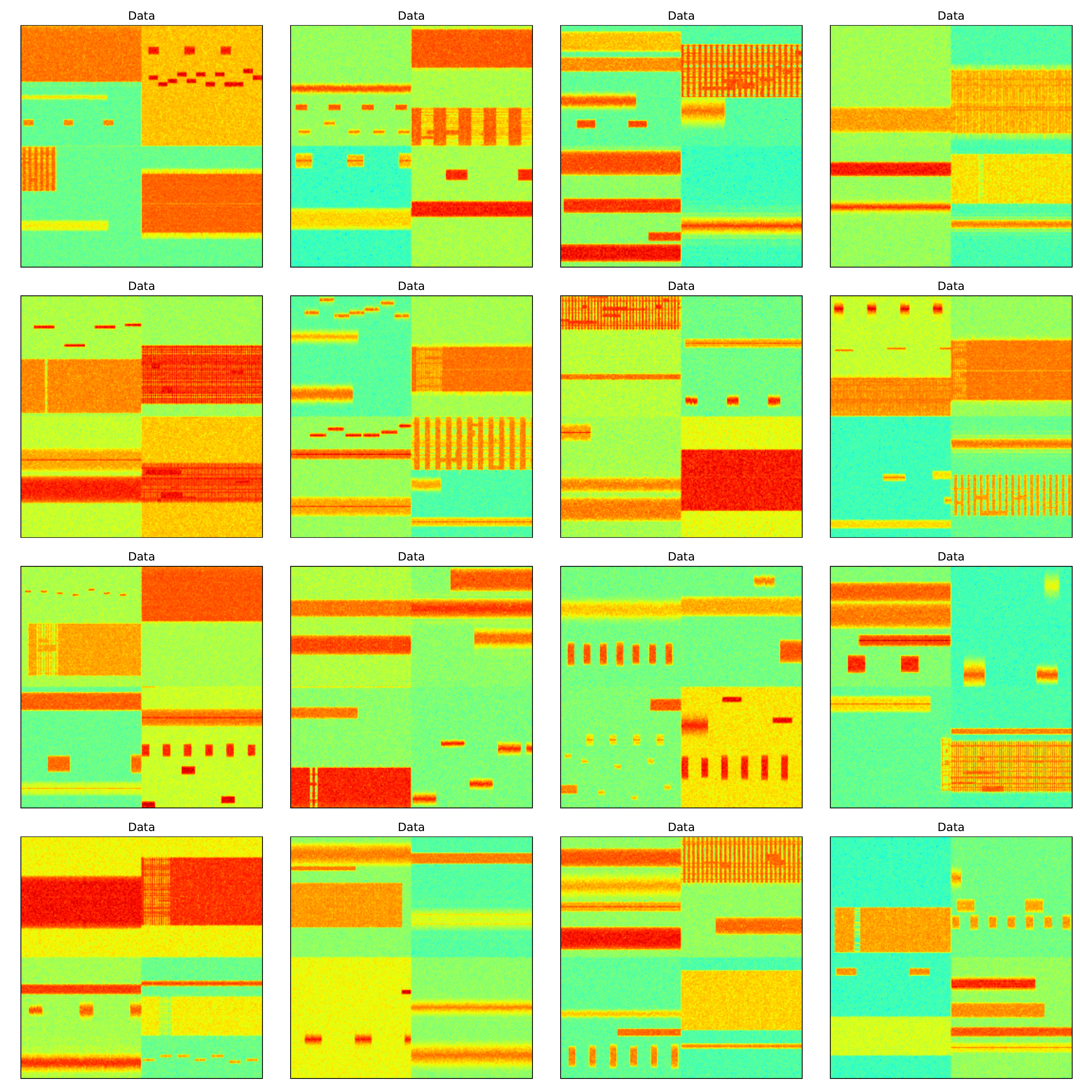}}
    \caption{Spectrogram Mosaic Downsample Augmentation}
    \label{fig:spec_mosaic_downsample}
\end{figure*}

\textbf{Spectrogram Mosaic Downsample.} The spectrogram mosaic downsample transform is similar to the spectrogram mosiac crop transform;
however, instead of cropping out a region, it downsamples the input such that all four spectrograms are fully contained at a lower resolution.
Note that this effect results in many smaller signals present in the transformed data.
\cref{fig:spec_mosaic_downsample} shows examples of this augmentation.
Note that the impaired data does retain all of the label information for each signal, 
but in the figure, the labels are omitted so the smaller signals can be more easily seen.

%% file: 12_model_appendix.tex
\clearpage
\subsection{Network Backgrounds}
\label{sec:appendix_model}

\subsubsection{You Only Look Once (YOLO)}

Prior to the introduction of YOLO, two stage detection methods detected possible object regions in an input image during the first stage and then (after feeding this information to another subsystem) classified the image in each of these regions during the second stage.  
In \cite{redmon2015yolo}, a new approach is introduced in which the neural network make bounding box predictions and class probabilities all at once.  
This algorithm produces SoTA results at much faster speeds.

The YOLOv1 algorithm breaks the input image into an array of $S \times S$ grid cells.  
Each grid cell produces $B$ bounding boxes and $C$ conditional class probabilities, 
where $C$ is the number of possible classes.  
Each bounding box contains $5$ predictions, representing its $(x,y)$ center location, width, height, and confidence score.  
This confidence score incorporates the probability that the bounding box contains an object as well as the accuracy of the bounding box location and size.  
Thus, the predictions produced by each input image are encoded as an output tensor of size $S \times S \times (B \times 5 + C)$.  
The system proposed in \cite{redmon2015yolo} assigns $S=7$, $B=2$ while processing images with $C=20$ classes.  
Thus, the output tensor has size: $7 \times 7 \times 30$.  
Non-maximal suppression is used to pare down the information from this output tensor to a final set of object classes and their respective locations.  
Note that every grid cell predicts only a single set of $C$ conditional class probabilities.  
Thus, objects of only one class are predicted from each grid cell.  
What if two different classes of objects have centers that fall into the same grid cell?  
This is addressed in YOLOv2.

YOLOv2 is introduced in \cite{redmon2016yolov2}, which has many improvements over YOLOv1.  
Changes are made to the network as well as the training procedures.  
A notable change is the replacement of bounding boxes with anchor boxes.  
The authors choose to break each input image into an array of $13 \times 13$ grid cells, with each grid cell containing 5 anchor boxes.  
The authors also introduced a method that makes use of the training data to determine the width and height of the anchor boxes.  
Another notable change in YOLOv2 is that it can handle input images of different sizes.  
The same authors published \cite{redmon2018yolov3} a couple of years later.  
Additional incremental improvements are made to create YOLOv3.  
More improvements are made to the YOLO algorithm in \cite{bochkovskiy2020yolov4} and \cite{jocher2022yolov5}, creating YOLOv4 and YOLOv5 respectivey.  

This effort uses YOLOv5 of various sizes obtained from the original authors' GitHub repository.  
These model are (from smallest to largest): YOLOv5n, YOLOv5s, YOLOv5m, YOLOv5l, and YOLOv5x.  
As these YOLO models get larger, the performance improves and the speed decreases.
We create a sixth model, called YOLOv5p (the ``pico'' model) which is smaller than the YOLOv5n model and expected to run much faster.
For this work, we explore the signal detection and signal recognition performance of YOLOv5p, YOLOv5n, and YOLOv5s on the WBSig53 impaired dataset.

\subsubsection{Detection Transformer (DETR)}

The DEtection TRansformer (DETR) was introduced in \cite{carion2020detr} as a direct set prediction technique 
that effectively removes the hand-crafted anchors and non-maximal suppression procedures of previous object detection approaches.
DETR accomplishes its goal of streamlining the detection pipeline by employing a convolutional backbone that feeds a transformer encoder-decoder architecture
and by forcing unique predictions via a bipartite matching set-based global loss function.
The convolutional backbone reduces the input dimensionality to a learned 2D representation.
The learned representation is then flattened, supplemented with positional encodings, and then passed as input to the transformer encoder.
The transformer decoder receives the encoded representation and inputs $N$ learnable object queries, 
where $N$ represents a number larger than the maximum number of objects expected in any input.
The output of the transformer decoder is $N$ learned embeddings that are passed to $N$ multi-layered perceptron (MLP) prediction heads.
These prediction heads infer the class and bounding box of unique objects present within the input.

During training, the outputs of the prediction heads are compared to the target labels by using a Hungarian matching loss function \cite{kuhn1955hungarian}
that jointly measures a linear combination of the class loss and localization loss.
The Hungarian matching function forces unique pairs of predicted objects with target objects, 
where inputs with fewer than $N$ objects are padded to $N$ with $\phi$ (no object).
The class loss is a negative log-likelihood score between class prediction and targets.
The localization loss is a weighted sum of the commonly-used $l_1$ loss with a scale-invariant generalized IoU loss \cite{rezatofighi2019giou}.

In the original work, the DETR authors experimented with ResNet-50 and ResNet-101 backbones \cite{he2015deep} with a vanilla transformer \cite{vaswani2017attention}.
In our work, we adapt this architecture slightly by using EfficientNet backbones \cite{tan2019efficientnet} with XCiT transformers \cite{el2021xcit}.
By making these slight modifications, our networks are significantly more data efficient and run faster, 
primarily due to the XCiT transformer's linear complexity with respect to the input length.
Similar to YOLOv5, we experiment with three scales of this architecture.
We experiment with the following backbones: EfficientNet-B0, EfficientNet-B2, and EfficientNet-B4.
For all architectures, we leave the transformer constant with the XCiT-Nano scale.
We refer to these DETR architectures as: DETR-B0-Nano, DETR-B2-Nano, and DETR-B4-Nano.

\subsubsection{Pyramid Scheme Parsing Network (PSPNet)}

The Pyramid Scene Parsing Network (PSPNet) is a fully convolutional neural network for semantic segmentation tasks.
PSPNet consists of an encoder that compresses the input into a learned feature map and a decoder that is built of a pyramid pooling module.
The pyramid pooling module inspects and fuses features under four different scales before upsampling and concatenating to form the final feature representation,
which contains local and global contextual information.
As a final step, the representation is fed into a convolutional layer that outputs the pixel-by-pixel class predictions.

In the original work, the authors experiment with various scale ResNet encoders.
In our work, we adapt this architecture slightly by replacing the ResNet encoders with EfficientNet encoders.
Once more, we conduct our experiments using multiple scales of encoders with: EfficientNet-B0, EfficientNet-B2, and EfficientNet-B4.
We refer to these networks as: PSPNet-B0, PSPNet-B2, and PSPNet-B4, respectively.

\subsubsection{Mask2Former}

\textbf{Mask2Former:} The Masked-attention Mask Transformer (Mask2Former) is an architecture capable of performing semantic, instance, or panoptic segmentation.
At the time of its release, Mask2Former set SoTA performance across all three segmentation categories, outperforming models specialized in each category.
Mask2Former builds on the architecture introduced in MaskFormer \cite{cheng2021maskformer}.
This architecture consists of a backbone, a pixel decoder, and a transformer decoder.
Mask2Former improves upon the original MaskFormer architecture by introducing masked attention in the transformer decoder.
The masked attention modulates the standard cross-attention matrix 
by imposing a binarized mask of the previous transformer decoder layer's mask predictions to the current key-query pairs.
The effect of this masking is that the attention only occurs within the foreground region of the predicted mask for each query, rather than the full feature map.
Mask2Former also uses high-resolution features from the pixel decoder at multi-scales by feedings one scale to differing transformer decoder layers one at a time.
Additionally, Mask2Former inverts the order of the self and masked attention, makes the transformer decoder's queries learnable, and removes dropout for computational effectiveness.
All of these improvements over MaskFormer results in a flexible neural network that trains significantly faster with higher performance.

In the original work, the authors conduct experiments with ResNet and Swin-L \cite{liu2021swin} backbones pretrained on ImageNet-22K \cite{russakovsky2015imagenet}.
The pixel decoder in their experiments is a multi-scale deformable attention transformer (MSDeformAttn) \cite{zhu2020deformable}.
The transformer decoder is a vanilla transformer decoder with 9 layers, 100 queries, and auxiliary losses at each layer.
The loss function is a weighted combination of a classification loss and a mask loss, where the mask loss consists of a binary cross-entropy loss and a dice loss \cite{milletari2016dice}.

In our work, we explore the Mask2Former as implemented by the original authors; however, we modify the backbone.
To better mirror our other networks' backbones/encoders, we use various scale EfficientNets for the Mask2Former backbone.
Once again, we use EfficientNet-B0, EfficientNet-B2, and EfficientNet-B4, and we refer to these networks as Mask2Former-B0, Mask2Former-B2, and Mask2Former-B4, respectively.
Unlike the original authors, our backbones are not pretrained on any task prior to their use within the full Mask2Former architecture.